\documentclass[english]{article}
\usepackage{lmodern}

\usepackage[T1]{fontenc}
\usepackage[latin9]{inputenc}
\usepackage{color}
\usepackage{babel}
\usepackage{verbatim}
\usepackage{float}
\usepackage{amsthm}
\usepackage{amsmath}
\usepackage{amssymb}
\usepackage{graphicx}
\usepackage{setspace}
\usepackage{esint}
\usepackage[unicode=true,
 bookmarks=true,bookmarksnumbered=false,bookmarksopen=false,
 breaklinks=false,pdfborder={0 0 1},backref=false,colorlinks=true]
 {hyperref}
\hypersetup{pdftitle={Derivative Free Estimation of the Score Vector},
 pdfauthor={Doucet Jacob Rubenthaler}}

\makeatletter

  \theoremstyle{plain}
  \newtheorem{lem}{\protect\lemmaname}
\theoremstyle{plain}
\newtheorem{thm}{\protect\theoremname}
  \theoremstyle{remark}
  \newtheorem{rem}{\protect\remarkname}
 \theoremstyle{definition}
  \newtheorem{example}{\protect\examplename}

\usepackage{color}\usepackage{babel}
\usepackage{color}\usepackage{babel}
\usepackage{breakurl}\usepackage{color}\usepackage{babel}
\usepackage{breakurl}\usepackage{color}\usepackage{babel}
\usepackage{color}\usepackage{babel}
\usepackage{breakurl}\usepackage{color}\usepackage{babel}
\usepackage{breakurl}\usepackage{color}\usepackage{babel}
\usepackage{breakurl}\usepackage{color}\usepackage{babel}
\usepackage{breakurl}\usepackage{color}\usepackage{babel}
\usepackage{breakurl}\usepackage{color}\usepackage{babel}
\usepackage{breakurl}\usepackage{color}\usepackage{babel}
\usepackage{breakurl}\usepackage{color}\usepackage{babel}
\usepackage{color}\usepackage{babel}
\usepackage{breakurl}\usepackage{color}\usepackage{babel}
\usepackage{breakurl}\usepackage{color}\usepackage{babel}
\usepackage{color}\usepackage{color}\usepackage{amsfonts}\usepackage{vmargin}\usepackage{epstopdf}\usepackage{color}\usepackage{subfigure}\usepackage{babel}
\usepackage{babel}
\usepackage{babel}
\providecommand{\U}[1]{\protect\rule{.1in}{.1in}}

\@ifundefined{definecolor}{
}{}
\@ifundefined{definecolor}{
}{}
\@ifundefined{definecolor}{
}{}
\@ifundefined{definecolor}{
}{}
\@ifundefined{definecolor}{
}{}
\@ifundefined{definecolor}{
}{}
\@ifundefined{definecolor}{
}{}
\@ifundefined{definecolor}{
}{}
\@ifundefined{definecolor}{
}{}
\@ifundefined{definecolor}{
}{}
\@ifundefined{definecolor}{
}{}
\@ifundefined{definecolor}{
}{}
\@ifundefined{definecolor}{
}{}
\@ifundefined{definecolor}{
}{}
\@ifundefined{definecolor}{
}{}
\@ifundefined{definecolor}{
}{}
\@ifundefined{definecolor}{
}{}
\@ifundefined{definecolor}{
}{}
\@ifundefined{definecolor}{
}{}
\@ifundefined{definecolor}{
}{}
\@ifundefined{definecolor}{
}{}
\@ifundefined{definecolor}{
}{}
\@ifundefined{definecolor}{
}{}
\@ifundefined{definecolor}{
}{}
\@ifundefined{definecolor}{
}{}
\@ifundefined{definecolor}{
}{}
\@ifundefined{definecolor}{
}{}
\@ifundefined{definecolor}{
}{}
\@ifundefined{definecolor}{
}{}
\@ifundefined{definecolor}{
}{}
\@ifundefined{definecolor}{
}{}
\@ifundefined{definecolor}{
}{}
\@ifundefined{definecolor}{
}{}
\@ifundefined{definecolor}{
}{}
\@ifundefined{definecolor}{
}{}
\@ifundefined{definecolor}{
}{}
\@ifundefined{definecolor}{
}{}
\@ifundefined{definecolor}{
}{}
\@ifundefined{definecolor}{
}{}
\@ifundefined{definecolor}{
}{}
\@ifundefined{definecolor}{
}{}
\@ifundefined{definecolor}{
}{}
\@ifundefined{definecolor}{
}{}
\@ifundefined{definecolor}{
}{}
\@ifundefined{definecolor}{
}{}
\@ifundefined{definecolor}{
}{}
\@ifundefined{definecolor}{
}{}
\@ifundefined{definecolor}{
}{}
\@ifundefined{definecolor}{
}{}
\@ifundefined{definecolor}{
}{}
\@ifundefined{definecolor}{
}{}
\@ifundefined{definecolor}{
}{}
\@ifundefined{definecolor}{
}{}
\@ifundefined{definecolor}{
}{}
\@ifundefined{definecolor}{
}{}
\@ifundefined{definecolor}{
}{}
\@ifundefined{definecolor}{
}{}
\@ifundefined{definecolor}{
}{}
\@ifundefined{definecolor}{
}{}
\@ifundefined{definecolor}{
}{}
\@ifundefined{definecolor}{
}{}
\@ifundefined{definecolor}{
}{}
\@ifundefined{definecolor}{
}{}
\@ifundefined{definecolor}{
}{}
\@ifundefined{definecolor}{
}{}
\@ifundefined{definecolor}{
}{}
\@ifundefined{definecolor}{
}{}
\@ifundefined{definecolor}{
}{}
\@ifundefined{definecolor}{
}{}
\@ifundefined{definecolor}{
}{}
\@ifundefined{definecolor}{
}{}
\@ifundefined{definecolor}{
}{}
\@ifundefined{definecolor}{
}{}
\@ifundefined{definecolor}{
}{}
\@ifundefined{definecolor}{
}{}
\@ifundefined{definecolor}{
}{}
\@ifundefined{definecolor}{
}{}
\@ifundefined{definecolor}{
}{}
\@ifundefined{definecolor}{
}{}
\@ifundefined{definecolor}{
}{}
\@ifundefined{definecolor}{
}{}
\@ifundefined{definecolor}{
}{}
\@ifundefined{definecolor}{
}{}
\@ifundefined{definecolor}{
}{}
\@ifundefined{definecolor}{
}{}
\@ifundefined{definecolor}{
}{}
\@ifundefined{definecolor}{
}{}
\@ifundefined{definecolor}{
}{}
\@ifundefined{definecolor}{
}{}
\@ifundefined{definecolor}{
}{}
\@ifundefined{definecolor}{
}{}
\@ifundefined{definecolor}{
}{}
\@ifundefined{definecolor}{
}{}
\@ifundefined{definecolor}{
}{}
\@ifundefined{definecolor}{
}{}
\@ifundefined{definecolor}{
}{}
\@ifundefined{definecolor}{
}{}
\@ifundefined{definecolor}{
}{}
\@ifundefined{definecolor}{
}{}
\@ifundefined{definecolor}{
}{}
\@ifundefined{definecolor}{
}{}
\@ifundefined{definecolor}{
}{}
\@ifundefined{definecolor}{
}{}
\@ifundefined{definecolor}{
}{}
\@ifundefined{definecolor}{
}{}
\@ifundefined{definecolor}{
}{}
\@ifundefined{definecolor}{\@ifundefined{definecolor}
{
}{}
}{}

\setmarginsrb{2cm}{2cm}{2cm}{3cm}{0cm}{0cm}{0cm}{1cm}

\makeatother

  \providecommand{\examplename}{Example}
  \providecommand{\lemmaname}{Lemma}
  \providecommand{\remarkname}{Remark}
\providecommand{\theoremname}{Theorem}

\begin{document}

\title{Derivative-Free Estimation of the Score Vector and Observed Information
Matrix with Application to State-Space Models}

\author{Arnaud Doucet$^{1}$, Pierre E. Jacob$^{2}$ and Sylvain Rubenthaler$^{3}$\\
$^{1}$Department of Statistics, University of Oxford, UK.\\
$^{2}$Department of Statistics, Harvard University, USA.\\
$^{3}$Département de Mathématiques, Université de Nice-Sophia Antipolis,
France.}
\maketitle
\begin{abstract}
Ionides et al. \cite{ionides2006,Ionides09} have recently introduced
an original approach to perform maximum likelihood parameter estimation
in state-space models which only requires being able to simulate the
latent Markov model according to its prior distribution. Their methodology
relies on an approximation of the score vector for general statistical
models based upon an artificial posterior distribution and bypasses
the calculation of any derivative. We show here that this score estimator
can be derived from a simple application of Stein's lemma and how
an additional application of this lemma provides an original derivative-free
estimator of the observed information matrix. We establish that these
estimators exhibit robustness properties compared to finite difference
estimators while their bias and variance scale as well as finite difference
type estimators, including simultaneous perturbations \cite{spall1992,spall2005},
with respect to the dimension of the parameter. For state-space models
where sequential Monte Carlo computation is required, these estimators
can be further improved. In this specific context, we derive original
derivative-free estimators of the score vector and observed information
matrix which are computed using sequential Monte Carlo approximations
of smoothed additive functionals associated with a modified version
of the original state-space model.

\end{abstract}

\section{Introduction\label{sec:Introduction}}

Consider a statistical model with parameter $\theta=\left(\theta_{1},\ldots,\theta_{d}\right)\in\mathbb{R}^{d}$
and likelihood function $\theta\mapsto\mathcal{L}(\theta)$, the dependence
of $\mathcal{L}(\theta)$ upon the observations being omitted from
the notation. Assuming that the corresponding log-likelihood function
$\theta\mapsto\ell(\theta)$ is twice differentiable, we are here
interested in calculating at a given parameter value $\theta^{\star}$
the score vector $\nabla\ell(\theta^{\star})$ and the observed information
matrix $-\nabla^{2}\ell(\theta^{\star})$ whose $r^{\mathrm{th}}$
component $\nabla_{r}\ell(\theta^{\star})$ and $(r,s)^{\mathrm{th}}$
component $-\nabla_{rs}^{2}\ell(\theta^{\star})$ are given for $r,s=1,\ldots$$,d$
by 
\begin{equation}
\nabla_{r}\ell(\theta^{\star})=\frac{\text{\ensuremath{\partial}}\ell(\theta^{\star})}{\text{\ensuremath{\partial\theta_{r}}}}\quad\mbox{ and }\quad-\nabla_{rs}^{2}\ell(\theta^{\star})=-\frac{\text{\ensuremath{\partial}}^{2}\ell(\theta^{\star})}{\partial\theta_{r}\partial\theta_{s}}.\label{eq:scoreobservedinformationmatrix}
\end{equation}

The score vector and observed information matrix are useful both algorithmically
and statistically. Algorithmically, they can be used to build efficient
maximum likelihood estimation techniques as in \cite{ionides2006,Ionides09}
or to build efficient Markov chain Monte Carlo proposals relying on
the local geometry of the target distribution \cite{Martin2012}.
Statistically, the observed information matrix can be used to estimate
the variance of the maximum likelihood estimate \cite{Efron1978}.

Exact calculations of the score vector and observed information matrix
are only possible for models where $\ell(\theta)$ can be evaluated
exactly for any $\theta$. For complex latent variable models, these
quantities are typically computed using Monte Carlo approximations
of the Fisher and Louis identities \cite{douc2014,poya2010}. However
there are many important scenarios where this is not a viable option.
For numerous state-space models arising in applied science, we are
able to obtain sample paths from the latent Markov process but we
have neither access to the expression of its transition kernel nor
of its derivatives \cite{ionides2006,Ionides09}. This prohibits the
numerical implementation of the Fisher and Louis identities. It is
thus useful to develop estimators of the score vector and observed
information matrix which, beyond the specification of the statistical
model, require a minimum amount of input from the user. These estimators
should be competitive with finite difference (FD) estimators \cite[chap. 7]{Asmuss}
and sophisticated variants such as simultaneous perturbation (SP)
estimators \cite{spall1992,spall2005} which have found numerous applications
in high-dimensional stochastic optimization.

For the score vector, an alternative to FD estimators has been recently
proposed in \cite{ionides2006,Ionides09}. The main idea of the authors
is to introduce an artificial random parameter $\Theta$ with prior
centered around $\theta^{\star}$. They establish that the expectation
of $\Theta-\theta^{\star}$ with respect to the posterior associated
to this prior and the likelihood function $\mathcal{L}(\theta)$ has
components approximately proportional to the components of $\nabla\ell(\theta^{\star})$;
the approximation improving as the artificial prior shrinks around
$\theta^{\star}$. In a state-space context where sequential Monte
Carlo approximations are required, the direct application of this
idea provides a high variance estimator. The authors propose a lower
variance estimator which is computed using the optimal filter associated
to a modified version of the original state-space model where an artificial
random walk dynamics initialized at the parameter $\theta^{\star}$
is introduced.

In this paper, our contributions are three-fold. First, we show in
Section \ref{sec:derivativefree:generalcase} how the score estimator
proposed in \cite{ionides2006,Ionides09} can be derived using a simple
application of Stein's lemma \cite[Lemma 1]{Stein1981} when the artificial
prior on $\Theta$ is normal. Moreover, an additional application
of this lemma provides a novel estimator of the observed information
matrix, which is a simple function of the covariance of $\Theta$
under the artificial posterior. Second, we establish in Section \ref{sec:Monte-Carlo-Shift-Properties}
various theoretical results for the Monte Carlo approximation of these
estimators. In particular, we show that their bias and variance scale
similarly as SP type estimators with respect to the parameter dimension
$d$. Additionally, they exhibit robustness properties compared to
FD and SP estimators which are of significant practical interest.
Third, in the specific context of state-space models, we propose in
Section \ref{sec:Latent-variable-models} original estimators of the
score vector and observed information matrix.%

All proofs are postponed to the appendix.

\section{Derivative-free estimators of the score vector and observed information
matrix \label{sec:derivativefree:generalcase}}

\subsection{Notation}

The multivariate normal distribution with mean $\mu$ and covariance
$\Sigma$ is denoted by $\mathcal{N}\left(\mu,\Sigma\right)$, and
its probability density function is denoted $x\mapsto\mathcal{N}\left(x;\mu,\Sigma\right)$.
The $(i,j)$-th element of a $d\times d$ matrix $\Sigma$ is denoted
$\Sigma_{ij}$. The $i$-th column (respectively, row) of $\Sigma$
is denoted by $\Sigma_{\centerdot i}$ (respectively, $\Sigma_{i\centerdot}$).
For a differentiable function $f:\,\mathbb{R}^{d}\to\mathbb{R}$ and
for all $\theta\in\mathbb{R}^{d}$, we note $\nabla f(\theta)=(\partial f(\theta)/\partial\theta_{1},\ldots,\partial f(\theta)/\partial\theta_{d})^{T}$
the column-vector of partial first order derivatives evaluated at
$\theta$, and its $i$-th element is denoted by $\nabla_{i}f(\theta)$.
Similarly, we denote by $\nabla^{2}f(\theta)$ the $d\times d$ matrix
of partial second order derivatives, i.e. its $(i,j)$-th element
is $\partial^{2}f(\theta)/\partial\theta_{i}\partial\theta_{j}$,
also denoted by $\nabla_{ij}^{2}f(\theta)$. Similar notation is used
for higher-order derivatives. For a vector $\theta\in\mathbb{R}^{d}$
(or a random variable $\Theta$ in $\mathbb{R}^{d}$), we denote by
$\theta_{i}$ (and by $\Theta_{i}$) its $i$-th element; sometimes
we will also write $\left\{ \theta\right\} _{i}$. Vectors are understood
as columns. We introduce the basis vectors $\left(e_{1},\ldots,e_{d}\right)$
in $\mathbb{R}^{d}$, where the only non-zero element of $e_{i}$
is a ``$1$'' at the $i$-th position. The Euclidean norm of a $d$-dimensional
vector $\theta$ is denoted $\left|\left|\theta\right|\right|_{2}$.
Expectations are denoted by $\mathbb{E}$, variances by $\mathbb{V}$
and covariances by $\mathbb{C}$, hence we have $\mathbb{C}\left[X,X\right]=\mathcal{\mathbb{V}}\left[X\right]$
for any random vector $X$ .

\subsection{Stein's lemma and expected derivatives of the log-likelihood}

Stein's lemma \cite[Lemma 1]{Stein1981}, as described in the multivariate
setting in \cite[Lemma 1]{liu1994}, states that, for a $\mathbb{R}^{d}$-valued
normal random variable $\Theta\sim\mathcal{N}\left(\theta^{\star},\Sigma\right)$
with $\Sigma$ positive definite, we have 
\begin{equation}
\mathbb{E}\left[f(\Theta)\left(\Theta-\theta^{\star}\right)\right]=\Sigma\,\mathbb{E}\left[\nabla f(\Theta)\right]\label{eq:SteinMultivariate}
\end{equation}
for any differentiable function $f:\,\mathbb{R}^{d}\to\mathbb{R}$
such that $\mathbb{E}\left[\left|\nabla_{i}f\left(\Theta\right)\right|\right]<\text{\ensuremath{\infty}}$
for all $i\in\left\{ 1,\ldots,d\right\} $. Component-wise, this expression
reads:
\begin{equation}
\mathbb{E}\left[f(\Theta)\left(\Theta_{i}-\theta_{i}^{\star}\right)\right]=\sum_{j=1}^{d}\Sigma_{ij}\,\mathbb{E}\left[\nabla_{j}f\left(\Theta\right)\right].\label{eq:SteinMultivariate:componentwise}
\end{equation}

Let $\mathcal{L}:\,\mathbb{R}^{d}\rightarrow\mathbb{R}^{+}$ be a
likelihood function and $\ell$ the corresponding log-likelihood.
Assume that $\mathcal{L}$ is differentiable, $\mathcal{Z}=\mathbb{E}\left[\mathcal{L}(\Theta)\right]<\text{\ensuremath{\infty}}$
and that $\mathbb{E}\left[\left|\nabla_{i}\mathcal{L}\left(\Theta\right)\right|\right]<\text{\ensuremath{\infty}}$
for all $i\in\left\{ 1,\ldots,d\right\} $. By considering the function
$f:\theta\mapsto\mathcal{L}(\theta)/\mathcal{Z}$ and applying Eq.
\eqref{eq:SteinMultivariate} we obtain:
\begin{equation}
\mathbb{E}\left[\left(\Theta-\theta^{\star}\right)\frac{\mathcal{L}\left(\Theta\right)}{\mathcal{Z}}\right]=\Sigma\,\mathbb{\mathbb{E}}\left[\frac{\nabla\mathcal{L}\left(\Theta\right)}{\mathcal{Z}}\right].\label{eq:posteriorshift1}
\end{equation}
This identity has a Bayesian interpretation. If we denote by $\check{\mathbb{E}}$
expectations with respect to the ``posterior'' distribution induced
by the ``prior'' $\mathcal{N}\left(\theta^{\star},\Sigma\right)$
and the likelihood function $\mathcal{L}$, then
\begin{equation}
\check{\mathbb{E}}\left[\varphi\left(\Theta\right)\right]=\frac{\mathbb{E}\left[\varphi\left(\Theta\right)\mathcal{L}(\Theta)\right]}{\mathbb{E}\left[\mathcal{L}(\Theta)\right]}\label{eq:bayesformula}
\end{equation}
for all test functions $\varphi$ such that this expectation is finite.
Hence we can rewrite Eq. \eqref{eq:posteriorshift1} as 
\begin{equation}
\check{\mathbb{E}}\left[\Theta-\theta^{\star}\right]=\Sigma\,\check{\mathbb{E}}\left[\nabla\ell(\Theta)\right].\label{eq:posteriorshift2}
\end{equation}
Pursuing the Bayesian analogy, this equation is a relationship between
the score and the shift of the prior mean $\theta^{\star}$ to the
posterior mean $\check{\mathbb{E}}\left[\Theta\right]$, when the
prior is normal.

We can similarly obtain a formula relating the second posterior moment
to the second order derivative of the log-likelihood. Assume now that
$\mathcal{L}$ is twice differentiable, and that $\mathbb{E}\left[\left|\nabla_{ij}^{2}\mathcal{L}\left(\Theta\right)\right|\right]<\infty$
for all $i,j\in\left\{ 1,\ldots,d\right\} $. We first apply Eq. \eqref{eq:SteinMultivariate}
to the function $\theta\mapsto\left(\theta_{i}-\theta_{i}^{\star}\right)\mathcal{L}(\theta)/\mathcal{Z}$,
for $i\in\left\{ 1,\ldots,d\right\} $, leading to 
\begin{eqnarray}
\mathbb{E}\left[\left(\Theta-\theta^{\star}\right)\left(\Theta_{i}-\theta_{i}^{\star}\right)\frac{\mathcal{L}(\Theta)}{\mathcal{Z}}\right] & = & \Sigma\,\mathbb{E}\left[\frac{\mathcal{L}(\Theta)}{\mathcal{Z}}e_{i}+\left(\Theta_{i}-\theta_{i}^{\star}\right)\nabla\ell(\Theta)\frac{\mathcal{L}\left(\Theta\right)}{\mathcal{Z}}\right],\nonumber \\
\text{i.e.}\quad\check{\mathbb{E}}\left[\left(\Theta-\theta^{\star}\right)\left(\Theta_{i}-\theta_{i}^{\star}\right)\right] & = & \Sigma\,e_{i}+\Sigma\,\mathbb{E}\left[\left(\Theta_{i}-\theta_{i}^{\star}\right)\nabla\ell(\Theta)\frac{\mathcal{L}\left(\Theta\right)}{\mathcal{Z}}\right].\label{eq:intermediate:posteriorshift}
\end{eqnarray}
The first term on the right hand side is $\Sigma e_{i}=\Sigma_{\centerdot i}$,
i.e. the $i$-th column of $\Sigma$. For the second term, we apply
Eq. \eqref{eq:SteinMultivariate:componentwise} with $g_{j}:\theta\mapsto\nabla_{j}\ell\left(\theta\right)\times\mathcal{L}\left(\theta\right)/\mathcal{Z}$,
for $j\in\left\{ 1,\ldots,d\right\} $. We obtain, for each $i,j\in\left\{ 1,\ldots,d\right\} $,
\[
\mathbb{E}\left[\left(\Theta_{i}-\theta_{i}^{\star}\right)\nabla_{j}\ell(\Theta)\frac{\mathcal{L}\left(\Theta\right)}{\mathcal{Z}}\right]=\sum_{k=1}^{d}\Sigma_{ik}\,\mathbb{E}\left[\nabla_{jk}^{2}\ell(\Theta)\frac{\mathcal{L}\left(\Theta\right)}{\mathcal{Z}}+\nabla_{j}\ell(\Theta)\nabla_{k}\ell(\Theta)\frac{\mathcal{L}\left(\Theta\right)}{\mathcal{Z}}\right],
\]
which can also be written
\[
\mathbb{E}\left[\left(\Theta_{i}-\theta_{i}^{\star}\right)\nabla_{j}\ell(\Theta)\frac{\mathcal{L}\left(\Theta\right)}{\mathcal{Z}}\right]=\check{\mathbb{E}}\left[D_{j\centerdot}(\Theta)\right]\,\Sigma_{\centerdot i},
\]
where $D(\theta)$ is the matrix $\nabla^{2}\ell(\theta)+\nabla\ell(\theta)\nabla\ell(\theta)^{T}$,
and where we have used the symmetry of $\Sigma$, that is $\Sigma_{\centerdot i}=\Sigma_{i\centerdot}$.
Performing this calculation for each $j\in\left\{ 1,\ldots,d\right\} $,
and stacking the results line by line, we obtain for each $i\in\left\{ 1,\ldots,d\right\} $,
\[
\mathbb{E}\left[\left(\Theta_{i}-\theta_{i}^{\star}\right)\nabla\ell(\Theta)\frac{\mathcal{L}\left(\Theta\right)}{\mathcal{Z}}\right]=\check{\mathbb{E}}\left[D(\Theta)\right]\,\Sigma_{\centerdot i},
\]
and thus, plugging this expression in Eq. \eqref{eq:intermediate:posteriorshift},
\[
\check{\mathbb{E}}\left[\left(\Theta-\theta^{\star}\right)\left(\Theta_{i}-\theta_{i}^{\star}\right)\right]=\Sigma_{\centerdot i}+\Sigma\,\check{\mathbb{E}}\left[D(\Theta)\right]\,\Sigma_{\centerdot i}.
\]
Finally, performing this calculation for each $i\in\left\{ 1,\ldots,d\right\} $,
and stacking the results column by column, we obtain the following
lemma that summarizes the results of this section.
\begin{lem}
\label{lemma:stein} Consider an \textup{$\mathbb{R}^{d}$-valued}
normal random variable $\Theta\sim\mathcal{N}\left(\theta^{\star},\Sigma\right)$
where $\Sigma$ is positive definite.%
{} Let $\mathcal{L}:\,\mathbb{R}^{d}\rightarrow\mathbb{R}^{+}$ be a
twice differentiable likelihood function, with logarithm $\ell$,
and assume that $\mathbb{E}\left[\left|\nabla_{i}\mathcal{L}\left(\Theta\right)\right|\right]<\infty$
and $\mathbb{E}\left[\left|\nabla_{ij}^{2}\mathcal{L}\left(\Theta\right)\right|\right]<\infty$
for $i,j\in\left\{ 1,\ldots,d\right\} $. The following identities
between posterior moments and derivatives of $\ell$ hold:
\begin{eqnarray*}
\check{\mathbb{E}}\left[\Theta-\theta^{\star}\right] & = & \Sigma\,\check{\mathbb{E}}\left[\nabla\ell(\Theta)\right],\\
\check{\mathbb{E}}\left[\left(\Theta-\theta^{\star}\right)\left(\Theta-\theta^{\star}\right)^{T}\right] & = & \Sigma+\Sigma\,\check{\mathbb{E}}\left[\nabla^{2}\ell(\Theta)+\nabla\ell(\Theta)\nabla\ell(\Theta)^{T}\right]\,\Sigma.
\end{eqnarray*}

\end{lem}

\subsection{Posterior expectations when the prior concentrates}

We now relate the posterior expectations $\check{\mathbb{E}}\left[\nabla\ell(\Theta)\right]$
and $\check{\mathbb{E}}\left[\nabla^{2}\ell(\Theta)\right]$ appearing
in Lemma \ref{lemma:stein} to $\nabla\ell(\theta^{\star})$ and $\nabla^{2}\ell(\theta^{\star})$.
We prove here that $\check{\mathbb{E}}\left[\nabla\ell\left(\Theta\right)\right]\approx\nabla\ell\left(\theta^{\star}\right)$
and $\check{\mathbb{E}}\left[\nabla^{2}\ell\left(\Theta\right)\right]\approx\nabla^{2}\ell\left(\theta^{\star}\right)$,
when the prior distribution concentrates around $\theta^{\star}$.
More precisely, we make the following assumptions.
\begin{itemize}
\item \textbf{A1}. The prior distribution is $\Theta\sim\mathcal{N}\left(\theta^{\star},\tau^{2}\Sigma\right)$
where $\theta^{\star}\in\mathbb{R}^{d}$, $\Sigma\in\mathbb{R}^{d\times d}$
is positive definite and $\tau>0$. We denote by $\mathbb{E}_{\tau}$
expectations (resp. $\mathbb{V}_{\tau}$ variances and $\mathbb{C}_{\tau}$
covariances) with respect to this prior distribution, $\check{\mathbb{E}}_{\tau}$
expectations (resp. $\check{\mathbb{V}}_{\tau}$ variances and $\check{\mathbb{C}}_{\tau}$
covariances) with respect to the corresponding posterior. Let $\Sigma^{-1/2}$
be a matrix such that $\Sigma^{-1/2}\left(\Sigma^{-1/2}\right)^{T}=\Sigma^{-1}$.
Let $B_{\Sigma}(\theta^{\star},\delta)=\left\{ \theta\in\mathbb{R}^{d}:\left|\left|\Sigma^{-1/2}\left(\theta-\theta^{\star}\right)\right|\right|_{2}\leq\delta\right\} $,
a level set of the prior distribution. 
\item \textbf{A2}. Let $\varphi:\mathbb{R}^{d}\to\mathbb{R}$ be a four
times continuously differentiable function. Assume that there exists
a constant $K<\infty$ and $\delta>0$ such that $\left|\nabla_{ijkl}^{4}\varphi(\theta)\right|\leq K$
for all $\theta\in B_{\Sigma}(\theta^{\star},\delta)$, all $i,j,k,l\in\left\{ 1,\ldots,d\right\} $.
We assume that the likelihood function $\mathcal{L}$ is such that
both $\theta\mapsto\mathcal{L}\left(\theta\right)$ and $\theta\mapsto\varphi\left(\theta\right)\times\mathcal{L}\left(\theta\right)$
satisfy the same assumption as $\varphi$.
\item \textbf{A3}. There exists $\tau_{\text{0}}>0$ such that the test
function $\varphi:\mathbb{R}^{d}\to\mathbb{R}$ satisfies $\mathbb{E}_{\tau_{0}}\left[\left|\varphi\left(\Theta\right)\right|\right]<\infty$,
$\mathbb{E}_{\tau_{0}}\left[\mathcal{L}\left(\Theta\right)\right]<\infty$
and $\mathbb{E}_{\tau_{0}}\left[\left|\varphi\left(\Theta\right)\right|\mathcal{L}\left(\Theta\right)\right]<\infty.$
\end{itemize}
The following lemma explains how posterior expectations behave when
$\tau\rightarrow0$, that is, when the prior distribution concentrates.
\begin{lem}
\label{lemma:posterior:expansion}Assume \textbf{A1-A2-A3} hold, then
we have:
\[
\check{\mathbb{E}}_{\tau}\left[\varphi\left(\Theta\right)\right]=\varphi\left(\theta^{\star}\right)+\frac{\tau^{2}}{2}\sum_{i=1}^{d}\sum_{j=1}^{d}\left(\nabla_{ij}^{2}\varphi\left(\theta^{\star}\right)+2\nabla_{i}\varphi\left(\theta^{\star}\right)\nabla_{j}\ell\left(\theta^{\star}\right)\right)\Sigma_{ij}+\mathcal{O}\left(\tau^{4}\right).
\]

\end{lem}
The proof is given in Section \ref{sub:proof:posteriorexpansions},
and relies on an expansion of prior moments given in Section \ref{sub:proof:prior:expansions}.

\subsection{Derivative-free estimators using posterior moments}

The combination of Lemmas \ref{lemma:stein} and \ref{lemma:posterior:expansion}
leads to approximations of the first two derivatives of the log-likelihood
at any point $\theta^{\star}$. Henceforth, we refer to these approximations
as the shift estimators.
\begin{thm}
\label{theorem:estimators} Assume \textbf{A1-A2-A3} hold whenever
the test function $\varphi$ is defined as $\theta\mapsto\nabla_{i}\ell(\theta)$,
$\theta\mapsto\nabla_{i}\ell(\theta)\nabla_{j}\ell(\theta)$ or $\theta\mapsto\nabla_{ij}^{2}\ell(\theta)$,
for any $i,j\in\left\{ 1,\ldots,d\right\} $. Then we have the following
approximations of the first two derivatives of the log-likelihood
$\ell$: 
\begin{eqnarray}
S_{\tau}^{(1)}\left(\theta^{\star}\right)=\tau^{-2}\Sigma^{-1}\,\check{\mathbb{E}}_{\tau}\left[\Theta-\theta^{\star}\right] & = & \nabla\ell(\theta^{\star})+\tau^{2}\mathcal{E}(\theta^{\star})+\mathcal{O}\left(\tau^{4}\right),\label{eq:score:approx}\\
S_{\tau}^{(2)}\left(\theta^{\star}\right)=\tau^{-4}\Sigma^{-1}\left(\check{\mathbb{V}}_{\tau}\left[\Theta\right]-\tau^{2}\Sigma\right)\Sigma^{-1} & = & \nabla^{2}\ell(\theta^{\star})+\tau^{2}\mathcal{F}(\theta^{\star})+\mathcal{O}\left(\tau^{4}\right),\label{eq:hessian:approx}
\end{eqnarray}
where $\mathcal{E}(\theta^{\star})$ is a $d$-dimensional vector
with $k$-th component defined by
\begin{equation}
\mathcal{E}_{k}(\theta^{\star})=\frac{1}{2}\sum_{i=1}^{d}\sum_{j=1}^{d}\left(\nabla_{ijk}^{3}\ell\left(\theta^{\star}\right)+2\nabla_{ik}^{2}\ell\left(\theta^{\star}\right)\nabla_{j}\ell\left(\theta^{\star}\right)\right)\Sigma_{ij},\label{eq:score:approx:error}
\end{equation}
and $\mathcal{F}(\theta^{\star})$ is a $d\times d$ matrix with $(k,l)$-th
component defined by 
\begin{align}
\mathcal{F}_{kl}(\theta^{\star})= & \frac{1}{2}\sum_{i=1}^{d}\sum_{j=1}^{d}\biggl(\nabla_{ijkl}^{4}\ell\left(\theta^{\star}\right)+2\nabla_{ikl}^{3}\ell\left(\theta^{\star}\right)\nabla_{j}\ell\left(\theta^{\star}\right)\label{eq:hessian:approx:error}\\
 & +\nabla_{ik}^{2}\ell\left(\theta^{\star}\right)\left(\nabla_{jl}^{2}\ell\left(\theta^{\star}\right)-2\nabla_{j}\ell\left(\theta^{\star}\right)\nabla_{l}\ell\left(\theta^{\star}\right)\right)+\nabla_{il}^{2}\ell\left(\theta^{\star}\right)\left(\nabla_{jk}^{2}\ell\left(\theta^{\star}\right)-2\nabla_{j}\ell\left(\theta^{\star}\right)\nabla_{k}\ell\left(\theta^{\star}\right)\right)\biggr)\Sigma_{ij}.\nonumber 
\end{align}

\end{thm}
The proof is given in Section \ref{sub:proof:estimators}. The expression
of these estimators pave the way to Monte Carlo approximations of
$\nabla\ell(\theta^{\star})$ and $\nabla^{2}\ell(\theta^{\star})$.
Indeed, if we can sample from the artificial posterior using a Monte
Carlo scheme such as Markov chain Monte Carlo or sequential Monte
Carlo, then Theorem \ref{theorem:estimators} states that we can approximate
the derivatives of the log-likelihood point-wise.

The results of Theorem \ref{theorem:estimators} could be obtained
for other prior distributions than the normal distribution. For example,
a bound in $\mathcal{O}\left(\tau\right)$ was established in \cite{Ionides09}
for the score estimator $S_{\tau}^{(1)}\left(\theta^{\star}\right)$,
for a broader class of priors. Furthermore, Theorem \ref{theorem:estimators}
is closely related to the asymptotic behavior of posterior moments,
which have been extensively studied \cite{Johnson70,gosh2003}. Usually
the number of observations goes to infinity, and the prior density
is fixed. Here the likelihood is fixed and the prior concentrates
in a deterministic manner, which allows a much simpler proof. Theorem
\ref{theorem:estimators} could also be extended to any higher order
derivative.
\begin{rem}
In a previous version of this report available on arXiv, we have established
bounds in $\mathcal{O}\left(\tau^{2}\right)$ for $S_{\tau}^{(1)}\left(\theta^{\star}\right)$
and $S_{\tau}^{(2)}\left(\theta^{\star}\right)$ for a larger class
of non-normal prior distributions. However the proofs are much more
intricate as we could not rely on Stein's lemma. As the prior is here
introduced for purely computational reasons, the normal assumption
is not restrictive but might require reparametrizing the model.
\end{rem}
\medskip{}

\begin{rem}
We note that there are connections between the first shift estimator
$S_{\tau}^{(1)}\left(\theta^{\star}\right)$ and proximal optimization
\cite{Moreau1962,Parikh2013}. For a function $\theta\mapsto\ell(\theta)$,
define for some $\gamma>0$ and some $\theta^{\star}\in\mathbb{R}^{d}$,
\[
\text{prox}_{\gamma}(\theta^{\star})=\text{argmax}_{u\in\mathbb{R}^{d}}\exp\left(\ell(u)-\frac{1}{2\gamma}\left|\left|u-\theta^{\star}\right|\right|_{2}^{2}\right).
\]
It is well known \cite{Parikh2013} that, under some regularity assumptions,
if $\nabla\ell(\theta^{\star})$ exists then 
\[
\mathcal{P}_{\gamma}(\theta^{\star})=\frac{\text{prox}_{\gamma}(\theta^{\star})-\theta^{\star}}{\gamma}\xrightarrow[\gamma\to0]{}\nabla\ell(\theta^{\star}).
\]
Therefore $\mathcal{P}_{\gamma}(\theta^{\star})$ is sometimes used
as a surrogate for $\nabla\ell(\theta^{\star})$, for instance in
optimization techniques, when $\nabla\ell(\theta^{\star})$ itself
is not available or not defined. The object $\mathcal{P}_{\gamma}(\theta^{\star})$
can be interpreted as a rescaled shift between the \emph{maximum a
posteriori} and the \emph{maximum a priori}, under a normal prior
centered at $\theta^{\star}$ and with diagonal covariance matrix
with diagonal elements equal to $\gamma$. When the prior concentrates
$(\gamma\to0$), the posterior becomes closer to a normal distribution
and thus considering the posterior mean or the \emph{maximum a posteriori}
does not make a difference, thus $\mathcal{P}_{\gamma}(\theta^{\star})$
and $S_{\tau}^{(1)}(\theta^{\star})$ behave very similarly.\end{rem}
\begin{example}
\label{example:gaussian} Consider a scenario where $\theta$ represents
a location parameter, and the observation $Y$ follows $\mathcal{N}(\theta,\Lambda_{y}^{-1})$,
for a fixed precision matrix $\Lambda_{y}$. The derivatives of the
log-likelihood at any $\theta^{\star}$ are
\[
\nabla\ell(\theta^{\star})=-\Lambda_{y}(\theta^{\star}-y)\quad\text{and}\quad\nabla^{2}\ell(\theta^{\star})=-\Lambda_{y}.
\]
Using a prior $\mathcal{N}\left(\theta^{\star},\tau^{2}\Sigma\right)$,
the posterior is normal:
\[
\Theta\mid\left(Y=y\right)\,\sim\,\mathcal{N}\left((\tau^{-2}\Sigma^{-1}+\Lambda_{y})^{-1}\left(\tau^{-2}\Sigma^{-1}\theta^{\star}+\Lambda_{y}y\right),\left(\tau^{-2}\Sigma^{-1}+\Lambda_{y}\right)^{-1}\right)
\]
 and the shift estimators are given by
\begin{align*}
S_{\tau}^{(1)}\left(\theta^{\star}\right) & =\tau^{-2}\Sigma^{-1}(\tau^{-2}\Sigma^{-1}+\Lambda_{y})^{-1}\left(-\Lambda_{y}(\theta^{\star}-y)\right),\\
S_{\tau}^{(2)}\left(\theta^{\star}\right) & =\tau^{-2}\Sigma^{-1}(\tau^{-2}\Sigma^{-1}+\Lambda_{y})^{-1}\left(-\Lambda_{y}\right).
\end{align*}
We see that they converge to $\nabla\ell(\theta^{\star})$ and $\nabla^{2}\ell(\theta^{\star})$,
respectively, when $\tau\to0$, and that the error is in $\mathcal{O}\left(\tau^{2}\right)$.
\end{example}
Note that the terminology of likelihood function, score vector, observed
information matrix and Bayesian inference is used to build up some
intuition, but that the results presented in this article are actually
generic and could be applied to any function $\ell$, for which we
would like to approximate the first and second derivatives.

\section{Monte Carlo shift estimators\label{sec:Monte-Carlo-Shift-Properties}}

In this section we consider Monte Carlo approximations of the shift
estimators defined in Theorem \ref{theorem:estimators}, which we
call Monte Carlo shift estimators. After introducing them, we proceed
to studying some of their properties and compare them to finite difference
(FD) type estimators, including simultaneous perturbations (SP).

\subsection{Monte Carlo shift estimators and finite difference type estimators
\label{sub:Monte-Carlo-Shift-Estimators}}

Assume that we have access to Monte Carlo estimators $\widehat{\mathcal{L}}(\theta)$
of $\mathcal{L}\left(\theta\right)$ for all $\theta\in\mathbb{R}^{d}$,
such that $\mathbb{E}[\widehat{\mathcal{L}}\left(\theta\right)]=\mathcal{L}(\theta)$
and $\mathbb{V}[\widehat{\mathcal{L}}(\theta)/\mathcal{L}(\theta)]=\upsilon_{M}(\theta)$,
for a function $\upsilon_{M}:\,\mathbb{R}^{d}\to\mathbb{R}^{+}$,
and a tuning parameter $M$ such that $\upsilon_{M}(\theta)\to0$
when $M\to\infty$, for all $\theta$; here expectation and variance
are with respect to the distribution of the likelihood estimator $\widehat{\mathcal{L}}(\theta)$,
the parameter value $\theta$ being fixed. We will assume that $\upsilon_{M}\left(\theta\right)$
is constant for all $\theta$ around $\theta^{\star}$, and equal
to $\upsilon_{M}\left(\theta^{\star}\right)$, which is reasonable
on a small neighborhood around any particular value of $\theta^{\star}$.
We consider the following procedure. Let $N\in\mathbb{N}$. First,
draw $\theta^{i}$ from $\mathcal{N}\left(\theta^{\star},\tau^{2}\Sigma\right)$
and $\hat{w}^{i}=\widehat{\mathcal{L}}\left(\theta^{i}\right)$ for
$i\in\left\{ 1,\ldots,N\right\} $. Then normalize the weights, by
defining $\hat{W}^{i}=\hat{w}^{i}/\sum_{j=1}^{N}\hat{w}^{j}$ for
each $i\in\left\{ 1,\ldots,N\right\} $. Finally, return

\begin{equation}
S_{N,\tau}^{(1)}\left(\theta^{\star}\right)=\tau^{-2}\Sigma^{-1}\left(\sum_{i=1}^{N}\hat{W}^{i}\theta^{i}-\theta^{\star}\right).\label{eq:first:shift:estimator}
\end{equation}
The estimator $S_{N,\tau}^{(1)}\left(\theta^{\star}\right)$ is a
normalized importance sampling estimator of $S_{\tau}^{(1)}(\theta^{\star})$
defined in Eq. \eqref{eq:score:approx}, using the prior distribution
as importance proposal, and random importance weights obtained by
approximating the likelihood $\mathcal{L}\left(\theta\right)$ by
$\widehat{\mathcal{L}}\left(\theta\right)$. This is an instance of
importance sampling squared \cite{Tran2013}. For the second order
derivative, we can similarly consider the following approximation
of $S_{\tau}^{(2)}(\theta^{\star})$ defined in Eq. \eqref{eq:hessian:approx},

\begin{eqnarray}
S_{N,\tau}^{(2)}\left(\theta^{\star}\right) & = & \tau^{-4}\Sigma^{-1}\left(\sum_{i=1}^{N}\hat{W}^{i}\left(\theta^{i}-\sum_{i=1}^{N}\hat{W}^{j}\theta^{j}\right)\left(\theta^{i}-\sum_{i=1}^{N}\hat{W}^{j}\theta^{j}\right)^{T}-\tau^{2}\Sigma\right)\Sigma^{-1}.\label{eq:second:shift:estimator}
\end{eqnarray}

For $d$-dimensional parameters, we can either directly estimate $\nabla\ell(\theta^{\star})$
using $S_{N,\tau}^{(1)}\left(\theta^{\star}\right)$, or we can estimate
the gradient component-wise. To do so, for each $k\in\left\{ 1,\ldots,d\right\} $,
we can introduce a univariate normal prior $\Theta_{k}\sim\mathcal{N}\left(\theta_{k}^{\star},\tau^{2}\Sigma_{kk}\right)$
for some $\Sigma_{kk}>0$ so that Theorem \ref{theorem:estimators}
yields
\[
\tau^{-2}\Sigma^{-1}\,\check{\mathbb{E}}_{\tau,k}\left[\Theta_{k}-\theta_{k}^{\star}\right]=\nabla_{k}\ell(\theta^{\star})+\mathcal{O}\left(\tau^{2}\right),
\]
where $\check{\mathbb{E}}_{\tau,k}\left[\Theta_{k}\right]$ refers
to the posterior expectation corresponding to the likelihood function
that maps $\theta_{k}$ to $\mathcal{L}(\theta_{1}^{\star},\ldots,\theta_{k-1}^{\star},\theta_{k},\theta_{k+1}^{\star},\ldots,\theta_{d}^{\star})$.
We then obtain an estimator for each component, that can be stacked
in a $d$-dimensional vector denoted by $S_{N,\tau}^{(1)\otimes}(\theta^{\star})$.
Likewise, the matrix $\nabla^{2}\ell(\theta^{\star})$ can be estimated
either using $S_{N,\tau}^{(2)}\left(\theta^{\star}\right)$ or using
a component-wise version denoted by $S_{N,\tau}^{(2)\otimes}\left(\theta^{\star}\right)$.

The Monte Carlo shift estimators can be compared to FD type estimators,
which are the standard approaches to estimate derivatives of functions
that can only be evaluated with some noise \cite{Asmuss}. In one
dimension, the central FD estimator is
\begin{equation}
D_{h}^{(1)}\left(\theta^{\star}\right)=\frac{\log\widehat{\mathcal{L}}(\theta^{\star}+h)-\log\widehat{\mathcal{L}}(\theta^{\star}-h)}{2h},\label{eq:finitediff:score}
\end{equation}
for a perturbation parameter $h>0$, while, for the second order derivative,
it is given by 
\begin{equation}
D_{h}^{(2)}\left(\theta^{\star}\right)=\frac{\log\widehat{\mathcal{L}}(\theta^{\star}+h)-2\log\widehat{\mathcal{L}}(\theta^{\star})+\log\widehat{\mathcal{L}}(\theta^{\star}-h)}{h^{2}}.\label{eq:finitediff:hessian}
\end{equation}
For functions of $d$-dimensional arguments, the FD estimators $D_{h}^{(1)}\left(\theta^{\star}\right)$
and $D_{h}^{(2)}\left(\theta^{\star}\right)$ above can be applied
component-wise. We denote by $D_{h}^{(1)\otimes}(\theta^{\star})$
the estimator of $\nabla\ell(\theta^{\star})$ obtained by defining
the $k$-th component as 
\[
\left\{ D_{h}^{(1)\otimes}(\theta^{\star})\right\} _{k}=\frac{\log\widehat{\mathcal{L}}(\theta^{\star}+e_{k}h)-\log\widehat{\mathcal{L}}(\theta^{\star}-e_{k}h)}{2h},
\]
where $e_{k}$ is the $k$-th basis vector introduced in Section \ref{sec:derivativefree:generalcase}.
Similarly, the second derivatives can be estimated using $d^{2}$
FD estimators as in Eq. \eqref{eq:finitediff:hessian}, and we denote
the resulting estimator by $D_{h}^{(2)\otimes}(\theta^{\star})$. 

Another popular FD type technique relies on simultaneous perturbations
(SP) \cite{spall1992}, and proceeds as follows. Introduce a positive
scalar $h$, and independent draws $\varepsilon^{i}=(\varepsilon_{1}^{i},\ldots,\varepsilon_{d}^{i})$,
for $i\in\left\{ 1,\ldots,N\right\} $, from a $d$-dimensional distribution
$p_{\varepsilon}$ with zero mean and some regularity conditions to
be commented on in Section \ref{sub:Comparison-with-finite-difference}.
For instance, each $\varepsilon^{i}$ is a vector of $d$ independent
draws from a uniform distribution on $\{-1,1\}$. The $k$-th component
of the SP score estimator takes the form
\begin{equation}
\left\{ D_{N,h}^{(1)}(\theta^{\star})\right\} _{k}=\frac{1}{N}\sum_{i=1}^{N}\frac{\log\widehat{\mathcal{L}}\left(\theta^{\star}+h\varepsilon^{i}\right)-\log\widehat{\mathcal{L}}\left(\theta^{\star}-h\varepsilon^{i}\right)}{2h\varepsilon_{k}^{i}}.\label{eq:spsa:score}
\end{equation}
A simple extension for second order derivatives is given in \cite{spall2005},
which we denote by $D_{N,h}^{(2)}\left(\theta^{\star}\right)$. In
one dimension, $D_{N,h}^{(1)}\left(\theta^{\star}\right)$ corresponds
to the central FD estimator $D_{h}^{(1)}\left(\theta^{\star}\right)$
for $N=1$. However, in multivariate settings, the estimator $D_{N,h}^{(1)}\left(\theta^{\star}\right)$
relies on joint perturbations of $\theta^{\star}$ instead of proceeding
component-wise, and thus might scale better with the dimension of
the parameter \cite{spall1992}.

Monte Carlo shift estimators ($S_{N,\tau}^{(1)}\left(\theta^{\star}\right)$
and $S_{N,\tau}^{(2)}\left(\theta^{\star}\right)$) and FD type estimators
($D_{N,h}^{(1)}\left(\theta^{\star}\right)$ and $D_{N,h}^{(2)}\left(\theta^{\star}\right)$)
rely on draws of the likelihood estimator $\widehat{\mathcal{L}}(\theta)$,
at $N$ different parameter values in the neighborhood of $\theta^{\star}$.
The performance of these estimators obviously depends on the quality
of the likelihood estimator $\widehat{\mathcal{L}}(\theta)$ for all
$\theta$ around $\theta^{\star}$, quantified here by its relative
variance $\upsilon_{M}\left(\theta^{\star}\right)$, the choice of
the perturbation parameters ($\tau$ and $h$), the number $N$ of
draws around $\theta^{\star}$, and the dimension $d$ of the parameter
space. The next sections provide results on the performance of Monte
Carlo shift estimators (Sections \ref{sub:Bias-and-variance} to \ref{sub:Robustness-to-noise}),
in terms of mean squared error, impact of the dimension and robustness
to high variance in the likelihood estimator. Section \ref{sub:Comparison-with-finite-difference}
states similar standard results for FD type estimators, for comparison.

\subsection{Bias, variance and mean squared error in one dimension\label{sub:Bias-and-variance}}

The Monte Carlo shift estimator $S_{N,\tau}^{(1)}\left(\theta^{\star}\right)$
converges to $S_{\tau}^{(1)}(\theta^{\star})$ when $N$ goes to infinity,
for any fixed $\tau$, by standard consistency of importance sampling.
Furthermore, $S_{\tau}^{(1)}(\theta^{\star})$ converges to $\nabla\ell(\theta^{\star})$
when $\tau\to0$ according to Theorem \ref{theorem:estimators}. In
this section, we study the bias and the variance of $S_{N,\tau}^{(1)}\left(\theta^{\star}\right)$
defined in Eq. \eqref{eq:first:shift:estimator} when both $N\to\infty$
and $\tau\to0$. We write $\tau_{N}$ for a sequence of non-negative
real values decreasing with $N$ and converging to zero; for instance
$\tau_{N}=N^{-\alpha}$ for some $\alpha>0$. We first consider the
case where the parameter is uni-dimensional, addressing the multivariate
case in Section \ref{sub:Effect-dimension}.

For any integer $N\geq1$ and $\tau_{N}>0$, we define 
\[
A_{N,\tau_{N}}=\frac{1}{N}\sum_{i=1}^{N}\widehat{\mathcal{L}}\left(\theta^{i}\right)\theta^{i},\text{ \ }B_{N,\tau_{N}}=\frac{1}{N}\sum_{i=1}^{N}\widehat{\mathcal{L}}\left(\theta^{i}\right),
\]
where $\theta^{i}$ is drawn from $\mathcal{N}\left(\theta^{\star},\tau_{N}^{2}\Sigma\right)$.
Thus the distribution of $\theta^{i}$ depends on $N$ through $\tau_{N}$
but this dependency is omitted from the notation. We introduce the
notation $\Delta X$ to generically refer to $\left(X-\mathbb{E}[X]\right)/\mathbb{E}[X]$
for a random variable $X$ with non-zero expectation, and finally
we denote by $\mathbb{E}_{\tau_{N}}$ (resp. $\mathbb{V}_{\tau_{N}}$
and $\mathbb{C}_{\tau_{N}}$) the expectation (resp. variance and
covariance) with respect to $\mathcal{N}\left(\theta^{\star},\tau_{N}^{2}\Sigma\right)$.
The other source of randomness comes from $\widehat{\mathcal{L}}(\theta)$
for any given $\theta$; we recall the assumed properties: $\mathbb{E}[\widehat{\mathcal{L}}\left(\theta\right)]=\mathcal{L}(\theta)$
and $\mathbb{V}[\widehat{\mathcal{L}}(\theta)/\mathcal{L}(\theta)]=\upsilon_{M}\left(\theta^{\star}\right)$
for $\theta$ in a neighborhood around $\theta^{\star}$.

We make the following assumptions.
\begin{itemize}
\item \textbf{B1}. The random variables $A_{N,\tau_{N}}$ and $B_{N,\tau_{N}}$
satisfy 
\[
\lim_{N\rightarrow\infty}\mathbb{E}_{\tau_{N}}\left[\left|\sqrt{N}A_{N,\tau_{N}}\right|^{\gamma}\right]<\infty,\quad\lim_{N\rightarrow\infty}\mathbb{E}_{\tau_{N}}\left[\left(\sqrt{N}B_{N,\tau_{N}}\right)^{\gamma}\right]<\infty
\]
for all $\gamma\geq1$.
\item \textbf{B2}. The ratio $A_{N,\tau_{N}}/B_{N,\tau_{N}}$ satisfies
\[
\lim_{N\to\infty}\mathbb{E}_{\tau_{N}}\left[\left|\frac{A_{N,\tau_{N}}}{B_{N,\tau_{N}}}\right|^{\gamma}\right]<\infty
\]
for all $\gamma\geq1$.\end{itemize}
\begin{lem}
\label{lemma:biasvarianceMCshift1} Let $\tau_{N}$ be a decreasing
sequence going to zero such that $N^{-1/4}=o(\tau_{N})$. Under Assumption\textup{s}
\textbf{B1-B2}, the bias and variance of $S_{N,\tau_{N}}^{(1)}\left(\theta^{\star}\right)$
satisfy 
\begin{align}
\mathbb{E}_{\tau_{N}}\left[S_{N,\tau_{N}}^{(1)}\left(\theta^{\star}\right)\right] & =\nabla\ell\left(\theta^{\star}\right)+\tau_{N}^{2}\Sigma\left(\frac{1}{2}\nabla^{3}\ell\left(\theta^{\star}\right)+\nabla^{2}\ell\left(\theta^{\star}\right)\nabla\ell\left(\theta^{\star}\right)\right)+o\left(\tau_{N}^{2}\right),\label{eq:MCshift:asymptoticbias}\\
\mathbb{V}_{\tau_{N}}\left[S_{N,\tau_{N}}^{(1)}\left(\theta^{\star}\right)\right] & =\frac{1}{\tau_{N}^{2}N}\Sigma^{-1}\left(1+\upsilon_{M}\left(\theta^{\star}\right)\right)+o\left(\frac{1}{\tau_{N}^{2}N}\right).\label{eq:MCshift:asymptoticvariance}
\end{align}
The mean squared error is thus optimized by choosing $\tau_{N}=N^{-1/6}$,
and is then of order $N^{-2/3}$.
\end{lem}
The proof is provided in Section \ref{sub:proof:shift:MSE}. It only
requires Assumption \textbf{B1} to hold with $\gamma\in(1,8+\delta)$,
for some $\delta>0$, and Assumption \textbf{B2} to hold for $\gamma\in(1,4+\delta)$,
for some $\delta>0$. For the Monte Carlo shift estimator of $\nabla^{2}\ell\left(\theta^{\star}\right)$,
we state the following result, with an informal proof at the end of
Section \ref{sub:proof:shift:MSE}.
\begin{lem}
\label{lemma:biasvarianceMCshift2} Let $\tau_{N}$ be a decreasing
sequence going to zero such that $N^{-1/4}=o(\tau_{N})$. Under Assumption\textup{s}
\textbf{B1-B2}, there exists a constant $C_{M}\left(\theta^{\star}\right)$
such that the bias and variance of $S_{N,\tau_{N}}^{(2)}\left(\theta^{\star}\right)$
satisfy: 
\begin{align*}
\mathbb{E}_{\tau_{N}}\left[S_{N,\tau_{N}}^{(2)}\left(\theta^{\star}\right)\right] & =\nabla^{2}\ell\left(\theta^{\star}\right)+\tau_{N}^{2}\mathcal{F}\left(\theta^{\star}\right)+o\left(\tau_{N}^{2}\right),\\
\mathbb{V}_{\tau_{N}}\left[S_{N,\tau_{N}}^{(2)}\left(\theta^{\star}\right)\right] & =\frac{C_{M}\left(\theta^{\star}\right)}{\tau_{N}^{4}N}+o\left(\frac{1}{\tau_{N}^{4}N}\right),
\end{align*}
where $\mathcal{F}\left(\theta^{\star}\right)$ was defined in Eq.
\eqref{eq:hessian:approx:error}. The mean squared error is thus optimized
by choosing $\tau_{N}=N^{-1/8}$, and is then of order $N^{-1/2}$.\end{lem}
\begin{example}
\label{example:latentgaussian} Let $\theta\in\mathbb{R}^{2}$ in
the context of Example \ref{example:gaussian}. We introduce some
latent variables $X$ with distribution $\mathcal{N}(\theta,\Lambda_{x}^{-1})$
for some precision matrix $\Lambda_{x}$, and some conditional distribution
$Y\mid\left(X=x\right)\sim\mathcal{N}(x,\Lambda_{y\mid x}^{-1})$,
for some precision matrix $\Lambda_{y\mid x}$ such that $\Lambda_{y}=(\Lambda_{x}^{-1}+\Lambda_{y\mid x}^{-1})^{-1}$.
Then, for any $\theta$, by sampling $X^{i}\sim\mathcal{N}(\theta,\Lambda_{x}^{-1})$
for $i\in\left\{ 1,\ldots,M\right\} $ and computing $\widehat{\mathcal{L}}\left(\theta\right)=M^{-1}\sum_{i=1}^{M}\mathcal{N}(y;\,X^{i},\,\Lambda_{y\mid x}^{-1})$,
we have $\mathbb{E}[\widehat{\mathcal{L}}\left(\theta\right)]=\mathcal{L}\left(\theta\right)$
and $\mathbb{V}[\widehat{\mathcal{L}}\left(\theta\right)/\mathcal{L}\left(\theta\right)]=\upsilon(\theta)/M$
for some function $\upsilon(\theta).$ We can thus implement the Monte
Carlo shift estimators. The mean squared error of $S_{N,\tau}^{(1)}\left(\theta^{\star}\right)$
as a function of $\tau$ is illustrated on Figure \ref{fig:Mean-squared-error-shift},
as well as the error of the component-wise FD estimator $D_{h}^{(1)\otimes}\left(\theta^{\star}\right)$
as a function of $h$. The bias and variance trade-off is similar
in spirit for both estimators. In this example, the FD estimator is
much more precise than the Monte Carlo shift estimator. Indeed, the
leading term of its bias is $\nabla^{3}\ell(\theta^{\star})$, which
happens to be zero for this model, for all $\theta^{\star}$.

\begin{figure}[H]
\begin{centering}
\includegraphics[width=0.45\textwidth]{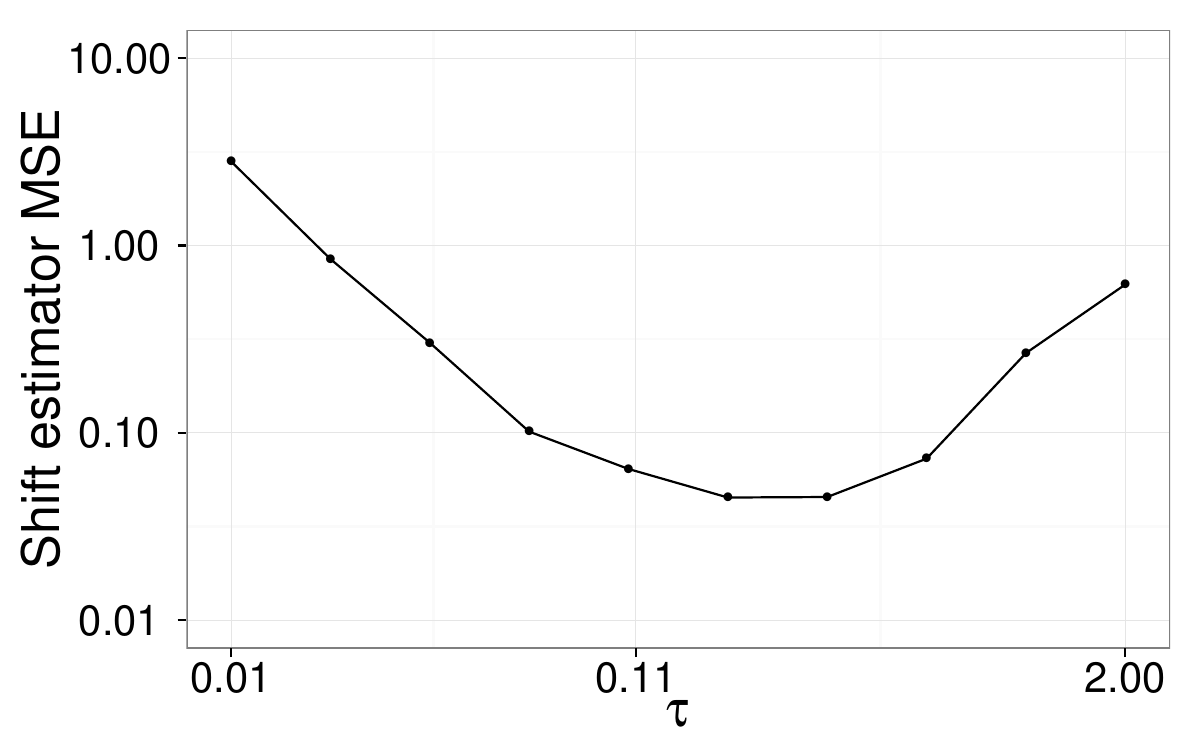}\includegraphics[width=0.45\textwidth]{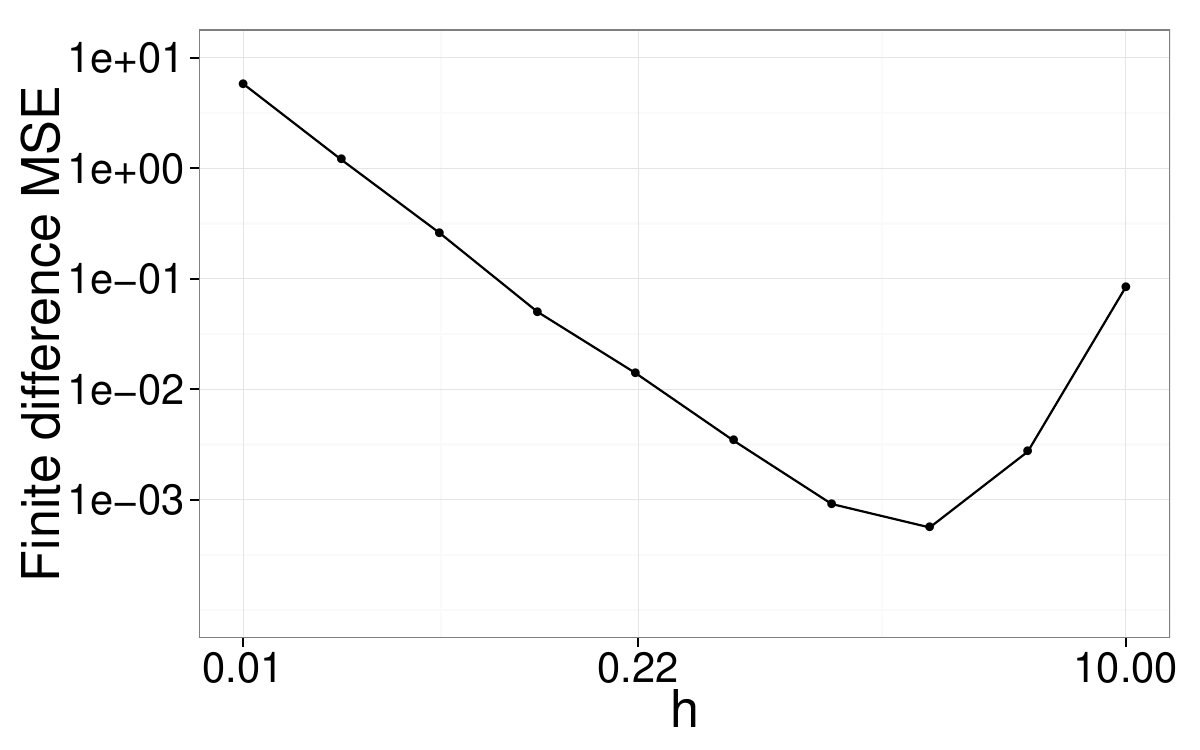}
\par\end{centering}

\protect\caption{Mean squared error of the Monte Carlo shift estimator (left), as a
function of $\tau$, and of the FD estimator (right) as a function
of $h$, on the model of Example \ref{example:latentgaussian}. These
have been obtained based on $100$ independent experiments, each using
$N=100$, $M=100$, $y=\left(0,0\right)$ and $\theta^{\star}=\left(1,1\right)$,
$\Lambda_{x}=\left(1,0.8,0.8,1\right)$ and $\Lambda_{y\mid x}=\left(0.8,0.4,0.4,1\right)$.
We see that the bias-variance trade-off leads to an optimal value
of $\tau$. On the right, the FD estimator exhibits a similar trade-off,
as a function of the perturbation parameter $h$. \label{fig:Mean-squared-error-shift}}
\end{figure}

\end{example}

\subsection{Bias and variance reduction\label{sub:Asymptotic-variance-and-control-variates}}

Lemma \ref{lemma:biasvarianceMCshift1} shows that the bias of the
estimator $S_{N,\tau_{N}}^{(1)}(\theta^{\star})$ is of order $\tau_{N}^{2}$.
We show here how a simple modification allows a substantial reduction
of the bias, at the cost of a significant variance inflation. The
proof is straightforward and thus omitted.
\begin{lem}
\label{lemma:biasreduction} Let $\tau_{N}$ be a decreasing sequence
going to zero such that $N^{-1/4}=o(\tau_{N})$. Under Assumption\textup{s}
\textbf{B1-B2}, the estimator $2S_{N,\tau_{N}/\sqrt{2}}^{\left(1\right)}(\theta^{\star})-S_{N,\tau_{N}}^{\left(1\right)}(\theta^{\star})$
satisfies
\[
\mathbb{E}_{\tau_{N}/\sqrt{2}}\left[2S_{N,\tau_{N}/\sqrt{2}}^{(1)}(\theta^{\star})\right]-\mathbb{E}_{\tau_{N}}\left[S_{N,\tau_{N}}^{(1)}(\theta^{\star})\right]=\nabla\ell\left(\theta^{\star}\right)+o\left(\tau_{N}^{2}\right).
\]
When \textup{$S_{N,\tau_{N}/\sqrt{2}}^{(1)}(\theta^{\star})$ and
$S_{N,\tau_{N}}^{(1)}(\theta^{\star})$ are statistically independent,
then the variance of this estimator is 
\[
\mathbb{V}_{\tau_{N}/\sqrt{2}}\left[2S_{N,\tau_{N}/\sqrt{2}}^{(1)}(\theta^{\star})\right]+\mathbb{V}_{\tau_{N}}\left[S_{N,\tau_{N}}^{(1)}(\theta^{\star})\right]=9\mathbb{V}_{\tau_{N}}\left[S_{N,\tau_{N}}^{(1)}(\theta^{\star})\right]+o\left(\frac{1}{\tau_{N}^{2}N}\right).
\]
}
\end{lem}
A similar reasoning can be done to reduce the bias of $S_{N,\tau_{N}}^{(2)}(\theta^{\star})$.
We now consider a simple variance reduction technique. We consider
the following modification of $S_{N,\tau_{N}}^{(1)}(\theta^{\star})$:

\[
\widetilde{S}_{N,\tau_{N}}^{(1)}(\theta^{\star})=\tau_{N}^{-2}\Sigma^{-1}\left(\sum_{i=1}^{N}\hat{W}^{i}\theta^{i}-\frac{1}{N}\sum_{i=1}^{N}\theta^{i}\right),
\]
where $\theta^{\star}$ has been replaced by the empirical average
$N^{-1}\sum_{i=1}^{N}\theta^{i}$, which acts as a control variate.
The bias of $\widetilde{S}_{N,\tau_{N}}^{(1)}(\theta^{\star})$ is
the same as the bias of $S_{N,\tau_{N}}^{(1)}(\theta^{\star})$. The
following lemma gives the variance of $\widetilde{S}_{N,\tau_{N}}^{(1)}(\theta^{\star})$
when $N\to\infty$.
\begin{lem}
\label{lemma:controlvariates} Let $\tau_{N}$ be a decreasing sequence
going to zero such that $N^{-1/4}=o(\tau_{N})$. Under Assumption\textup{s}
\textbf{B1-B2}, the variance of $\widetilde{S}_{N,\tau_{N}}^{(1)}(\theta^{\star})$
satisfies 
\begin{equation}
\mathbb{V}_{\tau_{N}}\left[\widetilde{S}_{N,\tau_{N}}^{(1)}(\theta^{\star})\right]=\frac{1}{\tau_{N}^{2}N}\Sigma^{-1}\upsilon_{M}\left(\theta^{\star}\right)+o\left(\frac{1}{\tau_{N}^{2}N}\right).\label{eq:MCcontrol:asymptoticvariance}
\end{equation}
 
\end{lem}
An informal proof is provided in Section \ref{sub:proof:controlvariates}.
From Eq. \eqref{eq:MCcontrol:asymptoticvariance}, we see that when
$\upsilon_{M}\left(\theta^{\star}\right)$ is small compared to $1$,
the variance of $\widetilde{S}_{N,\tau_{N}}^{(1)}(\theta^{\star})$
can be significantly smaller than the variance of $S_{N,\tau_{N}}^{(1)}(\theta^{\star})$.
On the other hand, if $\upsilon_{M}\left(\theta^{\star}\right)$ is
large compared to $1$, then both estimators have similar variances.
By a similar reasoning, we could consider reducing the variance of
$S_{N,\tau_{N}}^{(2)}(\theta^{\star})$, by replacing the prior variance
$\tau^{2}\Sigma$ in Eq. \eqref{eq:second:shift:estimator} by an
empirical counterpart computed from the sample $(\theta^{1},\ldots,\theta^{N})$. 
\begin{example}
In the context of Example \ref{example:latentgaussian}, we compare
$S_{N,\tau}^{(1)}\left(\theta^{\star}\right)$ and $\widetilde{S}_{N,\tau}^{(1)}\left(\theta^{\star}\right)$
for various values of $\tau$ and two values of $M$ on Figure \ref{fig:shift-controlvariate}.
The value of $M$ impacts the relative variance $\upsilon_{M}\left(\theta^{\star}\right)$
of the likelihood estimator $\widehat{\mathcal{L}}(\theta)$, for
$\theta$ around $\theta^{\star}$. We see that the control variates
within $\widetilde{S}_{N,\tau}^{(1)}\left(\theta^{\star}\right)$
make a significant improvement when the relative variance of $\widehat{\mathcal{L}}(\theta)$
is small ($M=1000$), but that their effect is barely noticeable when
the relative variance of $\widehat{\mathcal{L}}(\theta)$ is large
($M=1$).

\begin{figure}[H]
\begin{centering}
\includegraphics[width=0.8\textwidth]{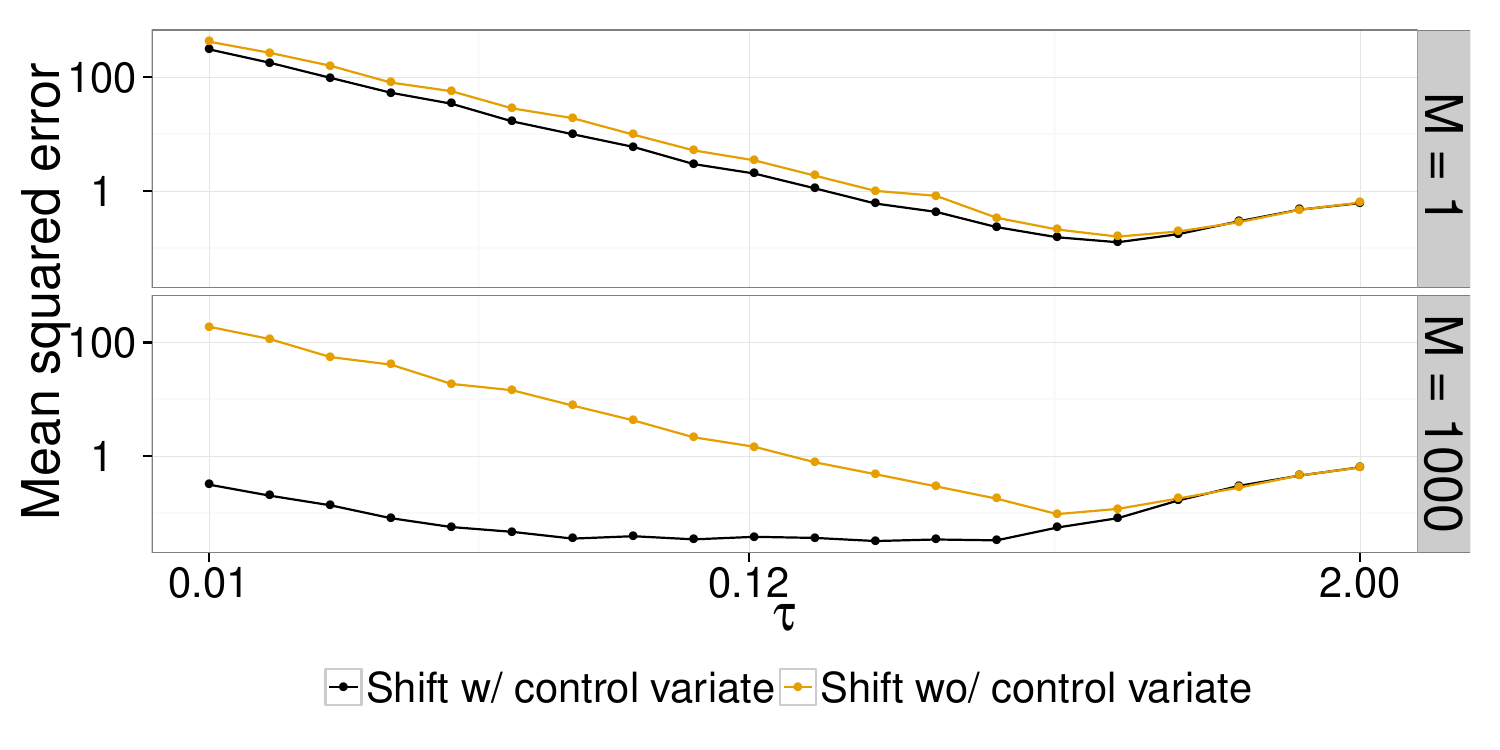}
\par\end{centering}

\protect\caption{Mean squared error of the Monte Carlo shift estimators $S_{N,\tau}^{(1)}\left(\theta^{\star}\right)$
and $\widetilde{S}_{N,\tau}^{(1)}\left(\theta^{\star}\right)$, respectively
without and with control variates, as a function of $\tau$, on the
model of Example \ref{example:latentgaussian}. The top panel compares
the estimators when $M=1$, i.e. the relative variance of the likelihood
estimator $\upsilon_{M}\left(\theta^{\star}\right)$ is large. The
bottom panel compares the estimators when $M=1000$, i.e. the relative
variance of the likelihood estimator $\upsilon_{M}\left(\theta^{\star}\right)$
is small. These have been obtained based on $100$ independent experiments,
each using $N=100$, $y=\left(0,0\right)$ and $\theta^{\star}=\left(1,1\right)$,
$\Lambda_{x}=\left(1,0.8,0.8,1\right)$ and $\Lambda_{y\mid x}=\left(0.8,0.4,0.4,1\right)$.
We see that the control variates improve the estimator by an order
of magnitude when $\upsilon_{M}\left(\theta^{\star}\right)$ is small,
but has no effect when $\upsilon_{M}\left(\theta^{\star}\right)$
is large. \label{fig:shift-controlvariate}}
\end{figure}

We next implement the bias reduction technique described in Lemma
\ref{lemma:biasreduction}, on top of the control variates. We thus
compare $\widetilde{S}_{N,\tau}^{(1)}\left(\theta^{\star}\right)$
and $2\widetilde{S}_{N,\tau/\sqrt{2}}^{(1)}\left(\theta^{\star}\right)-\widetilde{S}_{N,\tau}^{(1)}\left(\theta^{\star}\right)$.
The results are shown on Figure \ref{fig:shift-biasreduction}. We
see that the bias reduction technique leads to a decrease in the squared
bias for large values of $\tau$, where the systematic bias of $S_{\tau}^{(1)}(\theta^{\star})$
dominates the Monte Carlo bias. For smaller values of $\tau$, the
bias reduction technique appears detrimental. On the other hand, the
variance is always increased by a constant factor.

\begin{figure}[H]
\begin{centering}
\includegraphics[width=0.45\textwidth]{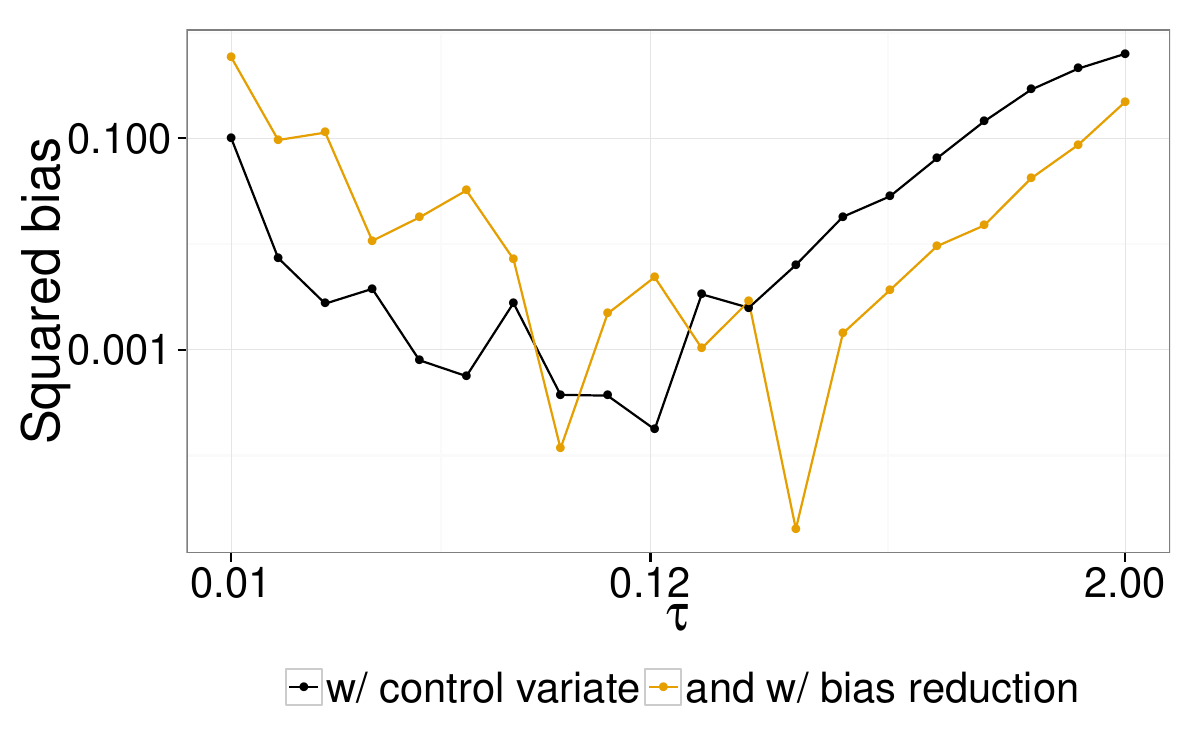}\includegraphics[width=0.45\textwidth]{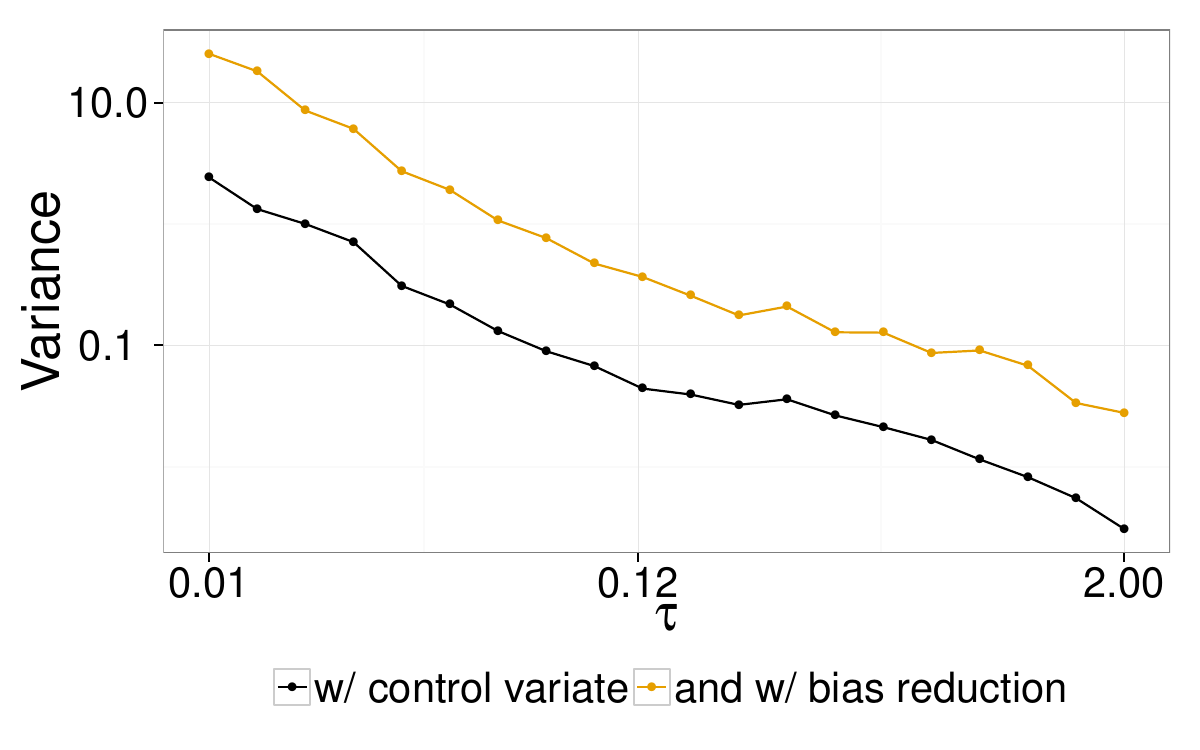}
\par\end{centering}

\begin{centering}
\includegraphics[width=0.8\textwidth]{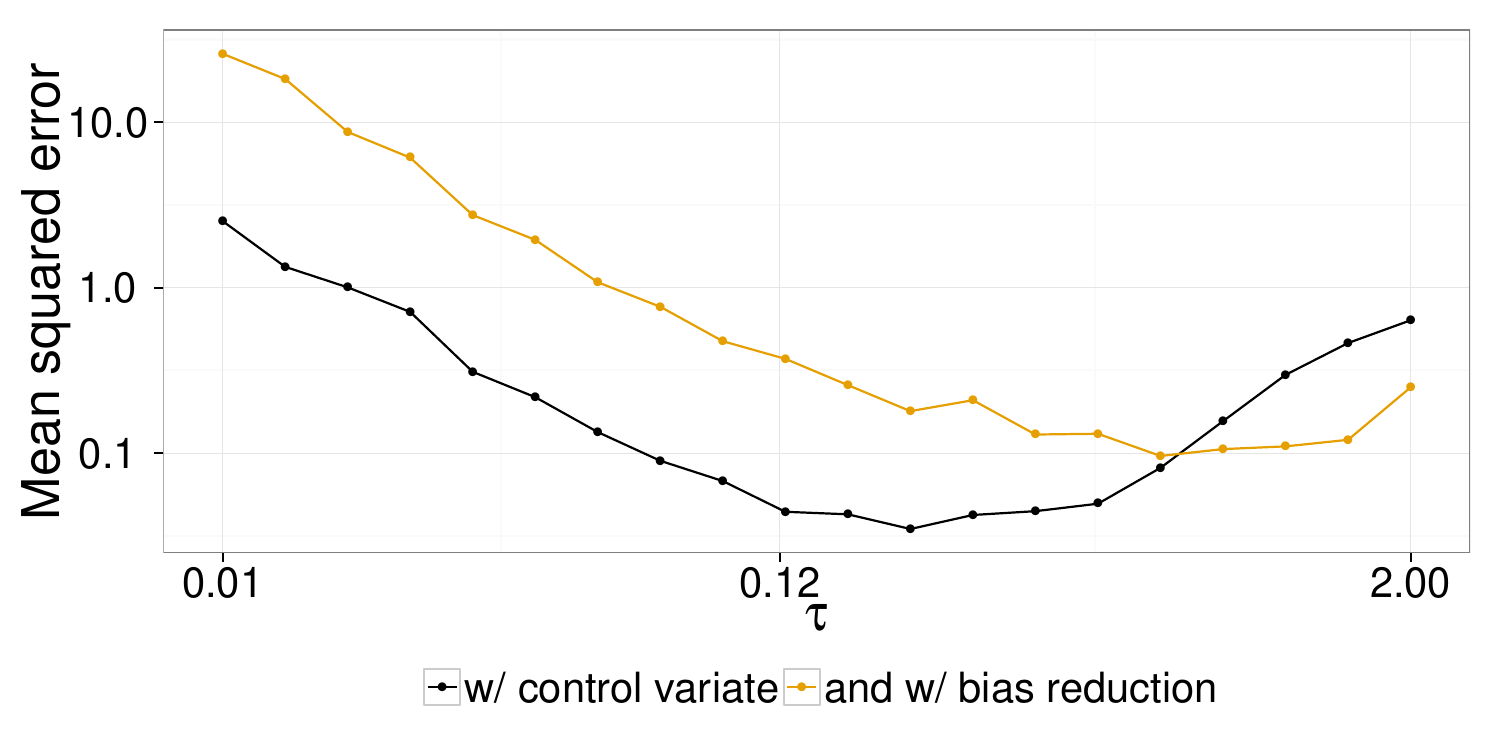}
\par\end{centering}

\protect\caption{Squared bias (top left) and variance (top right) and mean squared
error (bottom) of the Monte Carlo shift estimators, with control variates,
with and without bias reduction, as a function of $\tau$, on the
model of Example \ref{example:latentgaussian}. These have been obtained
based on $100$ independent experiments, each using $N=100$, $M=100$,
$y=\left(0,0\right)$ and $\theta^{\star}=\left(1,1\right)$, $\Lambda_{x}=\left(1,0.8,0.8,1\right)$
and $\Lambda_{y\mid x}=\left(0.8,0.4,0.4,1\right)$. We see that the
bias reduction technique allows a decrease of the bias for larger
values of $\tau$, where the systematic bias dominates the Monte Carlo
bias. On the other hand, it increases the variance by a constant factor.
In this example, the minimum mean squared error is achieved without
the bias reduction technique. \label{fig:shift-biasreduction}}
\end{figure}

\end{example}

\subsection{Effect of the dimension of the parameter \label{sub:Effect-dimension}}

We now consider a $d$-dimensional parameter space. As described in
Section \ref{sub:Monte-Carlo-Shift-Estimators}, we can either estimate
the derivatives jointly (using $S_{N,\tau_{N}}^{(1)}(\theta^{\star})$)
or component-wise (using $S_{N,\tau_{N}}^{(1)\otimes}(\theta^{\star})$).
We wonder whether to use $S_{N,\tau_{N}}^{(1)}\left(\theta^{\star}\right)$
or $S_{N,\tau_{N}}^{(1)\otimes}(\theta^{\star})$. For the latter,
Lemma \ref{lemma:biasvarianceMCshift1} leads to the bias and variance
expressions:
\begin{align}
\mathbb{E}_{\tau_{N},k}\left[\left\{ S_{N,\tau_{N}}^{(1)\otimes}(\theta^{\star})\right\} _{k}\right] & =\nabla_{k}\ell\left(\theta^{\star}\right)+\tau_{N}^{2}\Sigma_{kk}\left(\frac{1}{2}\nabla_{kkk}^{3}\ell\left(\theta^{\star}\right)+\nabla_{kk}^{2}\ell\left(\theta^{\star}\right)\nabla_{k}\ell\left(\theta^{\star}\right)\right)+o\left(\tau_{N}^{2}\right),\label{eq:MCshift:asymptoticbias-dimension-elementwise}\\
\mathbb{V}_{\tau_{N},k}\left[\left\{ S_{N,\tau_{N}}^{(1)\otimes}(\theta^{\star})\right\} _{k}\right] & =\frac{1}{\tau_{N}^{2}N}\Sigma_{kk}^{-1}\left(1+\upsilon_{M}\left(\theta^{\star}\right)\right)+o\left(\frac{1}{\tau_{N}^{2}N}\right).\label{eq:MCshift:asymptoticvariance-dimension-elementwise}
\end{align}
For comparison, we thus need a similar result for $S_{N,\tau}^{(1)}\left(\theta^{\star}\right)$.
The following result is a generalization of Lemma \ref{lemma:biasvarianceMCshift1}
in the $d$-dimensional setting. 
\begin{lem}
\label{lemma:biasvarianceMCshift-dimension}Let $\tau_{N}$ be a decreasing
sequence going to zero such that $N^{-1/4}=o(\tau_{N})$. Under Assumptions
\textbf{B1-B2}, the bias and variance of $S_{N,\tau_{N}}^{(1)}\left(\theta^{\star}\right)$
can be written, for $k\in\left\{ 1,\ldots,d\right\} $: 
\begin{align}
\mathbb{E}_{\tau_{N}}\left[\left\{ S_{N,\tau_{N}}^{(1)}\left(\theta^{\star}\right)\right\} _{k}\right] & =\nabla_{k}\ell\left(\theta^{\star}\right)+\tau_{N}^{2}\sum_{i=1}^{d}\sum_{j=1}^{d}\left(\frac{1}{2}\nabla_{ijk}^{3}\ell\left(\theta^{\star}\right)+\nabla_{ik}^{2}\ell\left(\theta^{\star}\right)\nabla_{j}\ell\left(\theta^{\star}\right)\right)\Sigma_{ij}+o\left(\tau_{N}^{2}\right),\label{eq:MCshift:asymptoticbias-dimension}\\
\mathbb{V}_{\tau_{N}}\left[\left\{ S_{N,\tau_{N}}^{(1)}\left(\theta^{\star}\right)\right\} _{k}\right] & =\frac{1}{\tau_{N}^{2}N}\Sigma_{kk}^{-1}\left(1+\upsilon_{M}\left(\theta^{\star}\right)\right)+o\left(\frac{1}{\tau_{N}^{2}N}\right).\label{eq:MCshift:asymptoticvariance-dimension}
\end{align}

\end{lem}
In the statement of the above lemma, Assumptions \textbf{B1-B2} need
to be interpreted component-wise. The proof is given in Section \ref{sub:proof:dimension}.
Note that the estimator $S_{N,\tau_{N}}^{(1)}\left(\theta^{\star}\right)$
has correlations between its components, whereas the components of
$S_{N,\tau_{N}}^{(1)\otimes}(\theta^{\star})$ are independent. The
correlations between components of $S_{N,\tau_{N}}^{(1)}\left(\theta^{\star}\right)$
are omitted from the above lemma for sake of brevity. Some guidelines
to evaluate these correlations are given in Section \ref{sub:proof:dimension}.

From Eq. \eqref{eq:MCshift:asymptoticbias-dimension-elementwise}
and Eq. \eqref{eq:MCshift:asymptoticbias-dimension}, we see that
the bias has fewer terms when the estimation is performed element-wise
rather than jointly. In particular, for general covariance matrices
$\Sigma$, the leading term in the bias expressed in Eq. \eqref{eq:MCshift:asymptoticbias-dimension}
is quadratic in $d$. If one uses a diagonal matrix for $\Sigma$,
then the bias is only linear in $d$. It could happen that these $d$
terms compensate each other, but in general, these equations indicate
that it is better to estimate the score element-wise in terms of bias.
Indeed, for the optimal choice $\tau_{N}=N^{-1/6}$, the bias is in
$N^{-1/3}$, and thus dividing the bias of $S_{N,\tau_{N}}^{(1)}\left(\theta^{\star}\right)$
by $d$ would cost more than a $d$-fold increase in computational
cost, that corresponds to the cost of $S_{N,\tau_{N}}^{(1)\otimes}(\theta^{\star})$
for the same values of $N$ and $M$. Note that we expect the use
of the bias reduction technique described in Lemma \ref{lemma:biasreduction}
to be more significant for $S_{N,\tau_{N}}^{(1)}\left(\theta^{\star}\right)$
than for $S_{N,\tau_{N}}^{(1)\otimes}(\theta^{\star})$. 

From Eq. \eqref{eq:MCshift:asymptoticvariance-dimension-elementwise}
and Eq. \eqref{eq:MCshift:asymptoticvariance-dimension}, we see that
the leading term in the variance is the same whether the estimation
is performed element-wise or jointly. Given that performing the estimation
jointly is $d$ times faster for fixed values of $N$ and $M$, it
is therefore advantageous to perform the estimation jointly, in terms
of variance. The term $\upsilon_{M}\left(\theta^{\star}\right)$ itself
is typically increasing with $d$, or conversely, $M$ has to be increased
with $d$ in order for $\upsilon_{M}\left(\theta^{\star}\right)$
to be stable. If we consider the simple case of a likelihood function
that factorizes into $d$ independent terms:
\[
\mathcal{L}\left(\theta\right)=\prod_{k=1}^{d}\mathcal{L}_{k}(\theta_{k}),
\]
then, for a given $\theta$, estimating each term $\mathcal{L}_{k}(\theta_{k})$
independently with $\widehat{\mathcal{L}}_{k}(\theta_{k})$ leads
to the variance
\begin{align*}
\mathbb{V}\left[\frac{\widehat{\mathcal{L}}(\theta)}{\mathcal{L}(\theta)}\right] & =\mathbb{E}\left[\left(\frac{\widehat{\mathcal{L}}(\theta)}{\mathcal{L}(\theta)}\right)^{2}\right]-1=\prod_{i=1}^{d}\left(1+\mathbb{V}\left[\frac{\widehat{\mathcal{L}}_{k}(\theta_{k})}{\mathcal{L}_{k}(\theta_{k})}\right]\right)-1\\
 & =\prod_{i=1}^{d}\left(1+\frac{\upsilon}{M}\right)-1\leq\exp\upsilon-1
\end{align*}
assuming that $\upsilon/M$ is the relative variance of each estimator
$\widehat{\mathcal{L}}_{k}(\theta_{k})$, that $M=d$, and noting
that $\left(1+\upsilon/d\right)^{d}\leq\exp\upsilon$ for all $d$.
Thus the relative variance of $\widehat{\mathcal{L}}(\theta)$ is
upper bounded by a constant when $M$ is chosen to increase linearly
in $d$. A similar behavior has been demonstrated for likelihood estimators
obtained by particle methods for specific models \cite{Cerou2011,delmoral2004}.
In general, the relative variance of the likelihood estimator is expected
to increase at least linearly with $d$. %

\subsection{Robustness to high variance in the likelihood estimator \label{sub:Robustness-to-noise}}

We now consider the behavior of the shift estimators when the relative
variance $\upsilon_{M}\left(\theta^{\star}\right)$ of the likelihood
estimator $\widehat{\mathcal{L}}(\theta)$ is large. Note that the
randomness of Monte Carlo shift estimators occurs in the form of weighted
averages of draws from $\mathcal{N}(\theta^{\star},\tau^{2}\Sigma)$,
with the likelihood estimator appearing only in the weights. When
the relative variance increases, the normalized weights $(\hat{W}^{1},\ldots\hat{W}^{N})$
used in the Monte Carlo shift estimators of Eq. \eqref{eq:first:shift:estimator}
and Eq. \eqref{eq:second:shift:estimator} become more and more unbalanced,
one of the normalized weights typically getting close to one while
all the others are nearing zero. As a result, the weighted average
$\sum_{i=1}^{N}\hat{W}^{i}\theta^{i}$ reduces to one draw $\theta^{i}$
that corresponds to the only significant normalized weight. Then the
difference between $\sum_{i=1}^{N}\hat{W}^{i}\theta^{i}$ and $\theta^{\star}$
is of order $\tau$ and the bias of the estimator $S_{N,\tau}^{(1)}(\theta^{\star})$
can then be bounded by a term of order $\tau^{-1}$, independently
of $\upsilon_{M}\left(\theta^{\star}\right)$. Similarly, the following
lemma gives an upper bound on the variance of $S_{N,\tau}^{(1)}(\theta^{\star})$.%

\begin{lem}
\label{lemma:robustness} Assume that the likelihood estimator $\widehat{\mathcal{L}}(\theta)$
is unbiased and has a relative variance equal to $\upsilon_{M}\left(\theta^{\star}\right)$.
Let $\tau$, $N$ and $M$ be fixed. There exists a constant $C$
independent of $\upsilon_{M}\left(\theta^{\star}\right)$ and $\tau$
such that
\[
\mathbb{V}_{\tau}\left[S_{N,\tau}^{(1)}(\theta^{\star})\right]\leq C\tau^{-4}.
\]

\end{lem}
A proof is provided in Section \ref{sub:proof:robustness}. The constant
$C$ depends implicitly on $N$, although we conjecture that a more
sophisticated proof might be able to remove this dependency. For the
second order derivative, the estimator $S_{N,\tau}^{(2)}(\theta^{\star})$
of Eq. \eqref{eq:second:shift:estimator} is very close to $0$ when
only one of the normalized weights is significant, and thus the variance
of $S_{N,\tau}^{(2)}(\theta^{\star})$ is very small. Since the real
posterior variance is of order $\tau^{2}$, the bias of $S_{N,\tau}^{(2)}(\theta^{\star})$
would then be of order $\tau^{-2}$, and thus $S_{N,\tau}^{(2)}(\theta^{\star})$
would have a mean squared error bounded by a term in $C\tau^{-4}$,
for another constant $C$ that does not depend on $\upsilon_{M}\left(\theta^{\star}\right)$.

Thus the Monte Carlo shift estimators benefit from some robustness
to the relative variance $\upsilon_{M}\left(\theta^{\star}\right)$
of the likelihood estimator. This will prove a significant advantage
over FD estimators, as illustrated in the following example. 
\begin{example}
\label{example:robustness} To simulate a setting where the relative
variance increases to infinity in the context of Example \ref{example:latentgaussian},
we consider the case where the variance of the observation distribution
$Y\mid X=x$, becomes smaller and smaller. We thus introduce a scaling
factor $\lambda\in\left(0,1\right)$ and $V_{y\mid x}=\lambda\Lambda_{y\mid x}^{-1},$
and we set $V_{x}=\Lambda_{y}^{-1}-V_{y\mid x}$. Hence the matrix
$\Lambda_{y}=(V_{x}+V_{y\mid x})^{-1}$ is the same for all $\lambda$,
and thus the score is unchanged, but the Monte Carlo procedure struggles
more and more as $\lambda$ approaches zero. Figure \ref{fig:shift-robustness}
represents the behavior of the shift estimator $S_{N,\tau}^{(1)}\left(\theta^{\star}\right)$,
the version with control variates $\widetilde{S}_{N,\tau}^{(1)}\left(\theta^{\star}\right)$,
and the FD estimator $D_{h}^{(1)}\left(\theta^{\star}\right)$, where
the computational effort has been matched, thus $D_{h}^{(1)}\left(\theta^{\star}\right)$
uses $N\times M/4$ samples for each log-likelihood estimate. We see
that the shift estimators get worse when $\lambda$ goes to very small
values, but that the errors are eventually bounded, whereas the mean
squared error of the FD estimator going to infinity.

\begin{figure}[H]
\begin{centering}
\includegraphics[width=0.8\textwidth]{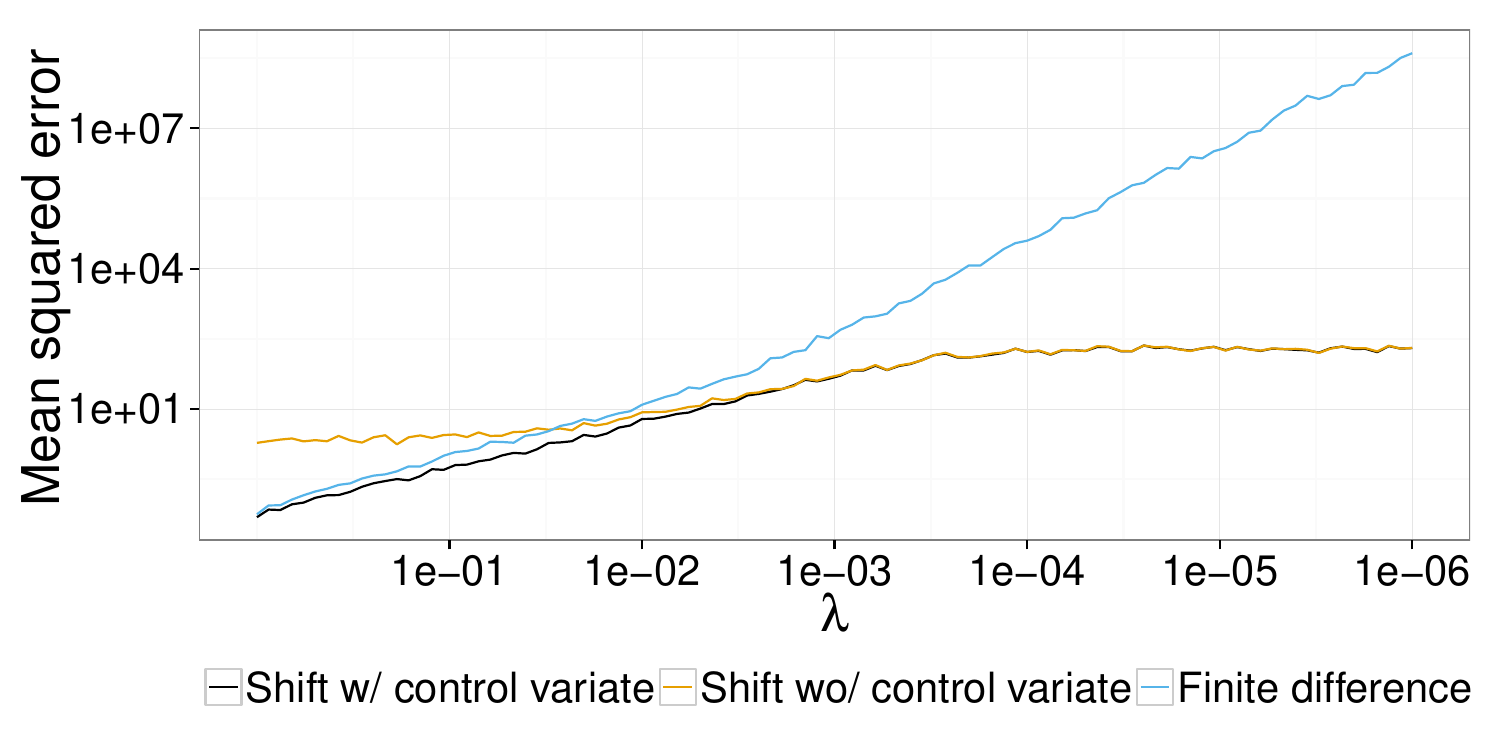}
\par\end{centering}

\protect\caption{Mean squared error of the Monte Carlo shift estimators with and without
covariates, and the FD estimator as a function of $\lambda$, which
parametrizes the signal to noise ratio in Example \ref{example:latentgaussian}.
The smaller the $\lambda$, the worse the estimation of the likelihood
estimator. We see that the shift estimators only degrade up to a certain
point when $\lambda$ decreases. On the other hand, the variance of
the FD estimator goes to infinity when $\lambda$ decreases. These
have been obtained based on $100$ independent experiments, each using
$N=100$, $M=100$, $y=\left(0,0\right)$ and $\theta^{\star}=\left(1,1\right)$,
$\Lambda_{x}=\left(1,0.8,0.8,1\right)$ and $\Lambda_{y\mid x}=\left(0.8,0.4,0.4,1\right)$.
The FD estimator uses $N\times M/4$ samples for each log-likelihood
estimate, so that the computational comparison is fair. The perturbation
parameters have been set to $\tau=0.1$ and $h=0.1$, arbitrarily.
\label{fig:shift-robustness}}
\end{figure}

\end{example}

\subsection{Comparison with finite difference type estimators \label{sub:Comparison-with-finite-difference}}

In order to compare Monte Carlo shift estimators with FD type estimators,
we first recall some of their standard properties. We will assume
the following properties of the log-likelihood estimator $\log\widehat{\mathcal{L}}(\theta)$,
for any $\theta$:
\begin{align*}
\mathbb{E}\left[\log\widehat{\mathcal{L}}(\theta)\right] & =\ell(\theta),\\
\mathbb{V}\left[\log\widehat{\mathcal{L}}(\theta)\right] & =\ell(\theta)^{2}U_{M}\left(\theta\right),
\end{align*}
where $U_{M}\left(\theta\right)$ thus quantifies the relative variance
of $\log\widehat{\mathcal{L}}(\theta)$ for all $\theta$, and $M$
is a tuning parameter. We further assume that $U_{M}\left(\theta\right)$
is equal to $U_{M}\left(\theta^{\star}\right)$ for all $\theta$
in a neighborhood of $\theta^{\star}$; see \cite{Asmuss} for similar
assumptions and results. We make the following assumption on the log-likelihood.
\begin{itemize}
\item \textbf{C1}. The log-likelihood $\ell$ is four times continuously
differentiable, and there exists $K<\infty$ and $\delta>0$ such
that $\left|\nabla_{ijkl}^{4}\ell(\theta)\right|\leq K$ for all $\theta$
such that $||\theta-\theta^{\star}||\leq\delta$, for all $i,j,k,l\in\left\{ 1,\ldots,d\right\} $.
\end{itemize}
The following results hold for the component-wise FD estimator $D_{h}^{(1)\otimes}(\theta^{\star})$.
\begin{lem}
\label{lemma:biasvarianceFD} Let $h_{M}$ be a decreasing sequence
going to zero. Under condition \textbf{C1}, for each $k\in\left\{ 1,\ldots,d\right\} $,
the $k$-th component of $D_{h}^{(1)\otimes}(\theta^{\star})$ satisfies

\begin{align}
\mathbb{E}\left[\left\{ D_{h_{M}}^{(1)\otimes}(\theta^{\star})\right\} _{k}\right] & =\nabla_{k}\ell\left(\theta^{\star}\right)+\frac{h_{M}^{2}}{6}\nabla_{kkk}^{3}\ell(\theta^{\star})+o(h_{M}^{2}),\label{eq:finitedifferencet:asymptoticbias}\\
\mathbb{V}\left[\left\{ D_{h_{M}}^{(1)\otimes}(\theta^{\star})\right\} _{k}\right] & =\frac{1}{4h_{M}^{2}}U_{M}\left(\theta^{\star}\right)\left(\ell(\theta^{\star}+h_{M})^{2}+\ell(\theta^{\star}-h_{M})^{2}\right).\label{eq:finitedifference:asymptoticvariance}
\end{align}
If we further assume that $U_{M}\left(\theta^{\star}\right)=U\left(\theta^{\star}\right)/M$
for some $U\left(\theta^{\star}\right)>0$, the mean squared error
can be optimized by choosing $h_{M}=M^{-1/6}$, and is then of order
$M^{-2/3}$.
\end{lem}
A proof is provided in Section \ref{sub:proof:Finite-difference}.
We thus obtain the same rate of convergence for the FD estimator when
$M\to\infty$ than for the shift estimator when $N\to\infty$ as in
Lemma \ref{lemma:biasvarianceMCshift1}.

We now recall the impact of the dimension on FD estimators. For the
component-wise FD estimator $D_{h_{M}}^{(1)\otimes}(\theta^{\star})$,
we obtain the same behavior as the component-wise Monte Carlo shift
estimator $S_{N,\tau_{N}}^{(1)\otimes}(\theta^{\star})$. For the
SP estimator $D_{N,h_{N}}^{(1)}(\theta^{\star})$ of Eq. \eqref{eq:spsa:score},
we make the following assumption.
\begin{itemize}
\item \textbf{C2}. The perturbation $\varepsilon^{i}=(\varepsilon_{1}^{i},\ldots,\varepsilon_{d}^{i})$,
for $i\in\left\{ 1,\ldots,N\right\} $, is such that the components
$\varepsilon_{k}^{i}$, for $k\in\left\{ 1,\ldots,d\right\} $, are
drawn independently from the uniform distribution on $\left\{ -1,1\right\} $.\end{itemize}
\begin{lem}
\label{lemma:biasvarianceSPSA-dimension} Let $h_{N}$ be a decreasing
sequence going to zero. Under Assumption\textup{s} \textbf{C1}-\textbf{C2},
the SP estimator $D_{N,h_{N}}^{(1)}(\theta^{\star})$ satisfies the
following properties, 
\begin{align}
\mathbb{E}\left[\left\{ D_{N,h_{N}}^{(1)}(\theta^{\star})\right\} _{k}\right] & =\nabla_{k}\ell\left(\theta^{\star}\right)+\frac{h_{N}^{2}}{6}\sum_{1\leq i_{1},i_{2},i_{3}\leq d}\nabla_{i_{1}i_{2}i_{3}}^{3}\ell(\theta^{\star})\mathbb{E}\left[\frac{\varepsilon_{i_{1}}\varepsilon_{i_{2}}\varepsilon_{i_{3}}}{\varepsilon_{k}}\right]+o(h_{N}^{2}),\label{eq:spsa:asymptoticbias}\\
\mathbb{V}\left[\left\{ D_{N,h_{N}}^{(1)}(\theta^{\star})\right\} _{k}\right] & =\frac{U_{M}\left(\theta^{\star}\right)}{2Nh_{N}^{2}}\ell(\theta^{\star})^{2}+o\left(\frac{1}{Nh_{N}^{2}}\right),\label{eq:spsa:asymptoticvariance}
\end{align}
and there is a constant $C(\theta^{\star})$ such that 
\[
\left|\sum_{1\leq i_{1},i_{2},i_{3}\leq d}\nabla_{i_{1}i_{2}i_{3}}^{3}\ell(\theta^{\star})\mathbb{E}\left[\frac{\varepsilon_{i_{1}}\varepsilon_{i_{2}}\varepsilon_{i_{3}}}{\varepsilon_{k}}\right]\right|\leq C(\theta^{\star})\times d.
\]

\end{lem}
A short proof is given in Section \ref{sub:proof:Finite-difference}.
Similar results could be obtained for other distributions of the perturbation
variable $\varepsilon$, as long as the distribution is symmetric,
with finite moments and finite inverse moments, which precludes the
normal distribution. From Lemma \ref{lemma:biasvarianceSPSA-dimension},
we see that the bias term is linear in $d$. On the other hand, the
variance term depends on $d$ only through the relative variance $U_{M}$
of the log-likelihood estimator. We thus conclude that the bias and
variance of $D_{N,h_{N}}^{(1)}(\theta^{\star})$ behave similarly
as those of the Monte Carlo shift estimator $S_{N,\tau_{N}}^{(1)}(\theta^{\star})$,
with respect to the dimension $d$ (compare Lemma \ref{lemma:biasvarianceSPSA-dimension}
and Lemma \ref{lemma:biasvarianceMCshift-dimension} with $\Sigma$
diagonal). Although omitted here, the estimators of the second derivatives,
$D_{N,h_{N}}^{(2)}(\theta^{\star})$ and $D_{N,h_{N}}^{(2)\otimes}(\theta^{\star})$,
could be studied similarly, and we would also find the same convergence
rates as for $S_{N,h_{N}}^{(2)}(\theta^{\star})$ and $S_{N,h_{N}}^{(2)\otimes}(\theta^{\star})$.
The Monte Carlo shift estimators thus exhibit the same convergence
rates in general as FD type estimators including simultaneous perturbations.

The non-asymptotic regime of both classes of estimators is different.
We have seen in Section \ref{sub:Robustness-to-noise} that the Monte
Carlo shift estimators always have a bounded variance, irrespective
of the relative variance $\upsilon_{M}\left(\theta^{\star}\right)$
of the likelihood estimator. On the other hand, the variance of FD
type estimators increases linearly with the relative variance $U_{M}$
of the log-likelihood estimator, possibly to infinity, as was illustrated
in Example \ref{example:robustness}.
\begin{rem}
In practice we typically either have access to an unbiased estimator
of the likelihood, or to an unbiased estimator of the log-likelihood.
Suppose that we have access to an unbiased estimator of the likelihood,
such as the one obtained by particle filters in the context of state-space
models. Taking the logarithm of the estimator yields a log-likelihood
estimator with a bias of order $M^{-1}$, where $M$ is, say, the
number of particles. We can see from the proof of Lemma \ref{lemma:biasvarianceFD}
that it does not change the overall convergence rate of the component-wise
FD estimator $D_{h_{M}}^{(1)\otimes}(\theta^{\star})$, as long as
$M^{-1}=o(h_{M}^{2})$, which ensures that the Monte Carlo bias vanishes
faster than the systematic bias coming from the Taylor expansion.
On the other hand, averaging over $N$ draws as in the SP estimator
$D_{N,h_{N}}^{(1)}(\theta^{\star})$ does not reduce the bias, which
would be of constant order $M^{-1}$ when $N\to\infty$. Thus, one
might want to decide which score estimator to use according to which
quantity can be unbiasedly estimated.
\end{rem}

\section{Monte Carlo shift estimators for latent variable models \label{sec:Latent-variable-models}}

\subsection{Extended likelihood function and shift estimators\label{sub:Extended-model}}

We discuss here an alternative to the shift estimators introduced
in Section \ref{sec:derivativefree:generalcase} which is applicable
whenever the log-likelihood function $\ell\left(\theta\right)$ arises
from multiple and/or multivariate observations. Given observations
$y_{1:T}=\left(y_{1},\ldots y_{T}\right)$, we can indeed always decompose
the log-likelihood as a sum of terms:
\begin{equation}
\ell(\theta)=\log\thinspace p(y_{1:T}\mid\theta)=p(y_{1}\mid\theta)+\sum_{t=2}^{T}\log\thinspace p(y_{t}\mid y_{1:t-1},\theta).\label{eq:predictivedecompositionloglikelihood}
\end{equation}
Directly applying the shift estimator yields the estimators $S_{\tau}^{(1)}\left(\theta^{\star}\right)$
and $S_{\tau}^{(2)}\left(\theta^{\star}\right)$ of Theorem \ref{theorem:estimators}.
An alternative exploiting the predictive decomposition of Eq. \eqref{eq:predictivedecompositionloglikelihood}
is possible, as advocated in \cite{ionides2006,Ionides09} for the
score vector. It proceeds as follows. We introduce a $T\times d$
dimensional parameter $\theta_{1:T}=\left(\theta_{1},\ldots\theta_{T}\right)$,
where $\theta_{t}\in\mathbb{R}^{d}$, and denote by $\theta^{[T]}=(\theta,\ldots,\theta)$
the vector made of $T$ copies of $\theta$. We then define the following
artificial log-likelihood function $\bar{\ell:}\,\mathbb{R}^{d\times T}\rightarrow\mathbb{R}$,
\begin{equation}
\bar{\ell}\left(\theta_{1:T}\right)=p(y_{1}\mid\theta_{1})+\sum_{t=2}^{T}\log p(y_{t}\mid y_{1:t-1},\theta_{t}).\label{eq:extendedlikelihoodobvious}
\end{equation}
which satisfies $\bar{\ell}\left(\theta^{[T]}\right)=\ell\left(\theta\right)$
for all $\theta$. Then the chain rule, applied to $\ell=\bar{\ell}\circ m_{T}$
where $m_{T}:\theta\mapsto\theta^{[T]}$, yields 
\begin{align*}
\nabla\ell\left(\theta\right) & =\sum_{t=1}^{T}\nabla_{t}\bar{\ell}\left(\theta^{[T]}\right),\\
\nabla^{2}\ell\left(\theta\right) & =\sum_{s=1}^{T}\sum_{t=1}^{T}\nabla_{st}^{2}\bar{\ell}\left(\theta^{[T]}\right).
\end{align*}
Therefore, an estimator of $\nabla\ell\left(\theta^{\star}\right)$,
respectively $\nabla^{2}\ell\left(\theta^{\star}\right)$, can be
obtained by summing the components of an estimator of the extended
gradient vector $\nabla\bar{\ell}(\theta^{\star[T]})$, respectively
of an estimator of the extended Hessian matrix $\nabla^{2}\bar{\ell}(\theta^{\star[T]})$.
Now, we can obtain such estimators by introducing a prior distribution
$\mathcal{N}\left(\theta^{\star[T]},\tau^{2}\bar{\Sigma}\right)$,
where $\bar{\Sigma}$ is a $dT\times dT$ covariance matrix, and using
Theorem \ref{theorem:estimators},
\begin{eqnarray}
\tau^{-2}\bar{\Sigma}^{-1}\left(\check{\mathbb{E}}_{\tau}^{[T]}\left[\bar{\Theta}\right]-\theta^{\star[T]}\right) & = & \nabla\bar{\ell}\left(\theta^{\star[T]}\right)+\mathcal{O}\left(\tau^{2}\right),\label{eq:score:approx_extendedscore}\\
\tau^{-4}\bar{\Sigma}^{-1}\left(\check{\mathbb{V}}_{\tau}^{[T]}\left[\bar{\Theta}\right]-\tau^{2}\bar{\Sigma}\right)\bar{\Sigma}^{-1} & = & \nabla^{2}\bar{\ell}\left(\theta^{\star[T]}\right)+\mathcal{O}\left(\tau^{2}\right).\label{eq:hessian:approx-extendedOIM}
\end{eqnarray}
where $\check{\mathbb{E}}_{\tau}^{[T]}\left[\bar{\Theta}\right]$
and $\check{\mathbb{V}}_{\tau}^{[T]}\left[\bar{\Theta}\right]$ refers
to the posterior mean and variance in the extended model. Summing
the $T$ $d$-dimensional blocks of the estimator of $\nabla\bar{\ell}\left(\theta^{\star[T]}\right)$,
respectively the $T^{2}$ $d\times d$-dimensional blocks of the estimator
of $\nabla^{2}\bar{\ell}(\theta^{\star[T]})$, we obtain 
\begin{eqnarray}
\bar{S}_{\tau}^{(1)}\left(\theta^{\star}\right) & = & \tau^{-2}\sum_{t=1}^{T}\left\{ \bar{\Sigma}^{-1}\left(\check{\mathbb{E}}_{\tau}^{[T]}\left[\bar{\Theta}\right]-\theta^{\star[T]}\right)\right\} _{t}=\nabla\ell\left(\theta^{\star}\right)+\mathcal{O}\left(\tau^{2}\right),\label{eq:extendedshiftestimator}\\
\bar{S}_{\tau}^{(2)}\left(\theta^{\star}\right) & = & \tau^{-4}\sum_{s=1}^{T}\sum_{t=1}^{T}\left\{ \bar{\Sigma}^{-1}\left(\check{\mathbb{V}}_{\tau}^{[T]}\left[\bar{\Theta}\right]-\tau^{2}\bar{\Sigma}\right)\bar{\Sigma}^{-1}\right\} _{st}=\nabla^{2}\ell(\theta^{\star})+\mathcal{O}\left(\tau^{2}\right),\label{eq:extendedshiftestimatorsecondderivative}
\end{eqnarray}
where here $\left\{ v\right\} _{t}$ denotes the $t$-th $d$-dimensional
block of a $dT$-dimensional vector and $\left\{ v\right\} _{st}$
denotes the $(s,t)$-th $d\times d$-dimensional block of a $dT\times dT$-dimensional
matrix.

\subsection{Independent latent variable models\label{sub:Independent-latent-variable}}

To motivate the introduction of the alternative shift estimators of
Eq. \eqref{eq:extendedshiftestimator} and Eq. \eqref{eq:extendedshiftestimatorsecondderivative},
consider the case where the log-likelihood satisfies 
\begin{equation}
\ell(\theta)=\sum_{t=1}^{T}\ell_{t}(\theta),\label{eq:likelihood:independentlatent}
\end{equation}
where $\ell_{t}(\theta)=\log\thinspace p(y_{t}\mid\theta)=\log\thinspace\mathcal{L}{}_{t}(\theta)$.
Assume that the covariance matrix $\bar{\Sigma}$, a $dT\times dT$
matrix, is chosen to be block diagonal, where each diagonal block
is equal to $\Sigma$, a $d\times d$ covariance matrix. That is,
the artificial prior assigned to $\Theta_{1:T}$ assumes that its
components $\Theta_{t}$ are independent and identically distributed
as $\mathcal{N}\left(\theta^{\star},\tau^{2}\Sigma\right)$. Then
the shift estimators of Eq. \eqref{eq:extendedshiftestimator} and
Eq. \eqref{eq:extendedshiftestimatorsecondderivative} are equal to
\begin{eqnarray}
\bar{S}_{\tau}^{(1)}\left(\theta^{\star}\right) & = & \tau^{-2}\Sigma^{-1}\sum_{t=1}^{T}\left(\check{\mathbb{E}}_{\tau,t}\left[\Theta\right]-\theta^{\star}\right),\label{eq:score:approx_extendedscore-1-1}\\
\bar{S}_{\tau}^{(2)}\left(\theta^{\star}\right) & = & \tau^{-4}\Sigma^{-1}\left(\sum_{t=1}^{T}\left(\check{\mathbb{V}}_{\tau,t}\left[\Theta\right]-\tau^{2}\Sigma\right)\right)\Sigma^{-1},\label{eq:extendedshiftOIMestimatorindependent}
\end{eqnarray}
where $\check{\mathbb{E}}_{\tau,t}$ and $\check{\mathbb{V}}_{\tau,t}$
denote the expectation and variance under the artificial posterior
associated to the prior $\mathcal{N}\left(\theta^{\star},\tau^{2}\Sigma\right)$
and the likelihood $\mathcal{L}_{t}(\theta)$.

We compare here the original shift estimator $S_{\tau}^{(1)}\left(\theta^{\star}\right)$
and its approximation $S_{N,\tau}^{(1)}\left(\theta^{\star}\right)$
to $\bar{S}_{\tau}^{(1)}\left(\theta^{\star}\right)$ and its approximation
$\bar{S}_{N,\tau}^{(1)}\left(\theta^{\star}\right)$%
. The bias of $S_{\tau}^{(1)}\left(\theta^{\star}\right)$ obtained
in Theorem \ref{theorem:estimators}, which is also the leading term
in the bias of $S_{N,\tau}^{(1)}\left(\theta^{\star}\right)$ obtained
in Lemma \ref{lemma:biasvarianceMCshift-dimension}, can be written,
for the $k$-th component of the gradient. 
\begin{align*}
 & \tau^{2}\sum_{i=1}^{d}\sum_{j=1}^{d}\left(\frac{1}{2}\nabla_{ijk}^{3}\ell\left(\theta^{\star}\right)+\nabla_{ik}^{2}\ell\left(\theta^{\star}\right)\nabla_{j}\ell\left(\theta^{\star}\right)\right)\Sigma_{ij}\\
= & \tau^{2}\sum_{i=1}^{d}\sum_{j=1}^{d}\left(\frac{1}{2}\sum_{t=1}^{T}\nabla_{ijk}^{3}\ell_{t}\left(\theta^{\star}\right)+\left(\sum_{t=1}^{T}\nabla_{ik}^{2}\ell_{t}\left(\theta^{\star}\right)\right)\left(\sum_{t=1}^{T}\nabla_{j}\ell_{t}\left(\theta^{\star}\right)\right)\right)\Sigma_{ij}
\end{align*}
where we have used the specific form of the log-likelihood of Eq.
\eqref{eq:likelihood:independentlatent}. It thus increases quadratically
with $T$. The variance of $S_{N,\tau}^{(1)}\left(\theta^{\star}\right)$,
according to Lemma \ref{lemma:biasvarianceMCshift-dimension}, is
led by $\tau^{-2}N^{-1}\Sigma_{kk}^{-1}(1+\upsilon_{M}\left(\theta^{\star}\right))$
for the $k$-th component, and thus increases with $T$ insofar as
$\upsilon_{M}\left(\theta^{\star}\right)$ does; typically $\upsilon_{M}\left(\theta^{\star}\right)$
would increase at least linearly with $T$. Thus, the mean squared
error would be dominated by the squared bias term in $T^{4}$ when
$T$ increases. 

On the other hand, the bias of $\bar{S}_{\tau}^{(1)}\left(\theta^{\star}\right)$
can be written
\[
\tau^{2}\sum_{t=1}^{T}\sum_{i=1}^{d}\sum_{j=1}^{d}\left(\frac{1}{2}\nabla_{ijk}^{3}\ell_{t}\left(\theta^{\star}\right)+\nabla_{ik}^{2}\ell_{t}\left(\theta^{\star}\right)\nabla_{j}\ell_{t}\left(\theta^{\star}\right)\right)\Sigma_{ij}
\]
which only increases linearly with $T$. The variance of $\bar{S}_{N,\tau}^{(1)}\left(\theta^{\star}\right)$
is led by the term $\tau^{-2}N^{-1}\Sigma_{kk}^{-1}\sum_{t=1}^{T}(1+\upsilon_{M,t})$,
where $\upsilon_{M,t}$ is the relative variance of the partial likelihood
estimator $\widehat{\mathcal{L}}_{t}(\theta)$, which can be assumed
to be constant. Thus the variance increases linearly with $T$, similarly
to the variance of $S_{N,\tau}^{(1)}\left(\theta^{\star}\right)$.
Overall the mean squared error of $\bar{S}_{N,\tau}^{(1)}\left(\theta^{\star}\right)$
is thus expected to become lower than that of $S_{N,\tau}^{(1)}\left(\theta^{\star}\right)$
when $T$ increases.

Furthermore, consider the case where one of the $\ell_{t}$ is particularly
hard to estimate, for instance because $y_{t}$ is an outlier. Denote
its index by $t$ and imagine the extreme scenario where the relative
variance of $\mathcal{\widehat{L}}_{t}(\theta^{\star})$ is infinite.
Then the relative variance of the full likelihood estimator $\widehat{\mathcal{L}}(\theta^{\star})$
would also be infinite. Using $S_{N,\tau}^{(1)}\left(\theta^{\star}\right)$
would certainly result in poor performances, although the reasoning
of Lemma \ref{lemma:robustness} guarantees a finite variance. Finite
difference type estimators would have an infinite variance in this
setting. On the other hand, if we use $\bar{S}_{N,\tau}^{(1)}\left(\theta^{\star}\right)$,
the term corresponding to $\ell_{t}$ would be poorly estimated, but
with a finite variance. If all the other terms are correctly estimated,
and if their norms are large enough to dominate the poorly estimated
term, then it is possible that $\bar{S}_{N,\tau}^{(1)}\left(\theta^{\star}\right)$
would still be satisfactory. This motivates the development of Monte
Carlo approximations of $\bar{S}_{\tau}^{(1)}\left(\theta^{\star}\right)$
in the next section, following \cite{ionides2006,Ionides09}, instead
of trying to approximate $S_{\tau}^{(1)}(\theta^{\star})$ directly
for state-space models, using particle Markov chain Monte Carlo \cite{andrieu2010}
or SMC$^{2}$ \cite{chopin2011}.
\begin{example}
We augment Example \ref{example:latentgaussian} with $T$ observations
$Y_{1},\ldots,Y_{T}$, and $T$ latent variables $X_{1},\ldots,X_{T}$,
independent and identically distributed. The derivatives of the log-likelihood
at $\theta^{\star}$ are then
\[
\nabla\ell(\theta^{\star})=-\Lambda_{y}\sum_{t=1}^{T}(\theta^{\star}-y_{t})\quad\text{and}\quad\nabla^{2}\ell(\theta^{\star})=-T\Lambda_{y}.
\]
The original shift estimators are given by 
\begin{align*}
S_{\tau}^{(1)}\left(\theta^{\star}\right) & =\tau^{-2}\Sigma^{-1}(\tau^{-2}\Sigma^{-1}+T\Lambda_{y})^{-1}\left(-\Lambda_{y}\sum_{t=1}^{T}(\theta^{\star}-y_{t})\right),\\
S_{\tau}^{(2)}\left(\theta^{\star}\right) & =\tau^{-2}\Sigma^{-1}(\tau^{-2}\Sigma^{-1}+T\Lambda_{y})^{-1}\left(-T\Lambda_{y}\right).
\end{align*}
In one dimension, we can check that the bias of $S_{\tau}^{(1)}\left(\theta^{\star}\right)$
would be equal to $\tau^{2}\Sigma T\Lambda_{y}\nabla\ell(\theta^{\star}),$
which is quadratic in $T$ because $\nabla\ell(\theta^{\star})$ is
equal to $-\Lambda_{y}\sum_{t=1}^{T}(\theta^{\star}-y_{t})$. On the
other hand, using the extended model, we obtain
\begin{align*}
\bar{S}_{\tau}^{(1)}\left(\theta^{\star}\right) & =\sum_{t=1}^{T}\tau^{-2}\Sigma^{-1}(\tau^{-2}\Sigma^{-1}+\Lambda_{y})^{-1}\left(-\Lambda_{y}(\theta^{\star}-y_{t})\right),\\
\bar{S}_{\tau}^{(2)}\left(\theta^{\star}\right) & =\sum_{t=1}^{T}\tau^{-2}\Sigma^{-1}(\tau^{-2}\Sigma^{-1}+\Lambda_{y})^{-1}\left(-\Lambda_{y}\right).
\end{align*}
The bias of $\bar{S}_{\tau}^{(1)}\left(\theta^{\star}\right)$ only
increases linearly with $T$. A similar bias comparison can be done
for the second order derivative.
\end{example}

\subsection{State-space models\label{sub:Hidden-Markov-models}}

We now focus on the class of state-space models, which generalizes
the model of the previous section. We propose alternative shift estimators
in this context and discuss their link to the score estimator proposed
in \cite{Ionides09}. Let $\left(X_{t},Y_{t}\right)_{t\in\mathbb{N}}$
be a stochastic process such that $\left(X_{t},Y_{t}\right)$ takes
values in a measurable space $\mathcal{X\times Y}$. The model is
specified as follows: $\left(X_{t}\right)_{t\in\mathbb{N}}$ is a
latent Markov process of initial density $\nu\left(x;\theta\right)$
and homogeneous Markov transition density $f\left(\left.x\right\vert x^{\prime};\theta\right)$
whereas the observations $\left(Y_{t}\right)_{t\in\mathbb{N}}$ are
assumed to be conditionally independent given $\left(X_{t}\right)_{t\in\mathbb{N}}$
of conditional density $g\left(\left.y_{t}\right\vert x_{t};\theta\right)$
(with respect to suitable dominating measures) where $\theta\in\mathbb{R}^{d}$
; that is $X_{1}\sim\mu\left(\cdot;\theta\right)$ and for $t\geq1$,
\begin{equation}
\left.X_{t+1}\right\vert \left(X_{t}=x\right)\sim f\left(\left.\cdot\right\vert x_{t-1};\theta\right),\quad\left.Y_{t}\right\vert \left(X_{t}=x_{t}\right)\sim g\left(\left.\cdot\right\vert x_{t};\theta\right).\label{eq:evolobs}
\end{equation}
It follows that the joint density of $\left(X_{1:T},Y_{1:T}\right)$
is given by 
\begin{equation}
p\left(x_{1:T},y_{1:T};\theta\right)=\nu\left(x_{1};\theta\right)\prod\limits _{t=2}^{T}f\left(\left.x_{t}\right\vert x_{t-1};\theta\right)\prod\limits _{t=1}^{T}g\left(\left.y_{t}\right\vert x_{t};\theta\right).\label{eq:jointdensity}
\end{equation}
For a realization $Y_{1:T}=y_{1:T}$ of the observations, the log-likelihood
of function satisfies the decomposition of Eq. \eqref{eq:predictivedecompositionloglikelihood}
where 
\[
p(y_{t}\mid y_{1:t-1},\theta)=\int g\left(\left.y_{t}\right\vert x_{t};\theta\right)p\left(\left.x_{t}\right\vert y_{1:t-1};\theta\right)dx_{t},
\]
where $p\left(\left.x_{t}\right\vert y_{1:t-1};\theta\right)$ denotes
the posterior distribution of $X_{t}$ given observations $y_{1:t-1}$.
Consider a prior where the components $\Theta_{t}$ of $\Theta_{1:T}$
are assumed independent and identically distributed according to $\mathcal{N}\left(\theta^{\star},\tau^{2}\Sigma\right)$.
The shift estimators of Eq. \eqref{eq:extendedshiftestimator} and
Eq. \eqref{eq:extendedshiftestimatorsecondderivative} can be rewritten
after simple manipulations as
\begin{eqnarray}
\bar{S}_{\tau}^{(1)}\left(\theta^{\star}\right) & = & \tau^{-2}\Sigma^{-1}\sum_{t=1}^{T}\left(\mathbb{E_{\tau}}\left[\Theta_{t}\mid y_{1:T}\right]-\theta^{\star}\right),\label{eq:extendedshiftestimatorscoreSSM}\\
\bar{S}_{\tau}^{(2)}\left(\theta^{\star}\right) & = & \tau^{-4}\Sigma^{-1}\left(\sum_{s=1}^{T}\sum_{t=1}^{T}\mathbb{C_{\tau}}\left[\Theta_{s},\Theta_{t}\mid y_{1:T}\right]-\tau^{2}\Sigma T\right)\Sigma^{-1},\label{eq:extendedshiftestimatorOIMSSM}
\end{eqnarray}
where $\mathbb{E_{\tau}}\left[\Theta_{t}\mid y_{1:T}\right]$ and
$\mathbb{C_{\tau}}\left[\Theta_{s},\Theta_{t}\mid y_{1:T}\right]$
are the expectation of $\Theta_{t}$, respectively covariance of $\left(\Theta_{s},\Theta_{t}\right)$,
under the joint posterior smoothing distribution $p_{\tau}\left(\theta_{1:T},x_{1:T}\mid y_{1:T}\right)$
induced by the artificial state space model with initial distribution
$\Theta_{1}\sim\mathcal{N}\left(\theta^{\star},\tau^{2}\Sigma\right),~\left.X_{1}\right\vert \left(\Theta_{1}=\theta\right)\sim\mu\left(\cdot;\theta\right)$
and satisfying for $t\geq1$,
\begin{equation}
\Theta_{t+1}\sim\mathcal{N}\left(\theta^{\star},\tau^{2}\Sigma\right),\quad\left.X_{t+1}\right\vert \left(X_{t}=x_{t},\Theta_{t+1}=\theta_{t+1}\right)\sim f\left(\left.\cdot\right\vert x_{t};\theta_{t+1}\right),\quad\left.Y_{t}\right\vert \left(X_{t}=x_{t},\Theta_{t}=\theta_{t}\right)\sim g\left(\left.\cdot\right\vert x_{t};\theta_{t}\right).\label{eq:modifiedmodel}
\end{equation}

\begin{rem}
One could introduce other prior distributions on $\Theta_{1:T}$.
For example, one could select $\Theta_{1}\sim\mathcal{N}\left(\theta^{\star},\tau^{2}\Sigma\right)$
and for $t\geq1$
\begin{equation}
\Theta{}_{t+1}-\theta^{\star}=\rho\left(\Theta{}_{t}-\theta^{\star}\right)+V_{t+1},\quad V_{t+1}\sim\mathcal{N}\left(0,\upsilon^{2}\Sigma\right),\label{eq:priorSSM}
\end{equation}
where $\rho$ is a scalar, $\left|\rho\right|<1$ and $\tau^{2}=\upsilon^{2}/\left(1-\rho^{2}\right)$.
In Appendix \ref{AppendixC:ARpriors}, we provide for this prior distribution
the expressions of the shift estimators of Eq. \eqref{eq:extendedshiftestimator}
and Eq. \eqref{eq:extendedshiftestimatorsecondderivative}. When $\rho\rightarrow1^{-}$,
we retrieve informally the score estimator proposed in \cite{Ionides09}
as 
\begin{eqnarray*}
\underset{\rho\rightarrow1^{-}}{\lim}\bar{S}_{\tau}^{(1)}\left(\theta^{\star}\right) & \thickapprox & \tau^{-2}\Sigma^{-1}\mathbb{E_{\tau}}\left[\Theta_{T}\mid y_{1:T}\right],
\end{eqnarray*}
and similarly we have 
\begin{eqnarray*}
\underset{\rho\rightarrow1^{-}}{\lim}\bar{S}_{\tau}^{(2)}\left(\theta^{\star}\right) & \thickapprox & \tau^{-4}\Sigma^{-1}\left\{ \mathbb{V}_{\tau}\left[\Theta_{T}\mid y_{1:T}\right]-\tau^{2}\Sigma\right\} \Sigma^{-1}.
\end{eqnarray*}

\end{rem}

\subsection{Sequential Monte Carlo estimators\label{sub:Sequential-Monte-Carlo}}

The approximations of the score vector and observed information matrix
given in the previous section require computing $\mathbb{E_{\tau}}\left[\Theta_{t}\mid y_{1:T}\right]$
and $\mathbb{C_{\tau}}\left[\Theta_{s},\Theta_{t}\mid y_{1:T}\right]$
for $s,t\in\left\{ 1,\ldots,T\right\} $. These can be approximated
using sequential Monte Carlo methods applied to the modified state-space
model described in Eq. \eqref{eq:modifiedmodel}. Particle filters
provide an approximation of $p_{\tau}\left(\theta_{1:T},x_{1:T}\mid y_{1:T}\right)$,
and hence of its marginals $p_{\tau}\left(\theta_{t}\mid y_{1:T}\right)$
and $p_{\tau}\left(\theta_{s},\theta_{t}\mid y_{1:T}\right)$. However,
this approximation will be progressively impoverished as $T$ increases
because of the successive resampling steps. Eventually, $p_{\tau}\left(\theta_{t}\mid y_{1:T}\right)$
will be approximated by a single unique particle for $T-t$ sufficiently
large. Sequential Monte Carlo smoothing procedures have been developed
to obtain lower variance estimators \cite{Briers09,doucet2000}. However
these approaches are only applicable when we can evaluate $f\left(\left.x^{\prime}\right\vert x;\theta\right)$
point-wise and the primary motivation for this work is to address
scenarios where this is not possible. In this case, we can only use
the bootstrap particle filter \cite{gordon1993}. To decrease the
variance of the sequential Monte Carlo estimators of $\mathbb{E_{\tau}}\left[\Theta_{t}\mid y_{1:T}\right]$
and $\mathbb{C_{\tau}}\left[\Theta_{s},\Theta_{t}\mid y_{1:T}\right]$,
at the cost of a bias increase, we will rely on the fact that, when
the state-space model enjoys forgetting properties, we have 
\[
\mathbb{E_{\tau}}\left[\Theta_{t}\mid y_{1:T}\right]\approx\mathbb{E_{\tau}}\left[\Theta_{t}\mid y_{1:\left(t+\Delta\right)\wedge T}\right]
\]
for a lag $\Delta$ large enough. This fixed-lag approximation was
first proposed in \cite{kitagawa2001} and has been studied in \cite{olsson2008}.
Similarly, without loss of generality consider that for $t\geq s$
we have 
\[
\mathbb{C_{\tau}}\left[\Theta_{s},\Theta_{t}\mid y_{1:T}\right]\approx\mathbb{C_{\tau}}\left[\Theta_{s},\Theta_{t}\mid y_{1:\left(t+\Delta\right)\wedge T}\right],
\]
and for $t-s>\Delta$ 
\[
\mathbb{C_{\tau}}\left[\Theta_{s},\Theta_{t}\mid y_{1:T}\right]\approx0\text{.}
\]
Practically, we will thus use the bootstrap filter to compute fixed-lag
smoothing approximations $\bar{S}_{\tau,\Delta}^{(1)}\left(\theta^{\star}\right)$
and $\bar{S}_{\tau,\Delta}^{(2)}\left(\theta^{\star}\right)$ of $\bar{S}_{\tau}^{(1)}\left(\theta^{\star}\right)$
and $\bar{S}_{\tau}^{(2)}\left(\theta^{\star}\right)$, defined in
Eq. \eqref{eq:extendedshiftestimatorscoreSSM} and Eq. \eqref{eq:extendedshiftestimatorOIMSSM}.
These approximations can be written
\begin{eqnarray}
\bar{S}_{\tau,\Delta}^{(1)}\left(\theta^{\star}\right) & = & \tau^{-2}\Sigma^{-1}\sum_{t=1}^{T}\left(\mathbb{E_{\tau}}\left[\Theta_{t}\mid y_{1:\left(t+\Delta\right)\wedge T}\right]-\theta^{\star}\right),\label{eq:fixedlagscore}\\
\bar{S}_{\tau,\Delta}^{(2)}\left(\theta^{\star}\right) & = & \tau^{-4}\Sigma^{-1}\biggl(\sum_{t=1}^{T}\mathbb{V_{\tau}}\left[\Theta_{t}\mid y_{1:\left(t+\Delta\right)\wedge T}\right]\nonumber \\
 &  & +2\sum_{s=1}^{T}\sum_{t=s+1}^{\left(s+\Delta\right)\wedge T}\mathbb{C_{\tau}}\left[\Theta_{s},\Theta_{t}\mid y_{1:\left(t+\Delta\right)\wedge T}\right]\text{ }-\tau^{2}\Sigma T\biggr)\Sigma^{-1},\label{eq:fixedlagOIM}
\end{eqnarray}
with the convention that $\sum_{k=i}^{j}=0$ if $i>j$.

Under regularity assumptions on the transition and observation densities
ensuring exponential ergodicity of the optimal filter, it is possible
to obtain quantitative bounds on the bias between the fixed-lag estimators
$\bar{S}_{\tau,\Delta}^{(1)}\left(\theta^{\star}\right),\bar{S}_{\tau,\Delta}^{(2)}\left(\theta^{\star}\right)$
and $\bar{S}_{\tau}^{(1)}\left(\theta^{\star}\right)$, $\bar{S}_{\tau}^{(2)}\left(\theta^{\star}\right)$.
It is also possible to obtain quantitative bounds on the $\mathbb{L}_{p}$
error of the bootstrap filter approximations of $\bar{S}_{\tau,\Delta}^{(1)}\left(\theta^{\star}\right)$,
$\bar{S}_{\tau,\Delta}^{(2)}\left(\theta^{\star}\right)$ by generalizing
the techniques in \cite{olsson2008}.\footnote{Details can be found in the previous version of this manuscript available
on arXiv.}

\section{Discussion\label{sec:Discussion}}

We have shown here how the score estimator proposed in \cite{ionides2006,Ionides09}
can be derived using Stein's lemma. The connection to Stein's is not
only elegant but also fruitful. From a methodological point of view,
this suggests an original estimator of the observed information matrix
which can be computed using Bayesian computational tools. From a theoretical
point of view, this allows the derivation of sharp quantitative results
for these estimators. We have shown that these estimators are competitive
to finite difference type estimators and enjoy additional robustness
properties. Moreover, in the specific context of state-space models,
we have proposed original derivative-free estimators of the score
and the observed information matrix. These are obtained by solving
smoothing problems for a modified state-space model that differs from
the one proposed in \cite{ionides2006,Ionides09}.

Extensive numerical experiments comparing in practical situations
the various estimators discussed in the paper will be made available
shortly.

\section*{Acknowledgments}

Arnaud Doucet\textquoteright s research was supported by the Engineering
and Physical Sciences Research Council (grant EP/K000276/1, EP/K009850/1)
and by the Air Force Office of Scientific Research/Asian Office of
Aerospace Research and Development (AFOSR/AOARD) (grant AOARD-144042).
Pierre Jacob's research was supported by grant EP/K009362/1.

\appendix

\section{Proofs of properties of the shift estimators}

\subsection{Prior expectations when the prior concentrates \label{sub:proof:prior:expansions}}

Before looking at posterior expectations, we first study prior expectations
and will then retrieve posterior expectations as ratios of prior expectations,
using Bayes formula as in Eq. \eqref{eq:bayesformula}.
\begin{lem}
\label{lemma:prior:expansion}Under Assumptions \textbf{A1-A2-A3,
}we have the following asymptotic behavior of expectations with respect
to the prior:
\[
\mathbb{E}_{\tau}\left[\varphi\left(\Theta\right)\right]=\varphi\left(\theta^{\star}\right)+\frac{\tau^{2}}{2}\sum_{i=1}^{d}\sum_{j=1}^{d}\nabla_{ij}^{2}\varphi\left(\theta^{\star}\right)\Sigma_{ij}+\mathcal{O}\left(\tau^{4}\right).
\]

\end{lem}
\begin{proof}[Proof of Lemma \ref{lemma:prior:expansion}] Letting
$\delta$ be as in \textbf{A2}, we cut the integral as follows:
\begin{eqnarray*}
\mathbb{E}_{\tau}\left[\varphi\left(\Theta\right)\right] & = & \int_{B_{\Sigma}(\theta^{\star},\delta)}\varphi\left(\theta\right)p_{\tau}\left(\theta\right)d\theta\,+\,\int_{B_{\Sigma}^{c}(\theta^{\star},\delta)}\varphi\left(\theta\right)p_{\tau}\left(\theta\right)d\theta
\end{eqnarray*}
where $p_{\tau}$ denotes the probability density function of $\mathcal{N}\left(\theta^{\star},\tau^{2}\Sigma\right)$,
and $B_{\Sigma}^{c}(\theta^{\star},\delta)$ is the complement of
$B_{\Sigma}(\theta^{\star},\delta)$ in $\mathbb{R}^{d}$. The second
integral is shown to be negligible compared to the first as $\tau\to0$.
Indeed, let $\tau\in\left(0,\tau_{0}\right)$, where $\tau_{0}$ is
as in \textbf{A3}, and denote by $\left|\tau^{2}\Sigma\right|=\tau^{2d}\left|\Sigma\right|$
the determinant of $\tau^{2}\Sigma$, then we have 
\begin{eqnarray*}
 &  & \left|\int_{B_{\Sigma}^{c}(\theta^{\star},\delta)}\varphi\left(\theta\right)\frac{1}{\left(2\pi\right)^{d/2}\left|\tau^{2}\Sigma\right|^{1/2}}\exp\left(-\frac{1}{2\tau^{2}}\left(\theta-\theta^{\star}\right)^{T}\Sigma^{-1}\left(\theta-\theta^{\star}\right)\right)d\theta\right|\\
 & = & \left|\int_{B_{\Sigma}^{c}(\theta^{\star},\delta)}\varphi\left(\theta\right)\frac{\left(2\pi\tau_{0}^{2}\right)^{d/2}}{\left(2\pi\tau^{2}\right)^{d/2}}\exp\left(-\left(\frac{1}{2\tau^{2}}-\frac{1}{2\tau_{0}^{2}}\right)\left(\theta-\theta^{\star}\right)^{T}\Sigma^{-1}\left(\theta-\theta^{\star}\right)\right)p_{\tau_{\text{0}}}\left(d\theta\right)\right|\\
 & \leq & \left|\int_{B_{\Sigma}^{c}(\theta^{\star},\delta)}\varphi\left(\theta\right)p_{\tau_{0}}\left(d\theta\right)\right|\times\frac{\tau_{0}^{d}}{\tau^{d}}\exp\left(-\left(\frac{1}{2\tau^{2}}-\frac{1}{2\tau_{0}^{2}}\right)\delta^{2}\right).
\end{eqnarray*}
Bounding $\left|\int_{B_{\Sigma}^{c}(\theta^{\star},\delta)}\varphi\left(\theta\right)p_{\tau_{0}}\left(d\theta\right)\right|$
by $\mathbb{E}_{\tau_{0}}\left[\left|\varphi(\Theta)\right|\right]$,
this proves 
\begin{equation}
\mathbb{E}_{\tau}\left[\varphi\left(\Theta\right)1_{B_{\Sigma}^{c}(\theta^{\star},\delta)}(\Theta)\right]=o\left(\tau^{k}\right)\mbox{ for all }k\in\mathbb{N}.\label{eq:integral:outside:ball}
\end{equation}
The other integral is over the ball $B_{\Sigma}(\theta^{\star},\delta)$,
i.e. close to $\theta^{\star}$; hence we perform a Taylor expansion
of the integrand, using the multi-index notation: for $\alpha=(\alpha_{1},\ldots,\alpha_{d})\in\left\{ 0,1,2,\ldots\right\} ^{d}$,
$\left|\alpha\right|:=\sum_{i=1}^{d}\alpha_{i}$, $\partial^{\alpha}\varphi:=\partial^{\left|\alpha\right|}\varphi/\partial^{\alpha_{1}}\partial^{\alpha_{2}}\ldots\partial^{\alpha_{d}}$,
$\theta^{\alpha}:=\theta_{1}^{\alpha_{1}}\theta_{2}^{\alpha_{2}}\ldots\theta_{d}^{\alpha_{d}}$
for all $\theta\in\mathbb{R}^{d}$, and $\alpha!=\alpha_{1}!\alpha_{2}!\ldots\alpha_{d}!$.
The Taylor expansion of $\varphi$ around $\theta^{\star}$ to the
third order reads:
\begin{align*}
\forall\theta\in B_{\Sigma}(\theta^{\star},\delta)\quad\varphi(\theta) & =\varphi\left(\theta^{\star}\right)+\sum_{\left|\alpha\right|\leq3}\frac{\partial^{\alpha}\varphi\left(\theta^{\star}\right)}{\alpha!}\left(\theta-\theta^{\star}\right)^{\alpha}+R_{3}\left(\theta,\theta^{\star}\right)
\end{align*}
where $R_{3}(\theta,\theta^{\star})$ is the remainder term, for which
we use the Lagrange form: 
\[
\forall\theta\in B_{\Sigma}(\theta^{\star},\delta)\quad\exists c_{\theta}\in[0,1]\quad R_{3}\left(\theta,\theta^{\star}\right)=\sum_{\vert\alpha\vert=4}\partial^{\alpha}\varphi\left(\theta^{\star}+c_{\theta}\left(\theta-\theta^{\star}\right)\right)\frac{\left(\theta-\theta^{\star}\right)^{\alpha}}{\alpha!}.
\]

We now use the symmetry of the normal distribution $\mathcal{N}\left(\theta^{\star},\tau^{2}\Sigma\right)$
on the ball $B_{\Sigma}(\theta^{\star},\delta)$. For any function
$f$ on $B_{\Sigma}(\theta^{\star},\delta)$ such that $f(\theta-\theta^{\star})=-f(\theta^{\star}-\theta)$
for all $\theta\in B_{\Sigma}(\theta^{\star},\delta)$, the integral
of $f$ with respect to $\mathcal{N}\left(\theta^{\star},\tau^{2}\Sigma\right)$
on $B_{\Sigma}(\theta^{\star},\delta)$ is zero. Thus, odd powers
of $\left(\theta-\theta^{\star}\right)$ integrate to zero, leading
to 
\begin{eqnarray*}
\int_{B_{\Sigma}(\theta^{\star},\delta)}\sum_{\left|\alpha\right|=1}\frac{\partial^{\alpha}\varphi\left(\theta^{\star}\right)}{\alpha!}\left(\theta-\theta^{\star}\right)^{\alpha}p_{\tau}\left(\theta\right)d\theta & = & 0,\\
\int_{B_{\Sigma}(\theta^{\star},\delta)}\sum_{\left|\alpha\right|=3}\frac{\partial^{\alpha}\varphi\left(\theta^{\star}\right)}{\alpha!}\left(\theta-\theta^{\star}\right)^{\alpha}p_{\tau}\left(\theta\right)d\theta & = & 0.
\end{eqnarray*}
The second order term can be written
\begin{eqnarray*}
 &  & \int_{B_{\Sigma}(\theta^{\star},\delta)}\sum_{\left|\alpha\right|=2}\frac{\partial^{\alpha}\varphi\left(\theta^{\star}\right)}{\alpha!}\left(\theta-\theta^{\star}\right)^{\alpha}p_{\tau}\left(\theta\right)d\theta\\
 & = & \frac{1}{2}\sum_{i=1}^{d}\sum_{j=1}^{d}\nabla_{ij}^{2}\varphi\left(\theta^{\star}\right)\int_{B_{\Sigma}(\theta^{\star},\delta)}\left(\theta_{i}-\theta_{i}^{\star}\right)\left(\theta_{j}-\theta_{j}^{\star}\right)p_{\tau}\left(\theta\right)d\theta\\
 & = & \frac{1}{2}\sum_{i=1}^{d}\sum_{j=1}^{d}\nabla_{ij}^{2}\varphi\left(\theta^{\star}\right)\left(\tau^{2}\Sigma_{ij}-\int_{B_{\Sigma}^{c}(\theta^{\star},\delta)}\left(\theta_{i}-\theta_{i}^{\star}\right)\left(\theta_{j}-\theta_{j}^{\star}\right)p_{\tau}\left(\theta\right)d\theta\right),
\end{eqnarray*}
where we have integrated over $\mathbb{R}^{d}$ and subtracted the
integral over $B_{\Sigma}^{c}(\theta^{\star},\delta)$. Using Eq.
\eqref{eq:integral:outside:ball}, the integral over $B_{\Sigma}^{c}(\theta^{\star},\delta)$
is negligible, that is:
\[
\int_{B_{\Sigma}(\theta^{\star},\delta)}\sum_{\left|\alpha\right|=2}\frac{\partial^{\alpha}\varphi\left(\theta^{\star}\right)}{\alpha!}\left(\theta-\theta^{\star}\right)^{\alpha}p_{\tau}\left(\theta\right)d\theta=\frac{\tau^{2}}{2}\sum_{i=1}^{d}\sum_{j=1}^{d}\nabla_{ij}^{2}\varphi\left(\theta^{\star}\right)\Sigma_{ij}+o\left(\tau^{k}\right)\mbox{ for all }k\in\mathbb{N}.
\]
Finally, we deal with the remainder term using Assumption \textbf{A2}:
\begin{align*}
 & \left|\int_{B_{\Sigma}(\theta^{\star},\delta)}\sum_{\vert\alpha\vert=4}\partial^{\alpha}\varphi\left(\theta^{\star}+c_{\theta}\left(\theta-\theta^{\star}\right)\right)\frac{\left(\theta-\theta^{\star}\right)^{\alpha}}{\alpha!}p_{\tau}\left(\theta\right)d\theta\right|\\
 & \leq K\sum_{\vert\alpha\vert=4}\frac{1}{\alpha!}\int_{B_{\Sigma}(\theta^{\star},\delta)}\left|\left(\theta-\theta^{\star}\right)^{\alpha}\right|p_{\tau}\left(\theta\right)d\theta\\
 & \leq K\sum_{\vert\alpha\vert=4}\frac{1}{\alpha!}\int\left|\left(\theta-\theta^{\star}\right)^{\alpha}\right|p_{\tau}\left(\theta\right)d\theta
\end{align*}
Without computing this term exactly, we want to exhibit a constant
times $\tau^{4}$. Let us perform a change of variable $z_{i}=\tau^{-1}\left(\theta_{i}-\theta_{i}^{\star}\right)$
for all $i\in\left\{ 1,\ldots,d\right\} $. We obtain, for each $\alpha$
such that $\left|\alpha\right|=4$, 
\begin{eqnarray*}
 &  & \int\left|\left(\theta-\theta^{\star}\right)^{\alpha}\right|\frac{1}{\left(2\pi\right)^{d/2}\left|\tau^{2}\Sigma\right|^{1/2}}\exp\left(-\frac{1}{2\tau^{2}}\left(\theta-\theta^{\star}\right)^{T}\Sigma^{-1}\left(\theta-\theta^{\star}\right)\right)d\theta\\
 & = & \int\left|\left(\tau z\right)^{\alpha}\right|\frac{1}{\left(2\pi\right)^{d/2}\tau^{d}\left|\Sigma\right|^{1/2}}\exp\left(-\frac{1}{2}z^{T}\Sigma^{-1}z\right)\,\,\tau^{d}\,\,dz\\
 & = & \tau^{\left|\alpha\right|}\int\left|z\right|^{\alpha}\frac{1}{\left(2\pi\right)^{d/2}\left|\Sigma\right|^{1/2}}\exp\left(-\frac{1}{2}z^{T}\Sigma^{-1}z\right)dz.
\end{eqnarray*}
The fourth moments of a multivariate normal distribution are finite,
thus we can conclude 
\[
\left|\int_{B_{\Sigma}(\theta^{\star},\delta)}R_{3}\left(\theta,\theta^{\star}\right)p_{\tau}\left(\theta\right)d\theta\right|\leq\tau^{4}K\,C(\Sigma)
\]
for the finite constant $C(\Sigma)=\sum_{\vert\alpha\vert=4}\frac{1}{\alpha!}\int\left|z\right|^{\alpha}\left(2\pi\right)^{-d/2}\left|\Sigma\right|^{-1/2}\exp\left(-\frac{1}{2}z^{T}\Sigma^{-1}z\right)dz$.

Combining all the terms, we finally obtain
\begin{align*}
\mathbb{E}_{\tau}\left[\varphi\left(\Theta\right)1_{B_{\Sigma}(\theta^{\star},\delta)}(\Theta)\right] & =\varphi\left(\theta^{\star}\right)+\frac{\tau^{2}}{2}\sum_{i=1}^{d}\sum_{j=1}^{d}\nabla_{ij}^{2}\varphi\left(\theta^{\star}\right)\Sigma_{ij}+\mathcal{O}\left(\tau^{4}\right),
\end{align*}
which concludes the proof. \end{proof}

\subsection{Posterior expectations when the prior concentrates (Lemma \ref{lemma:posterior:expansion})
\label{sub:proof:posteriorexpansions}}

\begin{proof}[Proof of Lemma \ref{lemma:posterior:expansion}]

Lemma \ref{lemma:prior:expansion} gives an expansion of prior expectations
of test functions $\varphi$ when the prior variance parameter $\tau$
goes to zero. Posterior expectations are defined by Bayes formula
as in Eq. \eqref{eq:bayesformula}. We thus apply Lemma \ref{lemma:prior:expansion}
to two test functions, $\phi:\,\theta\mapsto\varphi(\theta)\times\mathcal{L}(\theta)$
and $\theta\mapsto\mathcal{L}(\theta)$. We have $\nabla\phi=\nabla\varphi\times\mathcal{L}+\varphi\times\nabla\mathcal{L}$
and thus $\nabla^{2}\phi=\nabla^{2}\varphi\times\mathcal{L}+2\nabla\varphi\nabla\mathcal{L}^{T}+\varphi\times\nabla^{2}\mathcal{L}$.
Furthermore we can write $\nabla\mathcal{L}=\nabla\ell\times\mathcal{L}$
and $\nabla^{2}\mathcal{L}=(\nabla^{2}\ell+\nabla\ell\nabla\ell^{T})\times\mathcal{L}$.
Thus, we have $\nabla^{2}\phi=(\nabla^{2}\varphi+2\nabla\varphi\nabla\ell^{T}+\varphi\times(\nabla^{2}\ell+\nabla\ell\nabla\ell^{T}))\times\mathcal{L}$,
and Lemma \ref{lemma:prior:expansion} yields, for the test function
$\phi$,
\begin{eqnarray*}
\mathbb{E}_{\tau}\left[\varphi\left(\Theta\right)\mathcal{L}\left(\Theta\right)\right] & = & \mathcal{L}(\theta^{\star})\left(\varphi\left(\theta^{\star}\right)+\frac{\tau^{2}}{2}\sum_{i=1}^{d}\sum_{j=1}^{d}\left\{ \nabla_{ij}^{2}\varphi+2\nabla_{i}\varphi\nabla_{j}\ell+\varphi\times(\nabla_{ij}^{2}\ell+\nabla_{i}\ell\nabla_{j}\ell)\right\} \left(\theta^{\star}\right)\Sigma_{ij}\right)\\
 &  & +\mathcal{O}\left(\tau^{4}\right),
\end{eqnarray*}
and for the test function $\mathcal{L}$, 
\begin{eqnarray*}
\mathbb{E}_{\tau}\left[\mathcal{L}\left(\Theta\right)\right] & = & \mathcal{L}(\theta^{\star})\left(1+\frac{\tau^{2}}{2}\sum_{i=1}^{d}\sum_{j=1}^{d}\left\{ \nabla_{ij}^{2}\ell+\nabla_{i}\ell\nabla_{j}\ell\right\} \left(\theta^{\star}\right)\Sigma_{ij}\right)+\mathcal{O}\left(\tau^{4}\right).
\end{eqnarray*}
The ratio of both expansions yields
\begin{eqnarray*}
\check{\mathbb{E}}_{\tau}\left[\varphi\left(\Theta\right)\right] & = & \varphi\left(\theta^{\star}\right)+\frac{\tau^{2}}{2}\sum_{i=1}^{d}\sum_{j=1}^{d}\left\{ \nabla_{ij}^{2}\varphi+2\nabla_{i}\varphi\nabla_{j}\ell+\varphi\times(\nabla_{ij}^{2}\ell+\nabla_{i}\ell\nabla_{j}\ell)\right\} \left(\theta^{\star}\right)\Sigma_{ij}\\
 &  & -\varphi\left(\theta^{\star}\right)\frac{\tau^{2}}{2}\sum_{i=1}^{d}\sum_{j=1}^{d}\left\{ \nabla_{ij}^{2}\ell+\nabla_{i}\ell\nabla_{j}\ell\right\} \left(\theta^{\star}\right)\Sigma_{ij}+\mathcal{O}\left(\tau^{4}\right)\\
 & = & \varphi\left(\theta^{\star}\right)+\frac{\tau^{2}}{2}\sum_{i=1}^{d}\sum_{j=1}^{d}\left(\nabla_{ij}^{2}\varphi\left(\theta^{\star}\right)+2\nabla_{i}\varphi\left(\theta^{\star}\right)\nabla_{j}\ell\left(\theta^{\star}\right)\right)\Sigma_{ij}+\mathcal{O}\left(\tau^{4}\right),
\end{eqnarray*}
which concludes the proof. \end{proof}

\subsection{Error of shift estimators when the prior concentrates (Theorem \ref{theorem:estimators})
\label{sub:proof:estimators}}

We now show that the shift estimators $S_{\tau}^{(1)}\left(\theta^{\star}\right)$
and $S_{\tau}^{(2)}\left(\theta^{\star}\right)$ defined in Theorem
\ref{theorem:estimators}, are consistent when $\tau\to0$, and that
the error is of order $\tau^{2}$.

\begin{proof}[Proof of Theorem \ref{theorem:estimators}] We have
from Lemma \ref{lemma:stein} 
\[
\check{\mathbb{E}}_{\tau}\left[\left(\Theta-\theta^{\star}\right)\right]=\tau^{2}\Sigma\,\check{\mathbb{E}}_{\tau}\left[\nabla\ell(\Theta)\right]
\]
and under \textbf{A1-A2-A3, }from Lemma \ref{lemma:posterior:expansion},
element-wise for each $k\in\left\{ 1,\ldots,d\right\} $: 
\[
\check{\mathbb{E}}_{\tau}\left[\nabla_{k}\ell(\Theta)\right]=\nabla_{k}\ell(\theta^{\star})+\frac{\tau^{2}}{2}\sum_{i=1}^{d}\sum_{j=1}^{d}\left(\nabla_{ij}^{2}\left\{ \nabla_{k}\ell\left(\theta^{\star}\right)\right\} +2\nabla_{i}\left\{ \nabla_{k}\ell\left(\theta^{\star}\right)\right\} \nabla_{j}\ell\left(\theta^{\star}\right)\right)\Sigma_{ij}+\mathcal{O}(\tau^{4}).
\]
Thus 
\[
\left\{ \tau^{-2}\Sigma^{-1}\,\check{\mathbb{E}}_{\tau}\left[\left(\Theta-\theta^{\star}\right)\right]\right\} _{k}=\nabla_{k}\ell(\theta^{\star})+\tau^{2}\mathcal{E}_{k}(\theta^{\star})+\mathcal{O}\left(\tau^{4}\right),
\]
where $\mathcal{E}_{k}(\theta^{\star})$ involves derivatives of the
log-likelihood $\ell$ and elements of $\Sigma$. This is Eq. \eqref{eq:score:approx}.

Now consider the second order derivative. We have, using Lemma \ref{lemma:stein},
\begin{eqnarray*}
\check{\mathbb{V}}_{\tau}\left[\Theta\right] & = & \check{\mathbb{E}}_{\tau}\left[\left(\Theta-\theta^{\star}\right)\left(\Theta-\theta^{\star}\right)^{T}\right]-\check{\mathbb{E}}_{\tau}\left[\left(\Theta-\theta^{\star}\right)\right]\check{\mathbb{E}}_{\tau}\left[\left(\Theta-\theta^{\star}\right)\right]^{T}\\
 & = & \tau^{2}\Sigma+\tau^{4}\Sigma\,\check{\mathbb{E}}_{\tau}\left[\nabla^{2}\ell(\Theta)+\nabla\ell(\Theta)\nabla\ell(\Theta)^{T}\right]\,\Sigma-\tau^{4}\Sigma\,\check{\mathbb{E}}_{\tau}\left[\nabla\ell(\Theta)\right]\check{\mathbb{E}}_{\tau}\left[\nabla\ell(\Theta)\right]^{T}\,\Sigma\\
 & = & \tau^{2}\Sigma+\tau^{4}\Sigma\,\left(\check{\mathbb{E}}_{\tau}\left[\nabla^{2}\ell(\Theta)+\nabla\ell(\Theta)\nabla\ell(\Theta)^{T}\right]-\check{\mathbb{E}}_{\tau}\left[\nabla\ell(\Theta)\right]\check{\mathbb{E}}_{\tau}\left[\nabla\ell(\Theta)\right]^{T}\right)\,\Sigma,
\end{eqnarray*}
so that 
\[
\tau^{-4}\Sigma^{-1}\left(\check{\mathbb{V}}_{\tau}\left[\Theta\right]-\tau^{2}\Sigma\right)\Sigma^{-1}=\left(\check{\mathbb{E}}_{\tau}\left[\nabla^{2}\ell(\Theta)+\nabla\ell(\Theta)\nabla\ell(\Theta)^{T}\right]-\check{\mathbb{E}}_{\tau}\left[\nabla\ell(\Theta)\right]\check{\mathbb{E}}_{\tau}\left[\nabla\ell(\Theta)\right]^{T}\right).
\]
Under \textbf{A1-A2-A3} we use Lemma \ref{lemma:posterior:expansion},
element-wise for $k,l\in\left\{ 1,\ldots,d\right\} $:
\begin{eqnarray*}
\check{\mathbb{E}}_{\tau}\left[\nabla_{k}\ell(\Theta)\right]\check{\mathbb{E}}_{\tau}\left[\nabla_{l}\ell(\Theta)\right] & = & \nabla_{k}\ell(\theta^{\star})\nabla_{l}\ell(\theta^{\star})+\mathcal{O}\left(\tau^{2}\right),\\
\check{\mathbb{E}}_{\tau}\left[\nabla_{kl}^{2}\ell(\Theta)+\nabla_{k}\ell(\Theta)\nabla_{l}\ell(\Theta)\right] & = & \nabla_{kl}^{2}\ell(\theta^{\star})+\nabla_{k}\ell(\theta^{\star})\nabla_{l}\ell(\theta^{\star})+\mathcal{O}\left(\tau^{2}\right),
\end{eqnarray*}
so that 
\[
\left\{ \tau^{-4}\Sigma^{-1}\left(\check{\mathbb{V}}_{\tau}\left[\Theta\right]-\tau^{2}\Sigma\right)\Sigma^{-1}\right\} _{kl}=\nabla_{kl}^{2}\ell(\theta^{\star})+\mathcal{O}\left(\tau^{2}\right).
\]
This is Eq. \eqref{eq:hessian:approx}, which concludes the proof.
\end{proof}

\section{Proofs of properties of Monte Carlo shift estimators}

\subsection{Identities for the expectation and variance of ratios of random variables\label{sub:Expectation-and-variance-ratio-rv}}

We provide identities for the expectation and variance of ratios of
random variables which are due to \cite{Koop1972}.
\begin{lem}
\label{lemma:koop}

Let $X$ and $Y$ be univariate random variables with finite two first
moments, and such that $X$ is almost surely non-negative. Let $\mu_{X}=\mathbb{E}X$
and $\mu_{Y}=\mathbb{E}Y$. Let $\Delta X=(X-\mu_{X})/\mu_{X}$ and
$\Delta Y=(Y-\mu_{Y})/\mu_{Y}$. Then we have 
\begin{equation}
\mathbb{E}\left(\frac{Y}{X}\right)=\frac{\mathbb{E}\left(Y\right)}{\mathbb{E}\left(X\right)}-\frac{1}{\mathbb{E}\left(X\right)}\mathbb{C}\left(\frac{Y}{X},X\right).\label{eq:koop:expectation}
\end{equation}
and 
\begin{align}
\mathbb{V}\left(\frac{Y}{X}\right) & =\left(\frac{\mu_{Y}}{\mu_{X}}\right)^{2}\left(\mathbb{V}\left(\Delta Y-\Delta X\right)+\mathbb{V}\left(\Delta X\Delta Y\right)+2\mathbb{C}\text{\text{ov}}\left(\Delta X-\Delta Y,\Delta X\Delta Y\right)\right)\nonumber \\
 & -\mathbb{V}\left(\frac{Y}{X}\left(\Delta X\right)^{2}\right)+2\mathbb{C}\left(\frac{Y}{X},\frac{Y}{X}\left(\Delta X\right)^{2}\right).\label{eq:koop:variance}
\end{align}

\end{lem}
\begin{proof}[Proof of Lemma \ref{lemma:koop}]We can write
\begin{eqnarray*}
\frac{Y}{X} & = & \frac{\mu_{Y}}{\mu_{X}}\frac{1+\Delta Y}{1+\Delta X}\\
 & = & \frac{\mu_{Y}}{\mu_{X}}\left(1+\Delta Y\right)-\frac{Y}{X}\Delta X,
\end{eqnarray*}
which yields Eq. \eqref{eq:koop:expectation}. We can also write

\begin{align*}
\frac{Y}{X} & =\frac{\mu_{Y}}{\mu_{X}}\frac{1+\Delta Y}{1+\Delta X}\\
 & =\frac{\mu_{Y}}{\mu_{X}}\left(1+\Delta Y\right)\left(1-\Delta X+\mu_{X}\frac{\left(\Delta X\right)^{2}}{X}\right)\\
 & =\frac{\mu_{Y}}{\mu_{X}}\left(1-\Delta X+\Delta Y-\Delta X\Delta Y\right)+\frac{Y}{X}\left(\Delta X\right)^{2}.
\end{align*}
Hence we have 
\[
\mathbb{V}\left(\frac{Y}{X}-\frac{Y}{X}\left(\Delta X\right)^{2}\right)=\left(\frac{\mu_{Y}}{\mu_{X}}\right)^{2}\left(\mathbb{V}\left(-\Delta X+\Delta Y\right)+\mathbb{V}\left(\Delta X\Delta Y\right)+2\mathbb{C}\left(\Delta X-\Delta Y,\Delta X\Delta Y\right)\right).
\]
This yields Eq. \eqref{eq:koop:variance}.

\end{proof}

\subsection{Mean squared error of Monte Carlo shift estimators (Lemmas \ref{lemma:biasvarianceMCshift1}
and \ref{lemma:biasvarianceMCshift2})\label{sub:proof:shift:MSE}}

\begin{proof}[Proof of Lemma \ref{lemma:biasvarianceMCshift1}]

By the Lindeberg-Feller theorem, we can obtain central limit theorems
for $A_{N,\tau_{N}}$ and $B_{N,\tau_{N}}$, for instance under the
condition that there exists $\delta>0$ such that 
\[
\lim_{N\rightarrow\infty}\mathbb{E}_{\tau_{N}}\left[\left\vert \widehat{\mathcal{L}}\left(\Theta\right)\Theta\right\vert ^{2+\delta}\right]<\infty,\quad\lim_{N\rightarrow\infty}\mathbb{E}_{\tau_{N}}\left[\widehat{\mathcal{L}}\left(\Theta\right)^{2+\delta}\right]<\infty.
\]
We could then obtain a central limit theorem for the ratio $A_{N,\tau_{N}}/B_{N,\tau_{N}}$
by the delta method. We would then find that the asymptotic variance
is equal to zero. In other words, 
\begin{equation}
\sqrt{N}\left(\frac{A_{N,\tau_{N}}}{B_{N,\tau_{N}}}-\theta^{\star}\right)\xrightarrow[N\to\infty]{\mathbb{P}}0.\label{eq:failed:deltamethod}
\end{equation}
Under moment assumptions such as \textbf{B2}, this yields $\mathbb{V}(\sqrt{N}A_{N,\tau_{N}}/B_{N,\tau_{N}})\to0$,
but does not inform on the rate at which this convergence happens,
for instance as a function of $\tau_{N}$. Thus the delta method is
too coarse to yield exact rates of convergence of the bias and variance
of $S_{N,\tau_{N}}^{(1)}\left(\theta^{\star}\right)$. An alternative
would be to perform a Taylor expansion of $1/B_{N,\tau_{N}}$, but
the remainder is then difficult to control, as it requires assumptions
on the inverse moments of $\widehat{\mathcal{L}}\left(\Theta\right)$.
Instead, we will use identities for the expectation and variance of
a ratio of random variables, due to \cite{Koop1972} and recalled
in Section \ref{sub:Expectation-and-variance-ratio-rv}.

We first look at the variance of $A_{N,\tau_{N}}/B_{N,\tau_{N}}$.
We use Lemma \ref{lemma:koop} to get an expression for the variance
of this ratio, and look at each term on the right hand side of Eq.
\eqref{eq:koop:variance}. The leading term will prove to be the term
written $\left(\mu_{Y}/\mu_{X}\right)^{2}\mathbb{V}\left(\Delta Y-\Delta X\right)$
in Eq. \eqref{eq:koop:variance}, that is, 
\begin{align*}
\left(\frac{\mathbb{E}_{\tau_{N}}\left[A_{N,\tau_{N}}\right]}{\mathbb{E}_{\tau_{N}}\left[B_{N,\tau_{N}}\right]}\right)^{2}\mathbb{V}_{\tau_{N}}\left(\Delta A_{N,\tau_{N}}-\Delta B_{N,\tau_{N}}\right)= & \left(\frac{\mathbb{E}_{\tau_{N}}\left[\mathcal{L}\left(\Theta\right)\Theta\right]}{\mathbb{E}_{\tau_{N}}\left[\mathcal{L}\left(\Theta\right)\right]}\right)^{2}\frac{1}{N}\mathbb{V}_{\tau_{N}}\left(\frac{\Theta\widehat{\mathcal{L}}(\Theta)}{\mathbb{E}_{\tau_{N}}\left[\Theta\mathcal{L}\left(\Theta\right)\right]}-\frac{\widehat{\mathcal{L}}(\Theta)}{\mathbb{E}_{\tau_{N}}\left[\mathcal{L}\left(\Theta\right)\right]}\right),
\end{align*}
with $Y=A_{N,\tau_{N}}$ and $X=B_{N,\tau_{N}}$. Indeed, using the
variance decomposition formula and $\mathbb{V}(\mathcal{\widehat{L}}(\theta)/\mathcal{L}(\theta))=\upsilon_{M}\left(\theta^{\star}\right)$,
we obtain 
\begin{align*}
\mathbb{V}_{\tau_{N}}\left(\frac{\Theta\widehat{\mathcal{L}}(\Theta)}{\mathbb{E}_{\tau_{N}}\left[\Theta\mathcal{L}\left(\Theta\right)\right]}-\frac{\widehat{\mathcal{L}}(\Theta)}{\mathbb{E}_{\tau_{N}}\left[\mathcal{L}\left(\Theta\right)\right]}\right) & =\mathbb{V}_{\tau_{N}}\left(\frac{\Theta\mathcal{L}(\Theta)}{\mathbb{E}_{\tau_{N}}\left[\Theta\mathcal{L}\left(\Theta\right)\right]}-\frac{\mathcal{L}(\Theta)}{\mathbb{E}_{\tau_{N}}\left[\mathcal{L}\left(\Theta\right)\right]}\right)\\
 & +\mathbb{E}_{\tau_{N}}\left(\left(\frac{\Theta}{\mathbb{E}_{\tau_{N}}\left[\Theta\mathcal{L}\left(\Theta\right)\right]}-\frac{1}{\mathbb{E}_{\tau_{N}}\left[\mathcal{L}\left(\Theta\right)\right]}\right)^{2}\upsilon_{M}\left(\theta^{\star}\right)\mathcal{L}\left(\Theta\right)^{2}\right)\\
 & =\mathbb{E}_{\tau_{N}}\left(\left(\frac{\Theta\mathcal{L}\left(\Theta\right)}{\mathbb{E}_{\tau_{N}}\left[\Theta\mathcal{L}\left(\Theta\right)\right]}-\frac{\mathcal{L}\left(\Theta\right)}{\mathbb{E}_{\tau_{N}}\left[\mathcal{L}\left(\Theta\right)\right]}\right)^{2}\right)\left(1+\upsilon_{M}\left(\theta^{\star}\right)\right).
\end{align*}
To study how this term varies with $\tau_{N}$, we use the prior expansion
of Lemma \ref{lemma:prior:expansion}, to obtain for two test functions
$\varphi$ and $\psi$ such that $\varphi$, $\psi$ and $\varphi\times\psi$
satisfy the assumptions of the lemma, the expansion
\begin{equation}
\frac{\mathbb{E}_{\tau_{N}}\left[\varphi(\Theta)\psi(\Theta)\right]}{\mathbb{E}_{\tau_{N}}\left[\varphi(\Theta)\right]\mathbb{E}_{\tau_{N}}\left[\psi(\Theta)\right]}=1+\tau_{N}^{2}\sum_{i=1}^{d}\sum_{j=1}^{d}\Sigma_{ij}\nabla_{i}\log\varphi\left(\theta^{\star}\right)\nabla_{j}\log\psi\left(\theta^{\star}\right)+\mathcal{O}\left(\tau_{N}^{4}\right).\label{eq:expansion:twotestfunctions}
\end{equation}
Thus, for test functions such as $\theta\mapsto\theta\mathcal{L}(\theta)$
and $\theta\mapsto\mathcal{L}(\theta)$, we obtain 
\begin{align*}
\mathbb{E}_{\tau_{N}}\left(\left(\frac{\Theta\mathcal{L}\left(\Theta\right)}{\mathbb{E}_{\tau_{N}}\left[\Theta\mathcal{L}\left(\Theta\right)\right]}-\frac{\mathcal{L}\left(\Theta\right)}{\mathbb{E}_{\tau_{N}}\left[\mathcal{L}\left(\Theta\right)\right]}\right)^{2}\right) & =\frac{\mathbb{E}_{\tau_{N}}\left[\Theta^{2}\mathcal{L}\left(\Theta\right)^{2}\right]}{\mathbb{E}_{\tau_{N}}\left[\Theta\mathcal{L}\left(\Theta\right)\right]^{2}}+\frac{\mathbb{E}_{\tau_{N}}\left[\mathcal{L}\left(\Theta\right)^{2}\right]}{\mathbb{E}_{\tau_{N}}\left[\mathcal{L}\left(\Theta\right)\right]^{2}}-2\frac{\mathbb{E}_{\tau_{N}}\left[\Theta\mathcal{L}\left(\Theta\right)^{2}\right]}{\mathbb{E}_{\tau_{N}}\left[\Theta\mathcal{L}\left(\Theta\right)\right]\mathbb{E}_{\tau_{N}}\left[\mathcal{L}\left(\Theta\right)\right]}\\
 & =\tau_{N}^{2}\Sigma\left(\frac{1}{\theta^{\star}}\right)^{2}+\mathcal{O}\left(\tau_{N}^{4}\right).
\end{align*}
Therefore, we obtain the following expression for the leading term
of the variance of $A_{N,\tau_{N}}/B_{N,\tau_{N}}$:
\begin{align}
\left(\frac{\mathbb{E}_{\tau_{N}}\left[A_{N,\tau_{N}}\right]}{\mathbb{E}_{\tau_{N}}\left[B_{N,\tau_{N}}\right]}\right)^{2}\mathbb{V}_{\tau_{N}}\left(\Delta A_{N,\tau_{N}}-\Delta B_{N,\tau_{N}}\right) & =\left(\theta^{\star}+\mathcal{O}\left(\tau_{N}^{2}\right)\right)^{2}\frac{1}{N}\left(\tau_{N}^{2}\Sigma\left(\frac{1}{\theta^{\star}}\right)^{2}+\mathcal{O}\left(\tau_{N}^{4}\right)\right)\left(1+\upsilon_{M}\left(\theta^{\star}\right)\right)\nonumber \\
 & =\frac{\tau_{N}^{2}}{N}\Sigma\left(1+\upsilon_{M}\left(\theta^{\star}\right)\right)+\mathcal{O}\left(\frac{\tau_{N}^{4}}{N}\right).\label{eq:asymptoticvariance:term1}
\end{align}

Next, we need to control the other terms, that is, $\mathbb{V}\left(\Delta X\Delta Y\right)$
, $\mathbb{C}\left(\Delta X-\Delta Y,\Delta X\Delta Y\right)$, $\mathbb{V}\left(\frac{Y}{X}\left(\Delta X\right)^{2}\right)$
and $\mathbb{C}\left(\frac{Y}{X},\frac{Y}{X}\left(\Delta X\right)^{2}\right)$
in Eq. \eqref{eq:koop:variance}. We proceed term by term. The term
$\mathbb{V}\left(\Delta X\Delta Y\right)$ can be written 
\begin{align*}
\mathbb{V}_{\tau_{N}}\left(\Delta A_{N,\tau_{N}}\Delta B_{N,\tau_{N}}\right) & =\frac{1}{N^{2}}\mathbb{V}_{\tau_{N}}\left(\sqrt{N}\Delta A_{N,\tau_{N}}\sqrt{N}\Delta B_{N,\tau_{N}}\right).
\end{align*}
Then we use the following formula, for two dependent variables $X$
and $Y$:
\[
\mathbb{V}\left(XY\right)=\mathbb{C}\left(X^{2},Y^{2}\right)+\left[\mathbb{V}\left(X\right)+\mathbb{E}\left(X\right)^{2}\right]\left[\mathbb{V}\left(Y\right)+\mathbb{E}\left(Y\right)^{2}\right]-\left[\mathbb{C}\left(X,Y\right)+\mathbb{E}\left(X\right)\mathbb{E}\left(Y\right)\right]^{2}.
\]
This yields 
\begin{align*}
\mathbb{V}_{\tau_{N}}\left(\sqrt{N}\Delta A_{N,\tau_{N}}\sqrt{N}\Delta B_{N,\tau_{N}}\right) & =\mathbb{C}_{\tau_{N}}\left(N\left(\Delta A_{N,\tau_{N}}\right)^{2},N\left(\Delta B_{N,\tau_{N}}\right)^{2}\right)\\
 & +\mathbb{V}_{\tau_{N}}\left(\sqrt{N}\Delta A_{N,\tau_{N}}\right)\mathbb{V}_{\tau_{N}}\left(\sqrt{N}\Delta B_{N,\tau_{N}}\right)\\
 & -\mathbb{C}_{\tau_{N}}\left(\sqrt{N}\Delta A_{N,\tau_{N}},\sqrt{N}\Delta B_{N,\tau_{N}}\right)^{2}.
\end{align*}
The term $\mathbb{V}_{\tau_{N}}\left(\sqrt{N}\Delta A_{N,\tau_{N}}\right)\mathbb{V}_{\tau_{N}}\left(\sqrt{N}\Delta B_{N,\tau_{N}}\right)\xrightarrow[N\to\infty]{}v^{2}$,
for some $v^{2}>0$, under \textbf{B1}, with $\gamma=2+\delta$ for
any $\delta$. Now by Cauchy-Schwartz, we have 
\[
\mathbb{C}_{\tau_{N}}\left(\sqrt{N}\Delta A_{N,\tau_{N}},\sqrt{N}\Delta B_{N,\tau_{N}}\right)^{2}\leq\mathbb{V}_{\tau_{N}}\left(\sqrt{N}\Delta A_{N,\tau_{N}}\right)\mathbb{V}_{\tau_{N}}\left(\sqrt{N}\Delta B_{N,\tau_{N}}\right)\xrightarrow[N\to\infty]{}v^{2}
\]
under similar assumptions. Finally, we have 
\[
\mathbb{C}_{\tau_{N}}\left(N\left(\Delta A_{N,\tau_{N}}\right)^{2},N\left(\Delta B_{N,\tau_{N}}\right)^{2}\right)\leq\left(\mathbb{V}_{\tau_{N}}\left(N\left(\Delta A_{N,\tau_{N}}\right)^{2}\right)\mathbb{V}_{\tau_{N}}\left(N\left(\Delta B_{N,\tau_{N}}\right)^{2}\right)\right)^{1/2}.
\]
By the continuous mapping theorem, $N\left(\Delta A_{N,\tau_{N}}\right)^{2}$
and $N\left(\Delta B_{N,\tau_{N}}\right)^{2}$ are asymptotically
distributed according to a chi-squared distribution, hence $\mathbb{V}_{\tau_{N}}\left(N\left(\Delta A_{N,\tau_{N}}\right)^{2}\right)=\mathcal{O}\left(1\right)$
and $\mathbb{V}_{\tau_{N}}\left(N\left(\Delta B_{N,\tau_{N}}\right)^{2}\right)=\mathcal{O}\left(1\right)$
under \textbf{B1} with $\gamma=4+\delta$ for any $\delta$. Thus
we obtain $\mathbb{V}_{\tau_{N}}\left(\Delta A_{N,\tau_{N}}\Delta B_{N,\tau_{N}}\right)=\mathcal{O}(N^{-2})$.

The term $\mathbb{C}_{\tau_{N}}\left(\Delta X-\Delta Y,\Delta X\Delta Y\right)$
in Eq. \eqref{eq:koop:variance} can be written 
\begin{align*}
 & \mathbb{C}_{\tau_{N}}\left(\Delta B_{N,\tau_{N}}-\Delta A_{N,\tau_{N}},\Delta A_{N,\tau_{N}}\Delta B_{N,\tau_{N}}\right)\\
 & =\frac{1}{N\sqrt{N}}\mathbb{C}_{\tau_{N}}\left(\sqrt{N}\left(\Delta B_{N,\tau_{N}}-\Delta A_{N,\tau_{N}}\right),\sqrt{N}\Delta A_{N,\tau_{N}}\sqrt{N}\Delta B_{N,\tau_{N}}\right)
\end{align*}
where 
\begin{align*}
 & \mathbb{C}_{\tau_{N}}\left(\sqrt{N}\left(\Delta B_{N,\tau_{N}}-\Delta A_{N,\tau_{N}}\right),\sqrt{N}\Delta A_{N,\tau_{N}}\sqrt{N}\Delta B_{N,\tau_{N}}\right)^{2}\\
 & \leq\mathbb{V}_{\tau_{N}}\left(\sqrt{N}\left(\Delta B_{N,\tau_{N}}-\Delta A_{N,\tau_{N}}\right)\right)\mathbb{V}_{\tau_{N}}\left(\sqrt{N}\Delta A_{N,\tau_{N}}\sqrt{N}\Delta B_{N,\tau_{N}}\right).
\end{align*}
We have already controlled $\mathbb{V}_{\tau_{N}}\left(\sqrt{N}\Delta A_{N,\tau_{N}}\sqrt{N}\Delta B_{N,\tau_{N}}\right)$,
which is $\mathcal{O}(1)$, and $\mathbb{V}_{\tau_{N}}\left(\sqrt{N}\left(\Delta B_{N,\tau_{N}}-\Delta A_{N,\tau_{N}}\right)\right)$
goes to zero under \textbf{B1}, with $\gamma=2+\delta$ for any $\delta$,
by a delta method argument as in Eq. \eqref{eq:failed:deltamethod}\textbf{.
}Thus we obtain 
\[
\mathbb{C}_{\tau_{N}}\left(\Delta B_{N,\tau_{N}}-\Delta A_{N,\tau_{N}},\Delta A_{N,\tau_{N}}\Delta B_{N,\tau_{N}}\right)=o\left(\frac{1}{N\sqrt{N}}\right).
\]

The term $\mathbb{V}_{\tau_{N}}\left(\frac{Y}{X}\left(\Delta X\right)^{2}\right)$
in Eq. \eqref{eq:koop:variance} can be written 
\begin{align*}
\mathbb{V}_{\tau_{N}}\left(\frac{A_{N,\tau_{N}}}{B_{N,\tau_{N}}}\left(\Delta B_{N,\tau_{N}}\right)^{2}\right) & =\frac{1}{N^{2}}\mathbb{V}_{\tau_{N}}\left(\frac{A_{N,\tau_{N}}}{B_{N,\tau_{N}}}\left(\sqrt{N}\Delta B_{N,\tau_{N}}\right)^{2}\right)
\end{align*}
and 
\begin{align*}
\mathbb{V}_{\tau_{N}}\left(\frac{A_{N,\tau_{N}}}{B_{N,\tau_{N}}}\left(\sqrt{N}\Delta B_{N,\tau_{N}}\right)^{2}\right) & =\mathbb{C}_{\tau_{N}}\left(\left(\frac{A_{N,\tau_{N}}}{B_{N,\tau_{N}}}\right)^{2},\left(\sqrt{N}\Delta B_{N,\tau_{N}}\right)^{4}\right)\\
 & +\mathbb{V}_{\tau_{N}}\left(\frac{A_{N,\tau_{N}}}{B_{N,\tau_{N}}}\right)\mathbb{V}_{\tau_{N}}\left(\left(\sqrt{N}\Delta B_{N,\tau_{N}}\right)^{2}\right)\\
 & -\mathbb{C}_{\tau_{N}}\left(\frac{A_{N,\tau_{N}}}{B_{N,\tau_{N}}},\left(\sqrt{N}\Delta B_{N,\tau_{N}}\right)^{2}\right)^{2}.
\end{align*}
We have $\mathbb{V}_{\tau_{N}}\left(\frac{A_{N,\tau_{N}}}{B_{N,\tau_{N}}}\right)\to0$
by Eq. \eqref{eq:failed:deltamethod} under \textbf{B2} with $\gamma=2+\delta$
for any $\delta$\textbf{. }We also have $\mathbb{V}_{\tau_{N}}\left(\left(\sqrt{N}\Delta B_{N,\tau_{N}}\right)^{2}\right)$
in $\mathcal{O}\left(1\right)$ under \textbf{B1}, with $\gamma=4+\delta$
for any $\delta$,\textbf{ }as $\left(\sqrt{N}\Delta B_{N,\tau_{N}}\right)^{2}$
converges towards a chi-squared distribution thanks to the continuous
mapping theorem. Hence we have $\mathbb{V}_{\tau_{N}}\left(\frac{A_{N,\tau_{N}}}{B_{N,\tau_{N}}}\right)\mathbb{V}_{\tau_{N}}\left(\left(\sqrt{N}\Delta B_{N,\tau_{N}}\right)^{2}\right)=o\left(1\right)$.
Now using Cauchy-Schwartz, we have 
\[
\mathbb{C}_{\tau_{N}}\left(\frac{A_{N,\tau_{N}}}{B_{N,\tau_{N}}},\left(\sqrt{N}\Delta B_{N,\tau_{N}}\right)^{2}\right)^{2}\leq\mathbb{V}_{\tau_{N}}\left(\frac{A_{N,\tau_{N}}}{B_{N,\tau_{N}}}\right)\mathbb{V}_{\tau_{N}}\left(\left(\sqrt{N}\Delta B_{N,\tau_{N}}\right)^{2}\right)=o\left(1\right).
\]
Finally we have 
\[
\mathbb{C}_{\tau_{N}}\left(\left(\frac{A_{N,\tau_{N}}}{B_{N,\tau_{N}}}\right)^{2},\left(\sqrt{N}\Delta B_{N,\tau_{N}}\right)^{4}\right)^{2}\leq\mathbb{V}_{\tau_{N}}\left(\left(\frac{A_{N,\tau_{N}}}{B_{N,\tau_{N}}}\right)^{2}\right)\mathbb{V}_{\tau_{N}}\left(\left(\sqrt{N}\Delta B_{N,\tau_{N}}\right)^{4}\right)
\]
Once more we have $\mathbb{V}_{\tau_{N}}\left(\left(\frac{A_{N,\tau_{N}}}{B_{N,\tau_{N}}}\right)^{2}\right)\xrightarrow[N\to\infty]{}0$
by Eq. \eqref{eq:failed:deltamethod} under \textbf{B2, }with $\gamma=4+\delta$
for any $\delta$,\textbf{ }and $\mathbb{V}_{\tau_{N}}\left(\left(\sqrt{N}\Delta B_{N,\tau_{N}}\right)^{4}\right)=\mathcal{O}\left(1\right)$
as $\left(\sqrt{N}\Delta B_{N,\tau_{N}}\right)^{4}$ converges to
a distribution which is the square of a chi-squared if $\mathbb{E}\left(\left(\sqrt{N}\Delta B_{N,\tau_{N}}\right)^{8+\delta}\right)<\infty$,
and hence, under \textbf{B1}. We have thus proved:
\[
\mathbb{V}_{\tau_{N}}\left(\frac{A_{N,\tau_{N}}}{B_{N,\tau_{N}}}\left(\Delta B_{N,\tau_{N}}\right)^{2}\right)=o\left(\frac{1}{N^{2}}\right).
\]

We have to control the last term $\mathbb{C}\left(\frac{Y}{X},\frac{Y}{X}\left(\Delta X\right)^{2}\right)$
in Eq. \eqref{eq:koop:variance}, which can be written 
\begin{align*}
 & \mathbb{C}_{\tau_{N}}\left(\frac{A_{N,\tau_{N}}}{B_{N,\tau_{N}}},\frac{A_{N,\tau_{N}}}{B_{N,\tau_{N}}}\left(\Delta B_{N,\tau_{N}}\right)^{2}\right)\\
 & =\frac{1}{N\sqrt{N}}\mathbb{C}_{\tau_{N}}\left(\sqrt{N}\frac{A_{N,\tau_{N}}}{B_{N,\tau_{N}}},\frac{A_{N,\tau_{N}}}{B_{N,\tau_{N}}}\left(\sqrt{N}\Delta B_{N,\tau_{N}}\right)^{2}\right).
\end{align*}
By Cauchy Schwartz, 
\[
\mathbb{C}_{\tau_{N}}\left(\sqrt{N}\frac{A_{N,\tau_{N}}}{B_{N,\tau_{N}}},\frac{A_{N,\tau_{N}}}{B_{N,\tau_{N}}}\left(\sqrt{N}\Delta B_{N,\tau_{N}}\right)^{2}\right)^{2}\leq\mathbb{V}_{\tau_{N}}\left(\sqrt{N}\frac{A_{N,\tau_{N}}}{B_{N,\tau_{N}}}\right)\mathbb{V}_{\tau_{N}}\left(\frac{A_{N,\tau_{N}}}{B_{N,\tau_{N}}}\left(\sqrt{N}\Delta B_{N,\tau_{N}}\right)^{2}\right).
\]
Now we have $\mathbb{V}_{\tau_{N}}\left(\sqrt{N}\frac{A_{N,\tau_{N}}}{B_{N,\tau_{N}}}\right)=o\left(1\right)$
by Eq. \eqref{eq:failed:deltamethod} under \textbf{B2} with $\gamma=2+\delta$.
We have already controlled $\mathbb{V}_{\tau_{N}}\left(\frac{A_{N,\tau_{N}}}{B_{N,\tau_{N}}}\left(\sqrt{N}\Delta B_{N,\tau_{N}}\right)^{2}\right)$.
Thus 
\[
\mathbb{C}_{\tau_{N}}\left(\frac{A_{N,\tau_{N}}}{B_{N,\tau_{N}}},\frac{A_{N,\tau_{N}}}{B_{N,\tau_{N}}}\left(\Delta B_{N,\tau_{N}}\right)^{2}\right)=o\left(\frac{1}{N\sqrt{N}}\right).
\]

Hence we can conclude that under the given assumptions, 
\[
\mathbb{V}_{\tau_{N}}\left(\frac{A_{N,\tau_{N}}}{B_{N,\tau_{N}}}\right)=\frac{\tau_{N}^{2}}{N}\Sigma\left(1+\upsilon_{M}\left(\theta^{\star}\right)\right)+\mathcal{O}\left(\frac{\tau_{N}^{4}}{N}\right)+o\left(\frac{1}{N\sqrt{N}}\right).
\]
To make sure that the leading term is indeed in $\mathcal{O}(\tau_{N}^{2}N^{-1})$,
we assume that $N^{-1/4}=o(\tau_{N})$, and thus $N^{-3/2}=o(\tau_{N}^{2}N^{-1})$.
Hence the variance of $A_{N,\tau_{N}}/B_{N,\tau_{N}}$ satisfies 
\begin{equation}
\mathbb{V}_{\tau_{N}}\left(\frac{A_{N,\tau_{N}}}{B_{N,\tau_{N}}}\right)=\frac{\tau_{N}^{2}}{N}\Sigma\left(1+\upsilon_{M}\left(\theta^{\star}\right)\right)+o\left(\frac{\tau_{N}^{2}}{N}\right).\label{eq:asymptoticvariance:AoverB}
\end{equation}
As a result, 
\[
\mathbb{V}_{\tau_{N}}\left[S_{N,\tau_{N}}^{(1)}\left(\theta^{\star}\right)\right]=\frac{1}{\tau_{N}^{2}N}\Sigma^{-1}\left(1+\upsilon_{M}\left(\theta^{\star}\right)\right)+o\left(\frac{1}{\tau_{N}^{2}N}\right),
\]
which is Eq. \eqref{eq:MCshift:asymptoticvariance}.

We now look at the expectation of $A_{N,\tau_{N}}/B_{N,\tau_{N}}$.
Using Lemma \ref{lemma:koop}, we can write 
\begin{align*}
\mathbb{E}\left(\frac{A_{N,\tau_{N}}}{B_{N,\tau_{N}}}\right) & =\frac{\mathbb{E}_{\tau_{N}}\left[\mathcal{L}\left(\Theta\right)\Theta\right]}{\mathbb{E}_{\tau_{N}}\left[\mathcal{L}\left(\Theta\right)\right]}-\frac{1}{N\mathbb{E}_{\tau_{N}}\left[\mathcal{L}\left(\Theta\right)\right]}\mathbb{C}_{\tau_{N}}\left(\sqrt{N}\frac{A_{N,\tau_{N}}}{B_{N,\tau_{N}}},\sqrt{N}B_{N,\tau_{N}}\right).
\end{align*}
We have by Cauchy-Schwartz
\begin{align*}
\left\vert \frac{1}{\mathbb{E}_{\tau_{N}}\left[\mathcal{L}\left(\Theta\right)\right]}\mathbb{C}_{\tau_{N}}\left(\sqrt{N}\frac{A_{N,\tau_{N}}}{B_{N,\tau_{N}}},\sqrt{N}B_{N,\tau_{N}}\right)\right\vert  & =\left\vert \mathbb{C}_{\tau_{N}}\left(\sqrt{N}\frac{A_{N,\tau_{N}}}{B_{N,\tau_{N}}},\sqrt{N}\Delta B_{N,\tau_{N}}\right)\right\vert \\
 & \leq\left(\mathbb{V}_{\tau_{N}}\left(\sqrt{N}\frac{A_{N,\tau_{N}}}{B_{N,\tau_{N}}}\right)\mathbb{V}_{\tau_{N}}\left(\sqrt{N}\Delta B_{N,\tau_{N}}\right)\right)^{1/2}
\end{align*}
where 
\[
\mathbb{V}_{\tau_{N}}\left(\sqrt{N}\frac{A_{N,\tau_{N}}}{B_{N,\tau_{N}}}\right)=\mathcal{O}\left(\tau_{N}^{2}\right),
\]
according to Eq. \eqref{eq:asymptoticvariance:AoverB}, and $\mathbb{V}_{\tau_{N}}\left(\sqrt{N}\Delta B_{N,\tau_{N}}\right)=\mathcal{O}(1)$.
Hence we can write 
\begin{align*}
\mathbb{E}_{\tau_{N}}\left(\frac{A_{N,\tau_{N}}}{B_{N,\tau_{N}}}\right) & =\frac{\mathbb{E}_{\tau_{N}}\left[\mathcal{L}\left(\Theta\right)\Theta\right]}{\mathbb{E}_{\tau_{N}}\left[\mathcal{L}\left(\Theta\right)\right]}+\mathcal{O}\left(\frac{\tau_{N}}{N}\right).
\end{align*}
Combining this with Theorem \ref{theorem:estimators}, we have 
\begin{align*}
\mathbb{E}_{\tau_{N}}\left[S_{N,\tau_{N}}^{(1)}\left(\theta^{\star}\right)\right]-\nabla\ell(\theta^{\star}) & =\tau_{N}^{-2}\Sigma^{-1}\left(\frac{\mathbb{E}_{\tau_{N}}\left[\mathcal{L}\left(\Theta\right)\Theta\right]}{\mathbb{E}_{\tau_{N}}\left[\mathcal{L}\left(\Theta\right)\right]}+\mathcal{O}\left(\frac{\tau_{N}}{N}\right)-\theta^{\star}\right)-\nabla\ell(\theta^{\star})\\
 & =\nabla\ell(\theta^{\star})+\tau_{N}^{2}\mbox{\ensuremath{\Sigma}}\left(\frac{1}{2}\nabla^{3}\ell(\theta^{\star})+2\nabla^{2}\ell(\theta^{\star})\nabla\ell(\theta^{\star})\right)-\nabla\ell(\theta^{\star})+\mathcal{O}\left(\frac{\tau_{N}^{-1}}{N}\right)+\mathcal{O}\left(\tau_{N}^{4}\right)\\
 & =\tau_{N}^{2}\mbox{\ensuremath{\Sigma}}\left(\frac{1}{2}\nabla^{3}\ell(\theta^{\star})+2\nabla^{2}\ell(\theta^{\star})\nabla\ell(\theta^{\star})\right)+\mathcal{O}\left(\frac{\tau_{N}^{-1}}{N}\right)+\mathcal{O}\left(\tau_{N}^{4}\right),
\end{align*}
where the Monte Carlo bias is in $\mathcal{O}(\tau_{N}^{-1}N^{-1})$
and the ``systematic bias'' is in $\mathcal{O}(\tau_{N}^{2})$.
To make sure that the leading term is in $\tau_{N}^{2}$, we need
$\tau_{N}^{-1}N^{-1}$ small against $\tau_{N}^{2}$, which is guaranteed
as long as $N^{-1/3}=o(\tau_{N})$. This gives the bias of $S_{N,\tau_{N}}^{(1)}\left(\theta^{\star}\right)$
as in Eq. \eqref{eq:MCshift:asymptoticbias}. The bias and the variance
lead to the mean squared error:
\[
\tau_{N}^{4}\left(\mbox{\ensuremath{\Sigma}}\left(\frac{1}{2}\nabla^{3}\ell(\theta^{\star})+2\nabla^{2}\ell(\theta^{\star})\nabla\ell(\theta^{\star})\right)\right)^{2}+\frac{\tau_{N}^{-2}}{N}\left(1+\upsilon_{M}\left(\theta^{\star}\right)\right)+\text{remainder}(\tau_{N},N),
\]
which is optimized by choosing $\tau_{N}=N^{-1/6}$, making both $\tau_{N}^{4}$
and $\tau_{N}^{2}N^{-1}$ of order $N^{-2/3}$.

\end{proof}

We now consider the bias and variance of the estimator $S_{N,\tau_{N}}^{(2)}\left(\theta^{\star}\right)$
of $\nabla^{2}\ell(\theta^{\star})$, as proposed in Eq. \eqref{eq:second:shift:estimator}.
A formal proof of Lemma \ref{lemma:biasvarianceMCshift2} would require
following the same steps as the proof of Lemma \ref{lemma:biasvarianceMCshift1},
with a number of terms that can be systematically controlled using
Cauchy-Schwartz inequalities and moment assumptions. Instead, we provide
a sketch of the main intermediate steps. 

\begin{proof}[Informal proof of Lemma \ref{lemma:biasvarianceMCshift2}]We
consider the variables
\[
E_{N,\tau_{N}}=\frac{1}{N}\sum_{i=1}^{N}\widehat{\mathcal{L}}(\theta^{i})\left(\theta^{i}-\frac{A_{N,\tau_{N}}}{B_{N,\tau_{N}}}\right)^{2},
\]
so that we can write 
\[
S_{N,\tau_{N}}^{(2)}\left(\theta^{\star}\right)=\tau_{N}^{-4}\Sigma^{-2}\left(\frac{E_{N,\tau_{N}}}{B_{N,\tau_{N}}}-\tau_{N}^{2}\Sigma\right).
\]
The intuition of the result is that the relative bias of normalized
importance sampling estimators is in $N^{-1}$, and that the object
being estimated here, which is the posterior variance, is of order
$\tau_{N}^{2}$. By following the same steps as Lemma \ref{lemma:biasvarianceMCshift1},
we should thus obtain an absolute bias of order $\tau_{N}^{2}/N$:
\[
\mathbb{E}_{\tau_{N}}\left(\frac{E_{N,\tau_{N}}}{B_{N,\tau_{N}}}\right)=\frac{\mathbb{E}_{\tau_{N}}\left(\left(\Theta-\frac{\mathbb{E}_{\tau_{N}}\left[\mathcal{L}\left(\Theta\right)\Theta\right]}{\mathbb{E}_{\tau_{N}}\left[\mathcal{L}\left(\Theta\right)\right]}\right)^{2}\mathcal{L}(\Theta)\right)}{\mathbb{E}_{\tau_{N}}\left(\mathcal{L}(\Theta)\right)}+\mathcal{O}\left(\frac{\tau_{N}^{2}}{N}\right),
\]
and thus by using Theorem \ref{theorem:estimators}, 
\[
\mathbb{E}_{\tau_{N}}S_{N,\tau_{N}}^{(2)}\left(\theta^{\star}\right)-\nabla^{2}\ell(\theta^{\star})=\tau_{N}^{2}\mathcal{F}\left(\theta^{\star}\right)+\mathcal{O}\left(\frac{\tau_{N}^{-2}}{N}\right),
\]
where $\mathcal{F}(\theta^{\star})$ is given by Eq. \eqref{eq:hessian:approx:error}.
This assumes that $\tau_{N}^{-2}N^{-1}$ is small in front of $\tau_{N}^{2}$,
i.e. $N^{-1/4}=o(\tau_{N})$. The relative variance of normalized
importance sampling estimators being typically in $N^{-1}$, we expect
to find an absolute variance of order $\tau_{N}^{4}/N$:
\[
\mathbb{V}_{\tau_{N}}\left(\frac{E_{N,\tau_{N}}}{B_{N,\tau_{N}}}\right)=\frac{\tau_{N}^{4}}{N}C_{M}\left(\theta^{\star}\right)+o\left(\frac{\tau_{N}^{4}}{N}\right),
\]
for some constant $C_{M}\left(\theta^{\star}\right)$ that does not
depend on $\tau_{N}$ nor on $N$. The mean squared error would then
be dominated by
\[
\tau_{N}^{4}\mathcal{F}(\theta^{\star})^{2}+\frac{\tau_{N}^{-4}}{N}C_{M}\left(\theta^{\star}\right)
\]
which is minimized in $\tau_{N}$ by choosing $\tau_{N}=N^{-1/8}$,
and yields a mean squared error of order $N^{-1/2}$. 

\end{proof}

\subsection{Variance reduction using control variates (Lemma \ref{lemma:controlvariates})
\label{sub:proof:controlvariates}}

\begin{proof}[Informal proof of Lemma \ref{lemma:controlvariates}]

We introduce the random variables
\[
C_{N,\tau_{N}}=\frac{1}{N}\sum_{i=1}^{N}\theta^{i},
\]
which satisfy $\mathbb{E}_{\tau_{N}}\left[C_{N,\tau_{N}}\right]=\theta^{\star}$
and $\mathbb{V}_{\tau_{N}}\left[C_{N,\tau_{N}}\right]=\tau_{N}^{2}\Sigma N^{-1}$.
We can now write
\[
\widetilde{S}_{N,\tau_{N}}^{(1)}(\theta^{\star})=\tau_{N}^{-2}\Sigma^{-1}\left(\frac{A_{N,\tau_{N}}}{B_{N,\tau_{N}}}-C_{N,\tau_{N}}\right).
\]
We are interested in the variance of this estimator. We can write
the variance of $A_{N,\tau_{N}}/B_{N,\tau_{N}}-C_{N,\tau_{N}}$ in
a similar expression as Eq. \eqref{eq:koop:variance} of Lemma \ref{lemma:koop}.
Indeed we can write, for variables $X$,$Y$ and $Z$, following the
proof of Lemma \ref{lemma:koop}, 
\begin{align*}
\frac{Y}{X}-Z= & \frac{\mu_{Y}}{\mu_{X}}+\left(\frac{\mu_{Y}}{\mu_{X}}\left(\Delta Y-\Delta X\right)-Z\right)-\frac{\mu_{Y}}{\mu_{X}}\Delta X\Delta Y+\frac{Y}{X}\left(\Delta X\right)^{2}
\end{align*}
and thus
\begin{align*}
\mathbb{V}\left(\frac{Y}{X}-Z-\frac{Y}{X}\left(\Delta X\right)^{2}\right)= & \mathbb{V}\left(\frac{\mu_{Y}}{\mu_{X}}\left(\Delta Y-\Delta X\right)-Z\right)+\mathbb{V}\left(\frac{\mu_{Y}}{\mu_{X}}\Delta X\Delta Y\right)\\
 & -2\mathbb{C}\left(\frac{\mu_{Y}}{\mu_{X}}\left(\Delta Y-\Delta X\right)-Z,\frac{\mu_{Y}}{\mu_{X}}\Delta X\Delta Y\right),
\end{align*}
and finally
\[
\mathbb{V}\left(\frac{Y}{X}-Z\right)=\mathbb{V}\left(\frac{\mu_{Y}}{\mu_{X}}\left(\Delta Y-\Delta X\right)-Z\right)+R(X,Y,Z).
\]
where $R(X,Y,Z)$ denotes all the terms required for the equality
to hold:
\begin{align*}
R(X,Y,Z)= & \mathbb{V}\left(\frac{\mu_{Y}}{\mu_{X}}\Delta X\Delta Y\right)-2\mathbb{C}\left(\frac{\mu_{Y}}{\mu_{X}}\left(\Delta Y-\Delta X\right)-Z,\frac{\mu_{Y}}{\mu_{X}}\Delta X\Delta Y\right),\\
 & -\mathbb{V}\left(\frac{Y}{X}\left(\Delta X\right)^{2}\right)+2\mathbb{C}\left(\frac{Y}{X}-Z,\frac{Y}{X}\left(\Delta X\right)^{2}\right).
\end{align*}
The addition of the term $Z\equiv C_{N,\tau_{N}}$ thus incurs more
terms to control. By-passing this tedious exercise, we directly assume
that the leading term in the variance is
\[
\mathbb{V}\left(\frac{\mu_{Y}}{\mu_{X}}\left(\Delta Y-\Delta X\right)-Z\right),
\]
that is, 
\[
\mathbb{V}_{\tau_{N}}\left(\left(\frac{\mathbb{E}_{\tau_{N}}\left[A_{N,\tau_{N}}\right]}{\mathbb{E}_{\tau_{N}}\left[B_{N,\tau_{N}}\right]}\right)\left(\Delta A_{N,\tau_{N}}-\Delta B_{N,\tau_{N}}\right)-C_{N,\tau_{N}}\right).
\]
This variance can be written

\begin{eqnarray*}
 &  & \left(\frac{\mathbb{E}_{\tau_{N}}\left[A_{N,\tau_{N}}\right]}{\mathbb{E}_{\tau_{N}}\left[B_{N,\tau_{N}}\right]}\right)^{2}\mathbb{V}_{\tau_{N}}\left(\Delta A_{N,\tau_{N}}-\Delta B_{N,\tau_{N}}\right)+\mathbb{V}_{\tau_{N}}\left(C_{N,\tau_{N}}\right)-2\frac{\mathbb{E}_{\tau_{N}}\left[A_{N,\tau_{N}}\right]}{\mathbb{E}_{\tau_{N}}\left[B_{N,\tau_{N}}\right]}\mathbb{C}_{\tau_{N}}\left(\Delta A_{N,\tau_{N}}-\Delta B_{N,\tau_{N}},C_{N,\tau_{N}}\right).
\end{eqnarray*}
We have already computed the first term in Eq. \eqref{eq:asymptoticvariance:term1}.
We also have $\mathbb{V}_{\tau_{N}}\left[C_{N,\tau_{N}}\right]=\tau_{N}^{2}\Sigma N^{-1}$.
Finally, for the third term, we note that 
\begin{align*}
\frac{\mathbb{E}_{\tau_{N}}\left[A_{N,\tau_{N}}\right]}{\mathbb{E}_{\tau_{N}}\left[B_{N,\tau_{N}}\right]} & =\left(\theta^{\star}+\mathcal{O}\left(\tau_{N}^{2}\right)\right),\\
\mathbb{C}_{\tau_{N}}\left(\Delta A_{N,\tau_{N}}-\Delta B_{N,\tau_{N}},C_{N,\tau_{N}}\right) & =\mathbb{E}_{\tau_{N}}\left(\frac{A_{N,\tau_{N}}C_{N,\tau_{N}}}{\mathbb{E}_{\tau_{N}}\left[A_{N,\tau_{N}}\right]}\right)-\mathbb{E}_{\tau_{N}}\left(\frac{B_{N,\tau_{N}}C_{N,\tau_{N}}}{\mathbb{E}_{\tau_{N}}\left[B_{N,\tau_{N}}\right]}\right).
\end{align*}
We compute first
\[
\mathbb{E}_{\tau_{N}}\left(A_{N,\tau_{N}}C_{N,\tau_{N}}\right)=\frac{1}{N}\mathbb{E}_{\tau_{N}}\left[\mathcal{L}\left(\Theta\right)\Theta^{2}\right]+\left(1-\frac{1}{N}\right)\mathbb{E}_{\tau_{N}}\left[\mathcal{L}\left(\Theta\right)\Theta\right]\mathbb{E}_{\tau_{N}}\left[\Theta\right],
\]
and then 
\[
\mathbb{E}_{\tau_{N}}\left(B_{N,\tau_{N}}C_{N,\tau_{N}}\right)=\frac{1}{N}\mathbb{E}_{\tau_{N}}\left[\mathcal{L}\left(\Theta\right)\Theta\right]+\left(1-\frac{1}{N}\right)\mathbb{E}_{\tau_{N}}\left[\mathcal{L}\left(\Theta\right)\right]\mathbb{E}_{\tau_{N}}\left[\Theta\right],
\]
so that 
\[
\mathbb{C}_{\tau_{N}}\left(\Delta A_{N,\tau_{N}}-\Delta B_{N,\tau_{N}},C_{N,\tau_{N}}\right)=\frac{1}{N}\left(\frac{\mathbb{E}_{\tau_{N}}\left[\mathcal{L}\left(\Theta\right)\Theta^{2}\right]}{\mathbb{E}_{\tau_{N}}\left[\mathcal{L}\left(\Theta\right)\Theta\right]}-\frac{\mathbb{E}_{\tau_{N}}\left[\mathcal{L}\left(\Theta\right)\Theta\right]}{\mathbb{E}_{\tau_{N}}\left[\mathcal{L}\left(\Theta\right)\right]}\right).
\]
We can then use Eq. \eqref{eq:expansion:twotestfunctions} to compute
\[
\mathbb{C}_{\tau_{N}}\left(\Delta A_{N,\tau_{N}}-\Delta B_{N,\tau_{N}},C_{N,\tau_{N}}\right)=\frac{1}{N}\left(\frac{\tau_{N}^{2}\Sigma}{\theta^{\star}}+\mathcal{O}\left(\tau_{N}^{4}\right)\right),
\]
which leads to the desired expression for the variance for $\widetilde{S}_{N,\tau_{N}}^{(1)}(\theta^{\star})$.

\end{proof}

\subsection{Effect of the dimension (Lemma \ref{lemma:biasvarianceMCshift-dimension})
\label{sub:proof:dimension}}

\begin{proof}[Proof of Lemma \ref{lemma:biasvarianceMCshift-dimension}]

We follow the proof of Lemma \ref{lemma:biasvarianceMCshift1} in
Section \ref{sub:proof:shift:MSE}. We work element-wise, for each
component of $S_{N,\tau_{N}}^{(1)}\left(\theta^{\star}\right)$. For
the bias, the leading term is the systematic bias of Theorem \ref{theorem:estimators},
which is given in Eq. \eqref{eq:score:approx:error}. This directly
yields Eq. \eqref{eq:MCshift:asymptoticbias-dimension}.

For the variance, element-wise, we see from the proof of Lemma \ref{lemma:biasvarianceMCshift1}
that we can compute the leading term as
\begin{align*}
 & \mathbb{E}_{\tau_{N}}\left(\left(\frac{\Theta_{k}\mathcal{L}(\Theta)}{\mathbb{E}_{\tau_{N}}\left[\Theta_{k}\mathcal{L}(\Theta)\right]}-\frac{\mathcal{L}(\Theta)}{\mathbb{E}_{\tau_{N}}\left[\mathcal{L}(\Theta)\right]}\right)^{2}\right)\\
= & \frac{\mathbb{E}_{\tau_{N}}\left[\Theta_{k}^{2}\mathcal{L}(\Theta)^{2}\right]}{\mathbb{E}_{\tau_{N}}\left[\Theta_{k}\mathcal{L}(\Theta)\right]^{2}}+\frac{\mathbb{E}_{\tau_{N}}\left[\mathcal{L}(\Theta)^{2}\right]}{\mathbb{E}_{\tau_{N}}\left[\mathcal{L}(\Theta)\right]^{2}}-2\frac{\mathbb{E}_{\tau_{N}}\left[\Theta_{k}\mathcal{L}(\Theta)^{2}\right]}{\mathbb{E}_{\tau_{N}}\left[\Theta_{k}\mathcal{L}(\Theta)\right]\mathbb{E}_{\tau_{N}}\left[\mathcal{L}(\Theta)\right]},
\end{align*}
where $k\in\left\{ 1,\ldots,d\right\} $ denotes a component index.
Using Eq. \eqref{eq:expansion:twotestfunctions}, we can compute 
\begin{align*}
\frac{\mathbb{E}_{\tau_{N}}\left[\Theta_{k}^{2}\mathcal{L}(\Theta)^{2}\right]}{\mathbb{E}_{\tau_{N}}\left[\Theta_{k}\mathcal{L}(\Theta)\right]^{2}}= & 1+\tau_{N}^{2}\sum_{i=1}^{d}\sum_{j=1}^{d}\Sigma_{ij}\left(\frac{\delta_{i=k}}{\theta_{k}^{\star}}+\nabla_{i}\ell(\theta^{\star})\right)\left(\frac{\delta_{j=k}}{\theta_{k}^{\star}}+\nabla_{j}\ell(\theta^{\star})\right)+\mathcal{O}\left(\tau_{N}^{4}\right),\\
\frac{\mathbb{E}_{\tau_{N}}\left[\mathcal{L}(\Theta)^{2}\right]}{\mathbb{E}_{\tau_{N}}\left[\mathcal{L}(\Theta)\right]^{2}}= & 1+\tau_{N}^{2}\sum_{i=1}^{d}\sum_{j=1}^{d}\Sigma_{ij}\nabla_{i}\ell(\theta^{\star})\nabla_{j}\ell(\theta^{\star})+\mathcal{O}\left(\tau_{N}^{4}\right),\\
\frac{\mathbb{E}_{\tau_{N}}\left[\Theta_{k}\mathcal{L}(\Theta)^{2}\right]}{\mathbb{E}_{\tau_{N}}\left[\Theta_{k}\mathcal{L}(\Theta)\right]\mathbb{E}_{\tau_{N}}\left[\mathcal{L}(\Theta)\right]}= & 1+\tau_{N}^{2}\sum_{i=1}^{d}\sum_{j=1}^{d}\Sigma_{ij}\left(\frac{\delta_{i=k}}{\theta_{k}^{\star}}+\nabla_{i}\ell(\theta^{\star})\right)\nabla_{j}\ell(\theta^{\star})+\mathcal{O}\left(\tau_{N}^{4}\right).
\end{align*}
In the above equations, $\delta_{i=k}$ equals one if $i=k$ and zero
otherwise. Thus we obtain
\begin{align*}
 & \mathbb{E}_{\tau_{N}}\left(\left(\frac{\Theta_{k}\mathcal{L}(\Theta)}{\mathbb{E}_{\tau_{N}}\left[\Theta_{k}\mathcal{L}(\Theta)\right]}-\frac{\mathcal{L}(\Theta)}{\mathbb{E}_{\tau_{N}}\left[\mathcal{L}(\Theta)\right]}\right)^{2}\right)\\
= & \tau_{N}^{2}\sum_{i=1}^{d}\sum_{j=1}^{d}\Sigma_{ij}\left(\frac{\delta_{i=k}\delta_{j=k}}{\theta_{k}^{\star2}}+\frac{\delta_{j=k}}{\theta_{k}^{\star}}\nabla_{i}\ell(\theta^{\star})-\frac{\delta_{i=k}}{\theta_{k}^{\star}}\nabla_{j}\ell(\theta^{\star})\right)\\
= & \tau_{N}^{2}\frac{\Sigma_{kk}}{\theta_{k}^{\star2}},
\end{align*}
which leads to Eq. \eqref{eq:MCshift:asymptoticvariance-dimension}.

For the covariance terms,
\[
\mathbb{C}_{\tau_{N}}\left(\left\{ S_{N,\tau_{N}}^{(1)}\left(\theta^{\star}\right)\right\} _{k},\left\{ S_{N,\tau_{N}}^{(1)}\left(\theta^{\star}\right)\right\} _{l}\right)
\]
for $k,l\in\left\{ 1,\ldots,d\right\} $, we could go back to the
proof of Lemma \ref{lemma:koop}, and we see that we can write, for
variables $X$,$Y_{1}$ and $Y_{2}$, 
\[
\mathbb{C}\left(\frac{Y_{1}}{X},\frac{Y_{2}}{X}\right)=\frac{\mu_{Y_{1}}}{\mu_{X}}\frac{\mu_{Y_{2}}}{\mu_{X}}\mathbb{C}\left(\Delta Y_{1}-\Delta X,\,\Delta Y_{2}-\Delta X\right)+R(X,Y_{1},Y_{2}).
\]
The remainder term $R(X,Y_{1},Y_{2})$ could be controlled as in the
proof of Lemma \ref{lemma:biasvarianceMCshift1}. The leading term
satisfies
\begin{align*}
\frac{\mu_{Y_{1}}}{\mu_{X}}\frac{\mu_{Y_{2}}}{\mu_{X}}\mathbb{C}\left(\Delta Y_{1}-\Delta X,\,\Delta Y_{2}-\Delta X\right) & =\frac{\mu_{Y_{1}}}{\mu_{X}}\frac{\mu_{Y_{2}}}{\mu_{X}}\mathbb{E}\left(\left(\Delta Y_{1}-\Delta X\right)\left(\Delta Y_{2}-\Delta X\right)\right)\\
 & =\frac{\mu_{Y_{1}}}{\mu_{X}}\frac{\mu_{Y_{2}}}{\mu_{X}}\left(\mathbb{E}\left(\left(\frac{Y_{1}}{\mu_{Y_{1}}}-\frac{X}{\mu_{X_{1}}}\right)\left(\frac{Y_{2}}{\mu_{Y_{2}}}-\frac{X}{\mu_{X_{1}}}\right)\right)\right).
\end{align*}
Denote by $\theta_{k}^{i}$ the $k$-th component of the draw $\theta^{i}$
from $\mathcal{N}(\theta^{\star},\tau_{N}^{2}\Sigma)$. With $Y_{1}=N^{-1}\sum_{i=1}^{N}\widehat{\mathcal{L}}(\theta^{i})\theta_{k}^{i}$,
$Y_{2}=N^{-1}\sum_{i=1}^{N}\widehat{\mathcal{L}}(\theta^{i})\theta_{l}^{i}$
and $X=N^{-1}\sum_{i=1}^{N}\widehat{\mathcal{L}}(\theta^{i})$, we
could use the above equation to obtain the expressions of the covariance
terms, following the same ideas as for the marginal variance terms.\end{proof}

\subsection{Robustness of Monte Carlo shift estimators (Lemma \ref{lemma:robustness})
\label{sub:proof:robustness}}

\begin{proof}[Proof of Lemma \ref{lemma:robustness}] We consider
the univariate setting for simplicity. Denote by $\theta^{1},\ldots,\theta^{N}$
the generated samples from $\mathcal{N}(\theta^{\star},\tau^{2}\Sigma)$.
Denote the convex hull of $\theta^{1},\ldots,\theta^{N}$ by $\text{Conv}\left(\theta^{1},\ldots,\theta^{N}\right)$,
and the diameter of the convex hull by $D(\theta^{1},\ldots,\theta^{N})$:
\[
D\left(\theta^{1},\ldots,\theta^{N}\right)=\sup_{x,y\in\text{Conv}\left(\theta^{1},\ldots,\theta^{N}\right)}\left|\left|x-y\right|\right|_{2},
\]
where $||x||_{2}$ denotes the Euclidean norm of a vector $x$. Conditional
upon $\theta^{1},\ldots,\theta^{N}$, since $(\hat{W}^{1},\ldots\hat{W}^{N})$
belongs to the $N$-dimensional simplex almost surely, then $\sum_{i=1}^{N}\hat{W}^{i}\theta^{i}\in\text{Conv}\left(\theta^{1},\ldots,\theta^{N}\right)$
and $\mathbb{E}\left[\sum_{i=1}^{N}\hat{W}^{i}\theta^{i}\mid\theta^{1},\ldots,\theta^{N}\right]$
is in $\text{Conv}\left(\theta^{1},\ldots,\theta^{N}\right)$ almost
surely. The average squared distance between two points being less
than the maximum squared distance, we have
\[
\mathbb{V}\left[\sum_{i=1}^{N}\hat{W}^{i}\theta^{i}\mid\theta^{1},\ldots,\theta^{N}\right]\leq D(\theta^{1},\ldots,\theta^{N})^{2},
\]
almost surely, which yields
\[
\mathbb{E}_{\tau}\left[\mathbb{V}\left[\sum_{i=1}^{N}\hat{W}^{i}\theta^{i}\mid\theta^{1},\ldots,\theta^{N}\right]\right]\leq\mathbb{E}_{\tau}\left[D(\theta^{1},\ldots,\theta^{N})^{2}\right],
\]
Furthermore, since $\mathbb{E}\left[\sum_{i=1}^{N}\hat{W}^{i}\theta^{i}\mid\theta^{1},\ldots,\theta^{N}\right]\in\text{Conv}\left(\theta^{1},\ldots,\theta^{N}\right)$
almost surely, its variance satisfies 
\[
\mathbb{V}_{\tau}\left[\mathbb{E}\left[\sum_{i=1}^{N}\hat{W}^{i}\theta^{i}\mid\theta^{1},\ldots,\theta^{N}\right]\right]\leq\mathbb{E}_{\tau}\left[D(\theta^{1},\ldots,\theta^{N})^{2}\right].
\]
Thus we obtain 
\[
\mathbb{V}\left[\sum_{i=1}^{N}\hat{W}^{i}\theta^{i}\right]\leq2\mathbb{E}_{\tau}\left[D\left(\theta^{1},\ldots,\theta^{N}\right)^{2}\right].
\]
We conclude by noting that $\mathbb{E}_{\tau}\left[D\left(\theta^{1},\ldots,\theta^{N}\right)^{2}\right]=\tau^{2}\bar{D}^{2}$,
where $\bar{D}^{2}$ is the expected squared diameter of the convex
hull of $N$ normal variables with unit variance.

\end{proof}

\subsection{Finite difference schemes (Lemmas \ref{lemma:biasvarianceFD} and
\ref{lemma:biasvarianceSPSA-dimension}) \label{sub:proof:Finite-difference}}

\begin{proof}[Informal proof of Lemma \ref{lemma:biasvarianceFD}]

Let $k\in\left\{ 1,\ldots,d\right\} $. Since the log-likelihood estimator
is unbiased, 
\[
\mathbb{E}\left[\frac{\log\widehat{\mathcal{L}}\left(\theta^{\star}+h_{M}e_{k}\right)-\log\widehat{\mathcal{L}}\left(\theta^{\star}-h_{M}e_{k}\right)}{2h_{M}}\right]=\frac{\ell(\theta^{\star}+h_{M}e_{k})-\ell(\theta^{\star}-h_{M}e_{k})}{2h_{M}}.
\]
Writing a Taylor expansion in the multi-index notation of Section
\ref{sub:proof:prior:expansions}, for all $h_{M}<\delta$ where $\delta$
is defined in Assumption \textbf{C1},
\begin{align*}
\ell(\theta^{\star}+h_{M}e_{k}) & =\ell(\theta^{\star})+\sum_{\left|\alpha\right|\leq3}\frac{\partial^{\alpha}\ell\left(\theta^{\star}\right)}{\alpha!}\left(h_{M}e_{k}\right)^{\alpha}+R_{3}(\theta^{\star}+h_{M}e_{k},\theta^{\star}),
\end{align*}
where we use the Lagrange form of the remainder:
\[
\forall\theta\in B(\theta^{\star},\delta)\quad\exists c_{\theta}\in\left[0,1\right]\quad R_{3}(\theta,\theta^{\star})=\sum_{\left|\alpha\right|=4}\partial^{\alpha}\ell\left(\theta^{\star}+c_{\theta}\left(\theta-\theta^{\star}\right)\right)\frac{\left(\theta-\theta^{\star}\right)^{\alpha}}{\alpha!},
\]
where $B(\theta^{\star},\delta)$ is the Euclidean ball of radius
$\delta$ around $\theta^{\star}$. Thus, we obtain 
\begin{align*}
\frac{\ell\left(\theta^{\star}+h_{M}e_{k}\right)-\ell\left(\theta^{\star}-h_{M}e_{k}\right)}{2h_{M}} & =\sum_{\left|\alpha\right|=1}\partial^{\alpha}\ell\left(\theta^{\star}\right)e_{k}^{\alpha}+h_{M}^{2}\sum_{\left|\alpha\right|=3}\frac{\partial^{\alpha}\ell\left(\theta^{\star}\right)}{\alpha!}e_{k}^{\alpha}\\
 & +\frac{R_{4}(\theta^{\star}+h_{M}e_{k},\theta^{\star})-R_{4}(\theta^{\star}-h_{M}e_{k},\theta^{\star})}{2h_{M}}.
\end{align*}
We control the remainder using Assumption \textbf{C1}, and conclude
that 
\[
\frac{\ell\left(\theta^{\star}+h_{M}e_{k}\right)-\ell\left(\theta^{\star}-h_{M}e_{k}\right)}{2h_{M}}=\nabla_{k}\ell\left(\theta^{\star}\right)+\frac{h_{M}^{2}}{6}\nabla_{kkk}^{3}\ell(\theta^{\star})+\mathcal{O}(h_{M}^{4}).
\]
This yields Eq. \eqref{eq:finitedifferencet:asymptoticbias}. The
variance is directly computed as
\[
\mathbb{V}\left[D_{h_{M}}^{(1)}\left(\theta^{\star}\right)\right]=\frac{1}{4h_{M}^{2}}U_{M}\left(\theta^{\star}\right)\left(\ell(\theta^{\star}+h_{M})^{2}+\ell(\theta^{\star}-h_{M})^{2}\right),
\]
since $\log\widehat{\mathcal{L}}\left(\theta^{\star}+h_{M}e_{k}\right)$
and $\log\widehat{\mathcal{L}}\left(\theta^{\star}-h_{M}e_{k}\right)$
are independent. Therefore, if $U_{M}\left(\theta^{\star}\right)=U\left(\theta^{\star}\right)/M$,
we obtain a squared bias in $h_{M}^{4}$ and a variance in $h_{M}^{-2}M^{-1}$,
leading to an optimal choice $h_{M}=M^{-1/6}$.

\end{proof}

\begin{proof}[Informal proof of Lemma \ref{lemma:biasvarianceSPSA-dimension}]

Let $k\in\left\{ 1,\ldots,d\right\} $, and $h_{M}<\delta$ where
$\delta$ is defined in Assumption \textbf{C1}. Let $\varepsilon$
be a random variable satisfying Assumption \textbf{C2}. Following
the proof of Lemma \ref{lemma:biasvarianceFD}, we obtain almost surely
the expansion 
\[
\frac{\ell\left(\theta^{\star}+h_{M}\varepsilon\right)-\ell\left(\theta^{\star}-h_{M}\varepsilon\right)}{2h_{M}\varepsilon_{k}}=\varepsilon_{k}^{-1}\sum_{\left|\alpha\right|=1}\partial^{\alpha}\ell\left(\theta^{\star}\right)\varepsilon^{\alpha}p+h_{M}^{2}\varepsilon_{k}^{-1}\sum_{\left|\alpha\right|=3}\frac{\partial^{\alpha}\ell\left(\theta^{\star}\right)}{\alpha!}\varepsilon^{\alpha}+\frac{R_{4}(\theta^{\star}+h_{M}\varepsilon,\theta^{\star})-R_{4}(\theta^{\star}-h_{M}\varepsilon,\theta^{\star})}{2h_{M}\varepsilon_{k}},
\]
where $\varepsilon_{k}$ denotes the $k$-th component of $\varepsilon$.
For the remainder, using Assumption\textbf{ C1} we can obtain a bound
\[
\left|R_{4}(\theta^{\star}+h_{M}\varepsilon,\theta^{\star})-R_{4}(\theta^{\star}-h_{M}\varepsilon,\theta^{\star})\right|\leq C\,h_{M}^{4},
\]
for some constant $C$ that depends on $K$, $\delta$ and $d$. Thus
we can write, for the $k$-th element of the estimator $D_{N,h_{N}}^{(1)}\left(\theta^{\star}\right)$:
\begin{align*}
\mathbb{E}\left[\frac{\log\widehat{\mathcal{L}}(\theta^{\star}+h_{N}\varepsilon)-\log\widehat{\mathcal{L}}(\theta^{\star}-h_{N}\varepsilon)}{2h_{N}\varepsilon_{k}}\right] & =\mathbb{E}\left[\frac{\ell(\theta^{\star}+h_{N}\varepsilon)-\ell(\theta^{\star}-h_{N}\varepsilon)}{2h_{N}\varepsilon_{k}}\right]\\
 & =\nabla_{k}\ell\left(\theta^{\star}\right)+\frac{h_{N}^{2}}{6}\mathbb{E}\left[\varepsilon_{k}^{-1}\sum_{1\leq i_{1},i_{2},i_{3}\leq d}\nabla_{i_{1}i_{2}i_{3}}^{3}\ell(\theta^{\star})\,\varepsilon_{i_{1}}\varepsilon_{i_{2}}\varepsilon_{i_{3}}\right]+\mathcal{O}(h_{N}^{3}),
\end{align*}
where we have used $\mathbb{E}\left[\varepsilon_{i}^{-1}\right]<\infty$
and $\mathbb{E}\left[\varepsilon_{i}\right]=0$ for all $i\in\left\{ 1,\ldots,d\right\} $.
Now we consider the bias term that involves the triple sum
\begin{align*}
\sum_{1\leq i_{1},i_{2},i_{3}\leq d}\nabla_{i_{1}i_{2}i_{3}}^{3}\ell(\theta^{\star})\mathbb{E}\left[\frac{\varepsilon_{i_{1}}\varepsilon_{i_{2}}\varepsilon_{i_{3}}}{\varepsilon_{k}}\right].
\end{align*}
In this triple sum, the only non zero terms correspond to $i_{1}=i_{2}=i_{3}=k$
and the indices satisfying $i_{1}=i_{2}$, $i_{3}=k$ and associated
permutations. Thus if we bound each term in $\nabla^{3}\ell(\theta^{\star})$
by a constant, then there is a constant $C$ such that 
\[
\left|\sum_{1\leq i_{1},i_{2},i_{3}\leq d}\nabla_{i_{1}i_{2}i_{3}}^{3}\ell(\theta^{\star})\mathbb{E}\left[\frac{\varepsilon_{i_{1}}\varepsilon_{i_{2}}\varepsilon_{i_{3}}}{\varepsilon_{k}}\right]\right|\leq C\,d.
\]

For the variance, we have by the decomposition formula,
\begin{align*}
 & \mathbb{V}\left[\left\{ D_{N,h_{N}}^{(1)}\left(\theta^{\star}\right)\right\} _{k}\right]\\
= & \frac{1}{N}\mathbb{V}\left[\frac{\log\widehat{\mathcal{L}}(\theta^{\star}+h_{N}\varepsilon)-\log\widehat{\mathcal{L}}(\theta^{\star}-h_{N}\varepsilon)}{2h_{N}\varepsilon_{k}}\right]\\
= & \frac{1}{N}\mathbb{V}\left[\frac{\ell(\theta^{\star}+h_{N}\varepsilon)-\ell(\theta^{\star}-h_{N}\varepsilon)}{2h_{N}\varepsilon_{k}}\right]+\frac{U_{M}\left(\theta^{\star}\right)}{4Nh_{N}^{2}}\mathbb{E}\left[\frac{\ell(\theta^{\star}+h_{N}\varepsilon)^{2}}{\varepsilon_{k}^{2}}+\frac{\ell(\theta^{\star}-h_{N}\varepsilon)^{2}}{\varepsilon_{k}^{2}}\right]\\
= & \frac{1}{4Nh_{N}^{2}}\mathbb{E}\left[\left(\ell(\theta^{\star}+h_{N}\varepsilon)-\ell(\theta^{\star}-h_{N}\varepsilon)\right)^{2}\right]-\frac{1}{N}\mathbb{E}\left[\frac{\left(\ell(\theta^{\star}+h_{N}\varepsilon)-\ell(\theta^{\star}-h_{N}\varepsilon)\right)}{2h_{N}\varepsilon_{k}}\right]^{2}\\
+ & \frac{U_{M}\left(\theta^{\star}\right)}{4Nh_{N}^{2}}\mathbb{E}\left[\ell(\theta^{\star}+h_{N}\varepsilon)^{2}+\ell(\theta^{\star}-h_{N}\varepsilon)^{2}\right]
\end{align*}
where we have used $\varepsilon_{k}^{2}=1$ almost surely. Since the
log-likelihood $\ell$ is continuous around $\theta^{\star}$, it
can be locally bounded and thus by the bounded convergence theorem
we have
\begin{align*}
\mathbb{E}\left[\left(\ell(\theta^{\star}+h_{N}\varepsilon)-\ell(\theta^{\star}-h_{N}\varepsilon)\right)^{2}\right]\xrightarrow[N\to\infty]{}0,\\
\mathbb{E}\left[\ell(\theta^{\star}+h_{N}\varepsilon)^{2}+\ell(\theta^{\star}-h_{N}\varepsilon)^{2}\right]\xrightarrow[N\to\infty]{}2\ell(\theta^{\star})^{2}.
\end{align*}
Using $1=o(h_{N}^{-2})$, we obtain
\[
\mathbb{V}\left[\left\{ D_{N,h_{N}}^{(1)}\left(\theta^{\star}\right)\right\} _{k}\right]=\frac{U_{M}\left(\theta^{\star}\right)}{2Nh_{N}^{2}}\ell(\theta^{\star})^{2}+o\left(\frac{1}{Nh_{N}^{2}}\right).
\]
We see that the leading term depends on the dimension $d$ through
$U_{M}\left(\theta^{\star}\right)$.

\end{proof}

\section{Autoregressive prior for state-space models\label{AppendixC:ARpriors}}

Assume one selects $\Theta_{1}\sim\mathcal{N}\left(\theta^{\star},\tau^{2}\Sigma\right)$
and for $t\geq1$
\[
\Theta{}_{t+1}-\theta^{\star}=\rho\left(\Theta{}_{t}-\theta^{\star}\right)+V_{t+1},\quad V_{t+1}\sim\mathcal{N}\left(0,\upsilon^{2}\Sigma\right),
\]
where $\rho$ is a scalar, $\left|\rho\right|<1$ and $\tau^{2}=\upsilon^{2}/\left(1-\rho^{2}\right)$.
In this case, $\Theta_{1:T}\sim\mathcal{N}\left(\theta^{\star[T]},\tau^{2}\bar{\Sigma}\right)$
where 
\[
\bar{\Sigma}=\left(\begin{tabular}{llllll}
 \ensuremath{\Sigma}  &  \ensuremath{\rho\Sigma}  &  \ensuremath{\cdots}  &  \ensuremath{\cdots}  &  \ensuremath{\cdots}  &  \ensuremath{\rho^{T-1}\Sigma}\\
\ensuremath{\rho\Sigma}  &  \ensuremath{\Sigma}  &  \ensuremath{\rho\Sigma}  &  \ensuremath{\cdots}  &  \ensuremath{\cdots}  &  \ensuremath{\rho^{T-2}\Sigma}\\
\ensuremath{\rho^{2}\Sigma}  &  \ensuremath{\rho\Sigma}  &  \ensuremath{\Sigma}  &  \ensuremath{\ddots}  &   &  \ensuremath{\rho^{T-3}\Sigma}\\
\ensuremath{\vdots}  &   &  \ensuremath{\ddots}  &  \ensuremath{\ddots}  &  \ensuremath{\ddots}  &  \ensuremath{\vdots}\\
\ensuremath{\rho^{T-2}\Sigma}  &   &   &  \ensuremath{\ddots}  &  \ensuremath{\Sigma}  &  \ensuremath{\rho\Sigma}\\
\ensuremath{\rho^{T-1}\Sigma}  &  \ensuremath{\cdots}  &  \ensuremath{\cdots}  &  \ensuremath{\cdots}  &  \ensuremath{\rho\Sigma}  &  \ensuremath{\Sigma} 
\end{tabular}\right)
\]
 and the shift estimators of Eq. \eqref{eq:extendedshiftestimator}
and Eq. \eqref{eq:extendedshiftestimatorsecondderivative} can be
rewritten after tedious manipulations as 
\begin{eqnarray*}
\bar{S}_{\tau}^{(1)}\left(\theta^{\star}\right) & = & \frac{\tau^{-2}\Sigma^{-1}}{1+\rho}\biggl[\left(1-\rho\right)\sum_{t=2}^{T-1}\mathbb{E_{\tau}}\left[\Theta_{t}\mid y_{1:T}\right]-\left(\left(1-\rho\right)T+2\rho\right)\theta^{\star}\\
 &  & +\,\mathbb{E}_{\tau}\left[\Theta_{1}\mid y_{1:T}\right]+\mathbb{E}_{\tau}\left[\Theta_{T}\mid y_{1:T}\right]\biggr]
\end{eqnarray*}
and 
\begin{eqnarray*}
\bar{S}_{\tau}^{(2)}\left(\theta^{\star}\right) & = & \frac{\tau^{-4}}{\left(1+\rho\right)^{2}}\Sigma^{-1}\biggl[\left(1-\rho\right)^{2}\left(\sum_{s=2}^{T-1}\sum_{t=2}^{T-1}\mathbb{C_{\tau}}\left[\Theta_{s},\Theta_{t}\mid y_{1:T}\right]-\tau^{2}\rho^{\left|t-s\right|}\Sigma\right)\\
 &  & +2\left(1-\rho\right)\left(\left(\sum_{t=2}^{T-1}\mathbb{C_{\tau}}\left[\Theta_{1},\Theta_{t}\mid y_{1:T}\right]-\rho^{t-1}\Sigma\right)+\left(\sum_{t=2}^{T-1}\mathbb{C_{\tau}}\left[\Theta_{t},\Theta_{T}\mid y_{1:T}\right]-\rho^{T-t}\Sigma\right)\right)\\
 &  & +\left(\mathbb{V}_{\tau}\left[\Theta_{1}\mid y_{1:T}\right]-\tau^{2}\Sigma\right)+\left(\mathbb{V}_{\tau}\left[\Theta_{T}\mid y_{1:T}\right]-\tau^{2}\Sigma\right)\\
 &  & +2\left(\mathbb{C_{\tau}}\left[\Theta_{1},\Theta_{T}\mid y_{1:T}\right]-\tau^{2}\rho^{T-1}\Sigma\right)\biggr]\Sigma^{-1}.
\end{eqnarray*}
We note that we have
\begin{eqnarray*}
\underset{\rho\rightarrow1^{-}}{\lim}\bar{S}_{\tau}^{(1)}\left(\theta^{\star}\right) & = & \frac{\tau^{-2}\Sigma^{-1}}{2}\left(\mathbb{E_{\tau}}\left[\Theta_{1}\mid y_{1:T}\right]+\mathbb{E_{\tau}}\left[\Theta_{T}\mid y_{1:T}\right]\right)\\
 & \thickapprox & \tau^{-2}\Sigma^{-1}\mathbb{E_{\tau}}\left[\Theta_{T}\mid y_{1:T}\right],
\end{eqnarray*}
as $\mathbb{E_{\tau}}\left[\Theta_{1}\mid y_{1:T}\right]\thickapprox\mathbb{E_{\tau}}\left[\Theta_{T}\mid y_{1:T}\right]$
because $\upsilon\rightarrow0^{+}$ when $\rho\rightarrow1^{-}$ as
$\tau^{2}=\upsilon^{2}/\left(1-\rho^{2}\right)$. Hence, we retrieve
informally the score estimator proposed in \cite{Ionides09}. Similarly,
we have
\begin{eqnarray*}
\underset{\rho\rightarrow1^{-}}{\lim}\bar{S}_{\tau}^{(2)}\left(\theta^{\star}\right) & = & \frac{\tau^{-4}}{4}\Sigma^{-1}\left[\mathbb{V}_{\tau}\left[\Theta_{1}\mid y_{1:T}\right]-\tau^{2}\Sigma+\mathbb{V}_{\tau}\left[\Theta_{T}\mid y_{1:T}\right]-\tau^{2}\Sigma\right.\\
 &  & \left.+2\left(\mathbb{C_{\tau}}\left[\Theta_{1},\Theta_{T}\mid y_{1:T}\right]-\tau^{2}\Sigma\right)\right]\Sigma^{-1}\\
 & \approx & \tau^{-4}\Sigma^{-1}\left(\mathbb{V}_{\tau}\left[\Theta_{T}\mid y_{1:T}\right]-\tau^{2}\Sigma\right)\Sigma^{-1}.
\end{eqnarray*}

\end{document}